\documentclass[1p]{elsarticle}
\usepackage{enumerate}
\usepackage{amssymb,latexsym,url,amsthm}
\usepackage{subfigure}
\usepackage{array}
\usepackage{enumitem,color}

\definecolor{subtler}{rgb}{1,0,0.1}    

\newcommand{\be}{\begin{equation}}
\newcommand{\ee}{\end{equation}}
\newcommand{\ba}{\begin{eqnarray}}
\newcommand{\ea}{\end{eqnarray}}
\newcommand{\bitem}{\begin{itemize}}
\newcommand{\eitem}{\end{itemize}}

\newcommand{\pnoise}{p_{\mathrm{noise}}}

\newcommand{\rhodl}{\rho_{\ulcorner}}
\newcommand{\rhodr}{\rho_{\urcorner}}
\newcommand{\Jd}{J_{\rm d}}
\newcommand{\rhoc}{\rho_{\rm c}}
\newcommand{\Jc}{J_{\rm c}}
\newcommand{\Jo}{J_{\rm o}}
\newcommand{\EE}{\mathbb{E}}
\newcommand{\alphac}{\alpha_{\rm c}}
\newcommand{\betac}{\beta_{\rm c}}

\newcommand{\ddw}{\rm DDW} 
\newcommand{\sdw}{\rm SDW} 

\def\mP{\mathcal{P}} 
\def\mC{{\cal C}}
\def\mM{{\cal M}}
\def\mU{{\cal U}}
\def\mA{{\cal A}}
\def\mB{{\cal B}}
\def\mF{{\cal F}}

\def\mL{L}
\def\mx{x}
\def\pr{{\mathbb P}}
\def\d{{\rm d}}
\def\pt{p_{\rm T}}

\def\smallest{0.13}
\def\smaller{0.3}
\def\oneup{0.33}

\def\domainwall				{\includegraphics[scale=0.21]{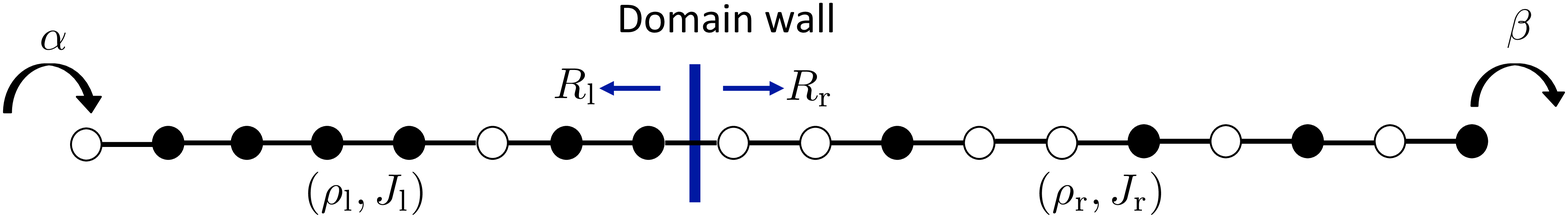}}
\def\phasediagram				{\includegraphics[scale=0.21]{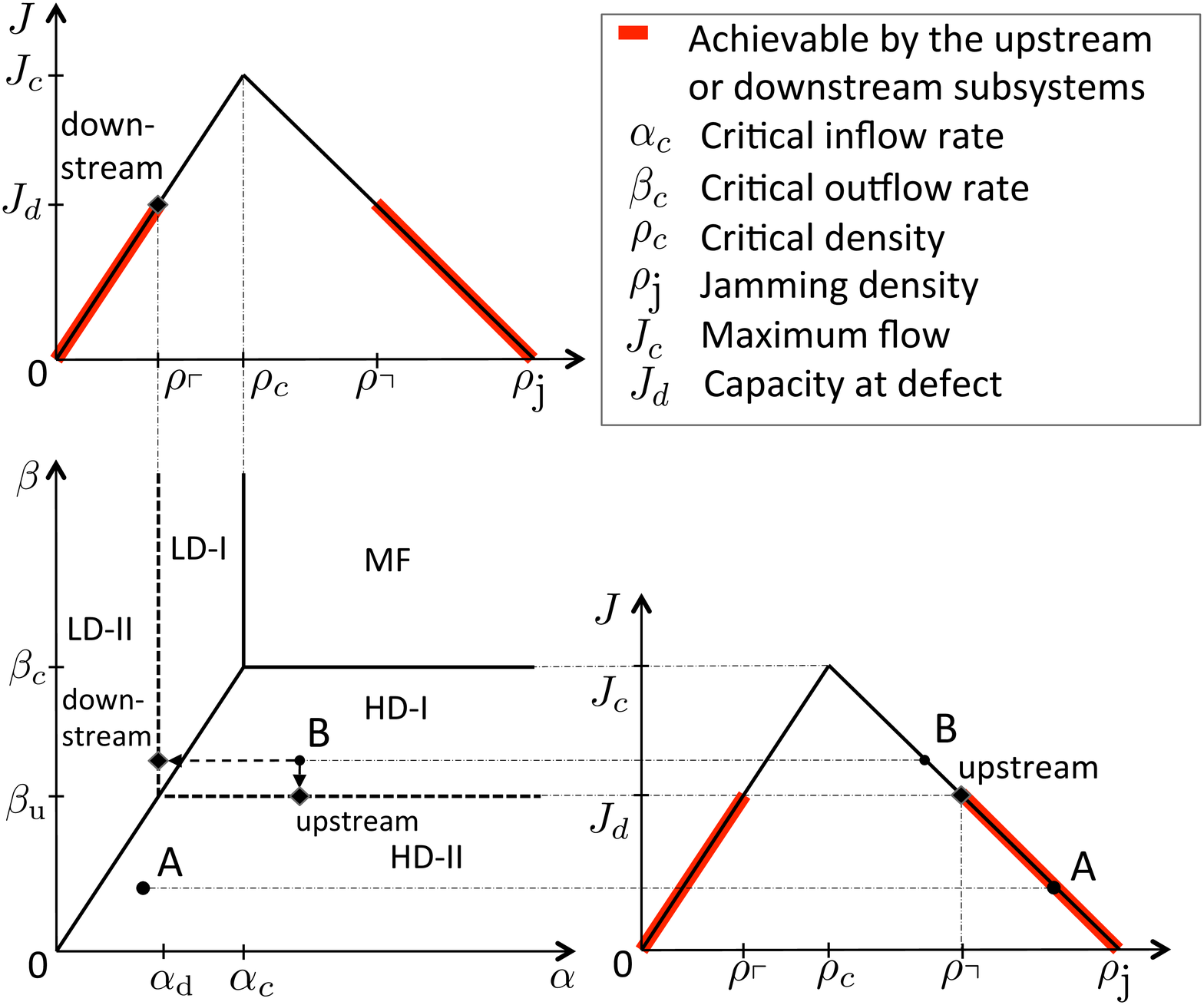}}
\def\LKfds                      {\includegraphics[scale=\oneup]{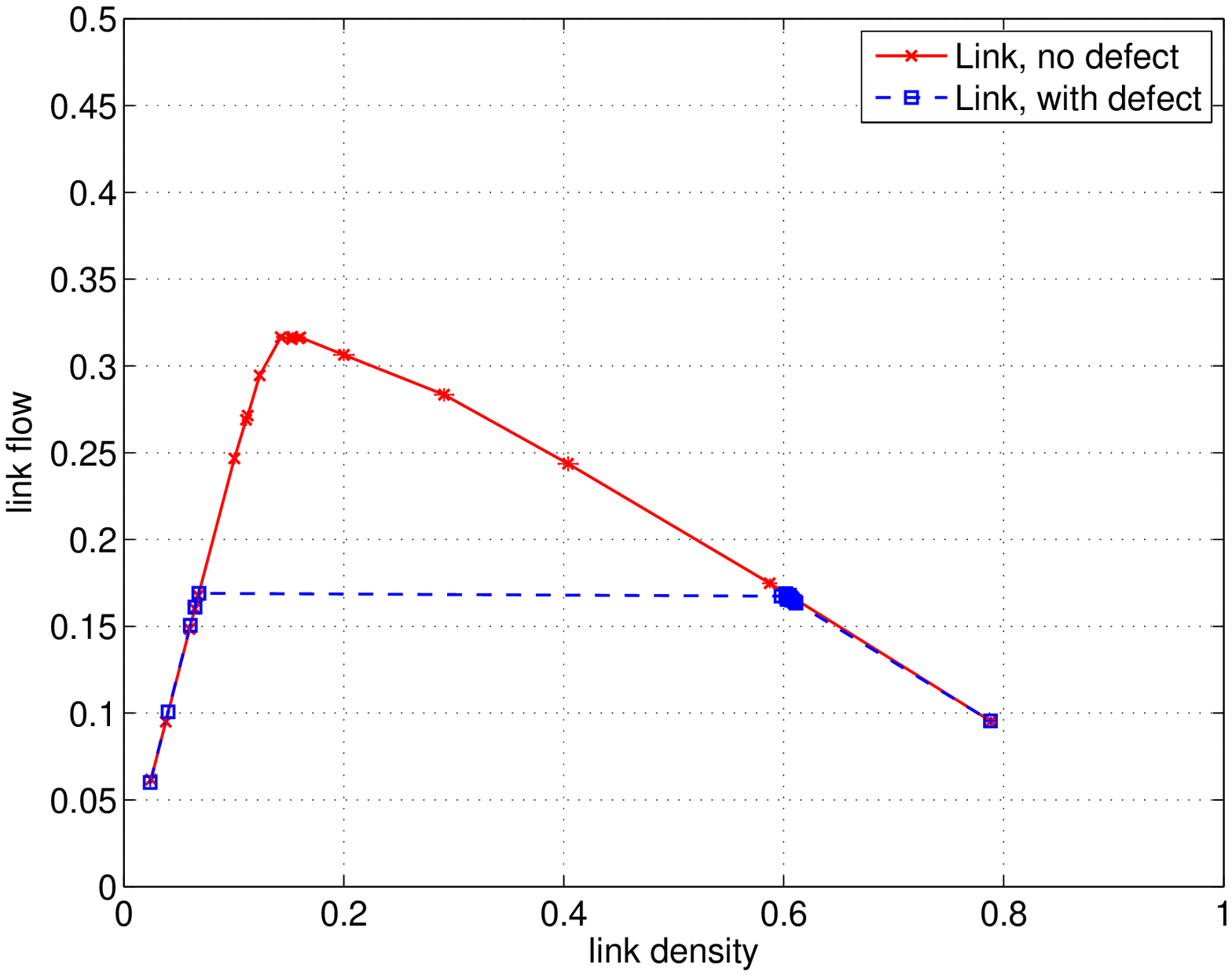}}
\def\LKregiondemo				{\includegraphics[scale=\smaller]{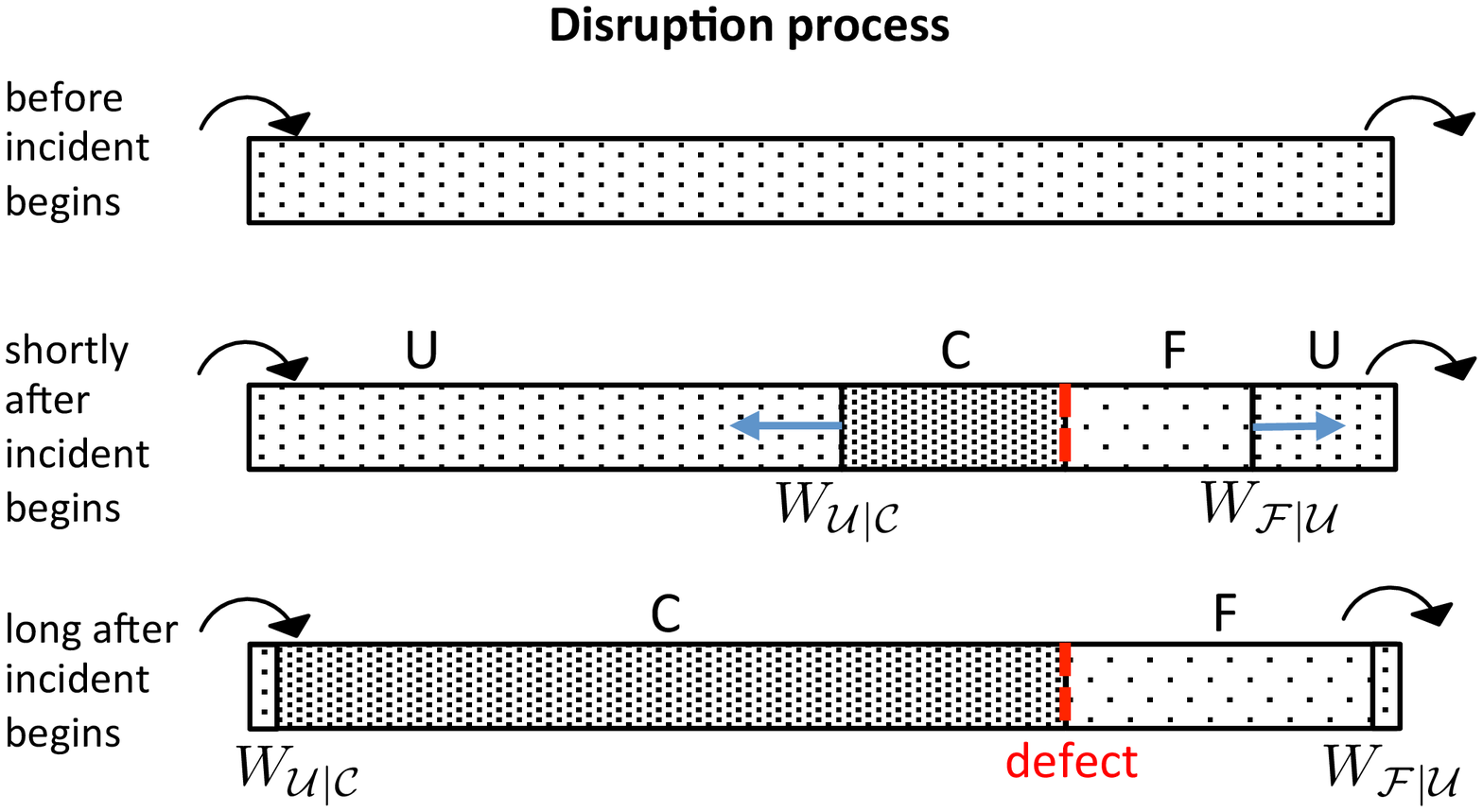}}
\def\LKrecoveryB				{\includegraphics[scale=\smaller]{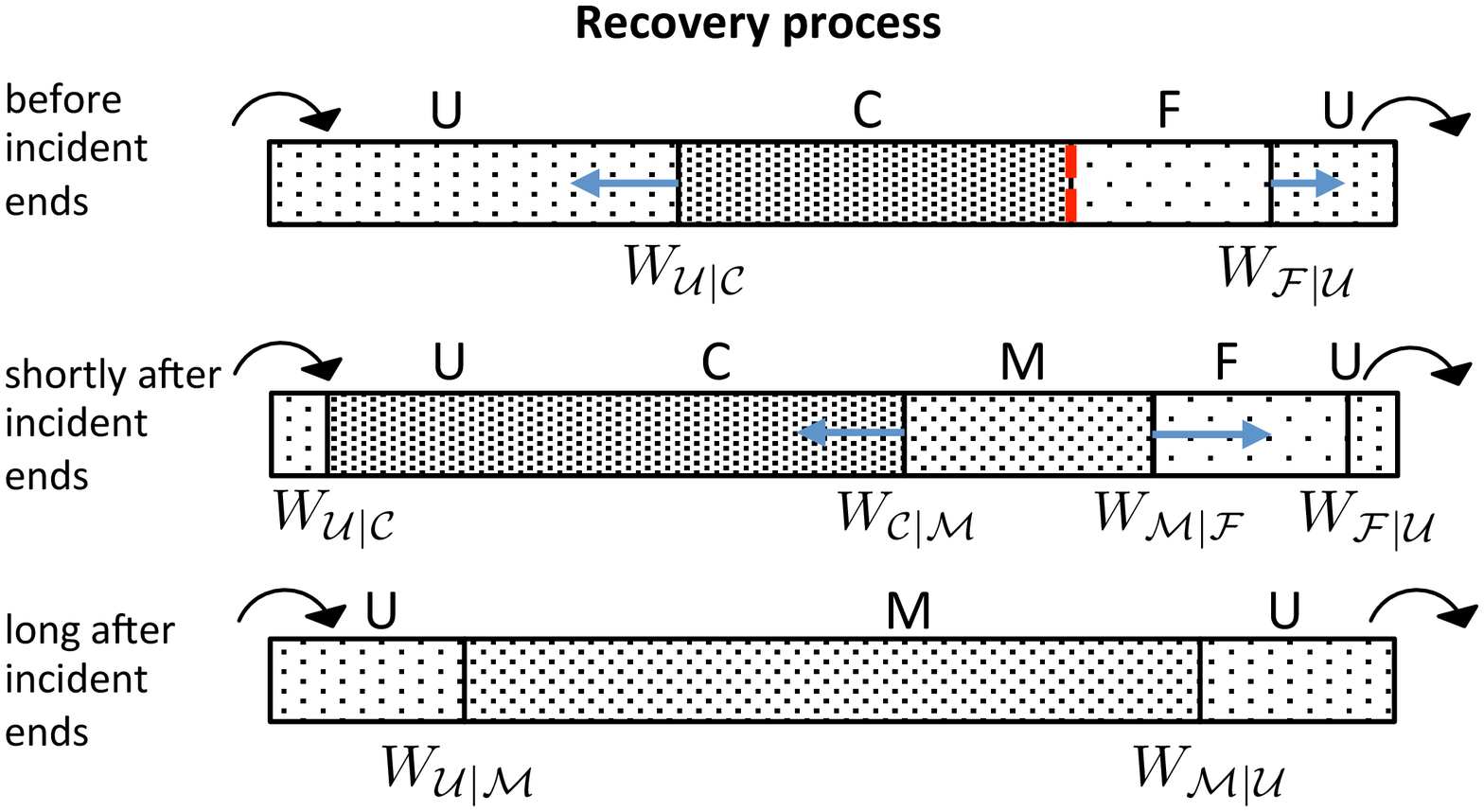}}

\def\dwmodelrouteflowLD			{\includegraphics[scale=\oneup]{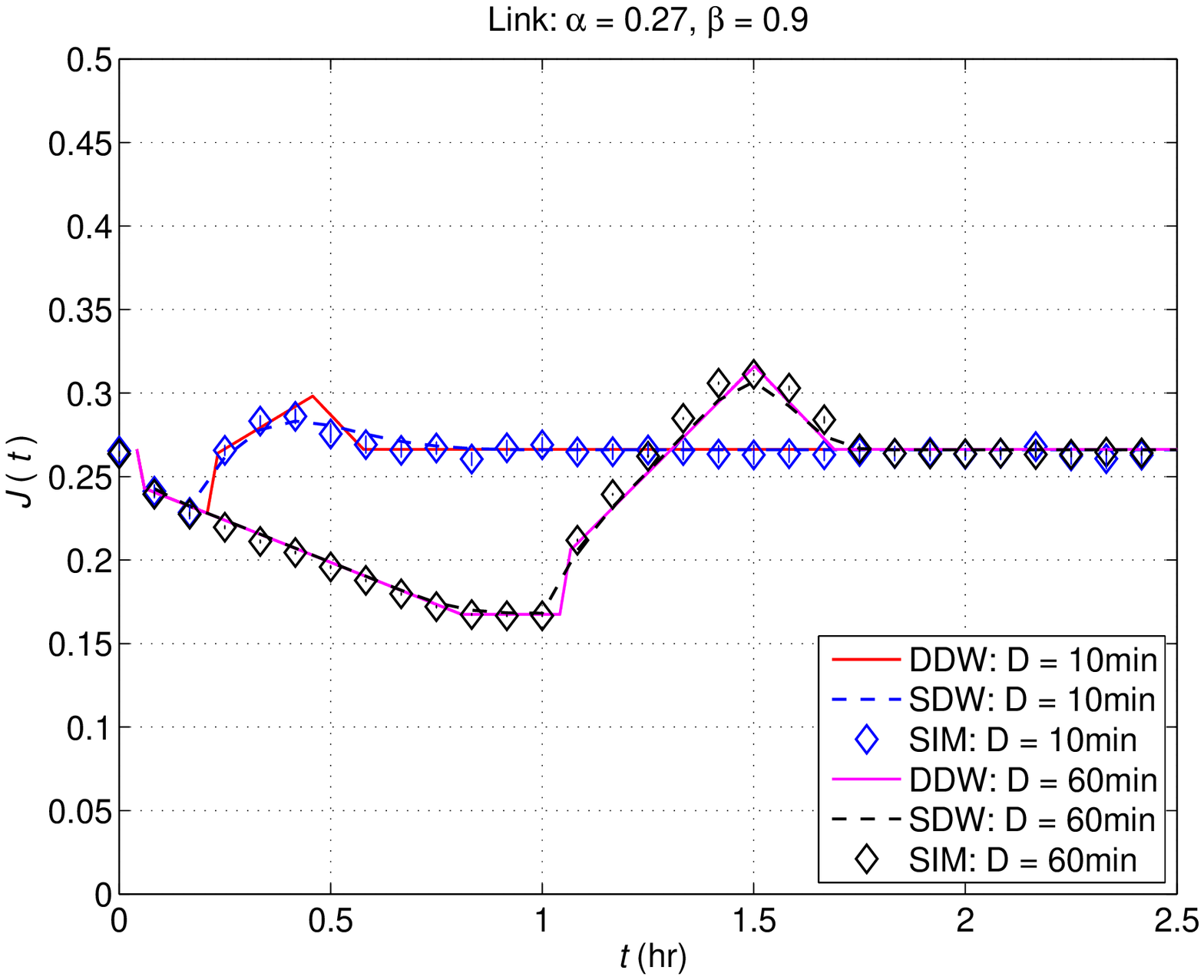}}
\def\dwmodelroutedensityLD		{\includegraphics[scale=\oneup]{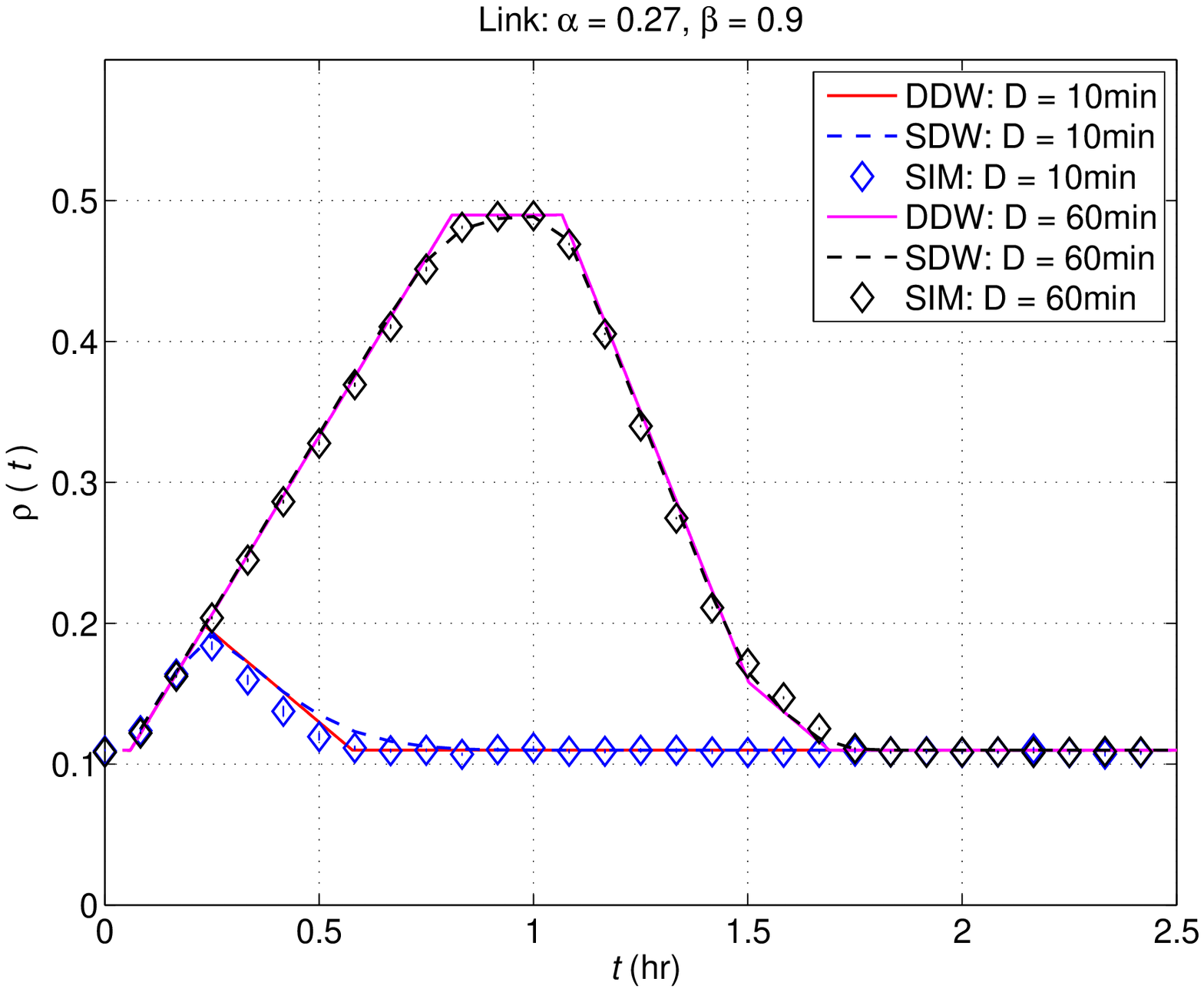}}
\def\dwmodelcelldensityLDL		{\includegraphics[scale=\oneup]{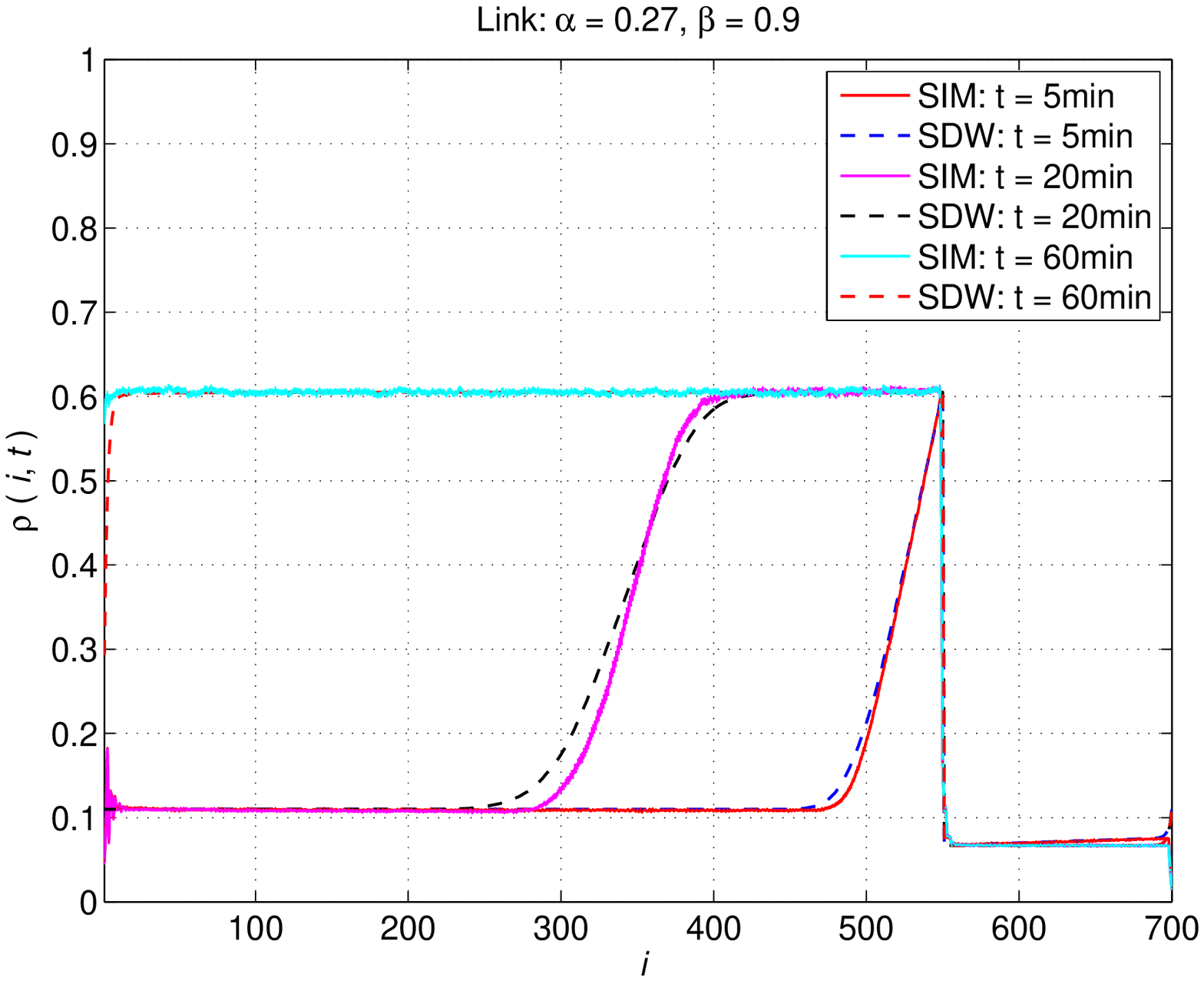}}
\def\dwmodelcelldensityLDR		{\includegraphics[scale=\oneup]{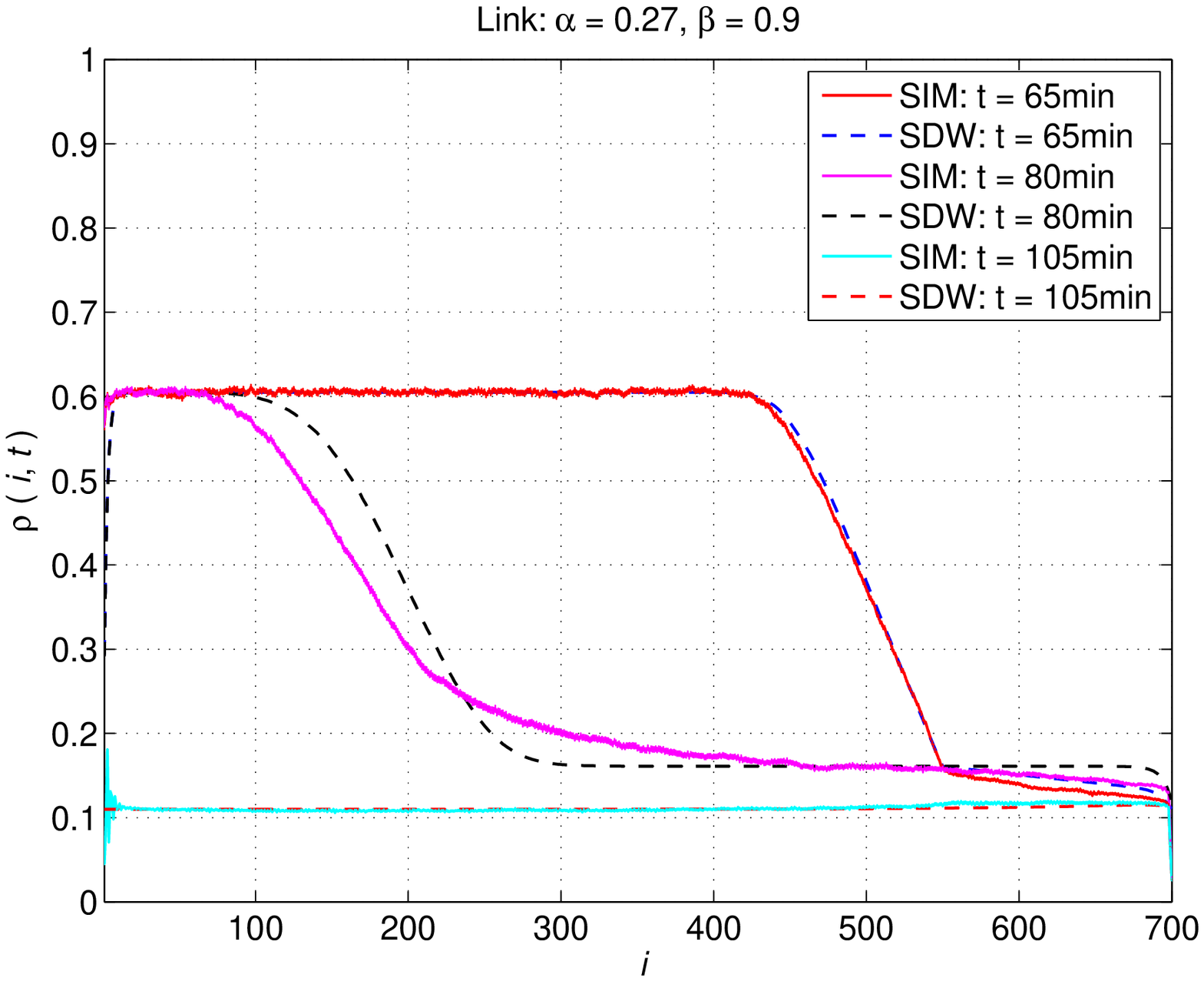}}
\def\dwmodelrouteflowcmp		{\includegraphics[scale=\oneup]{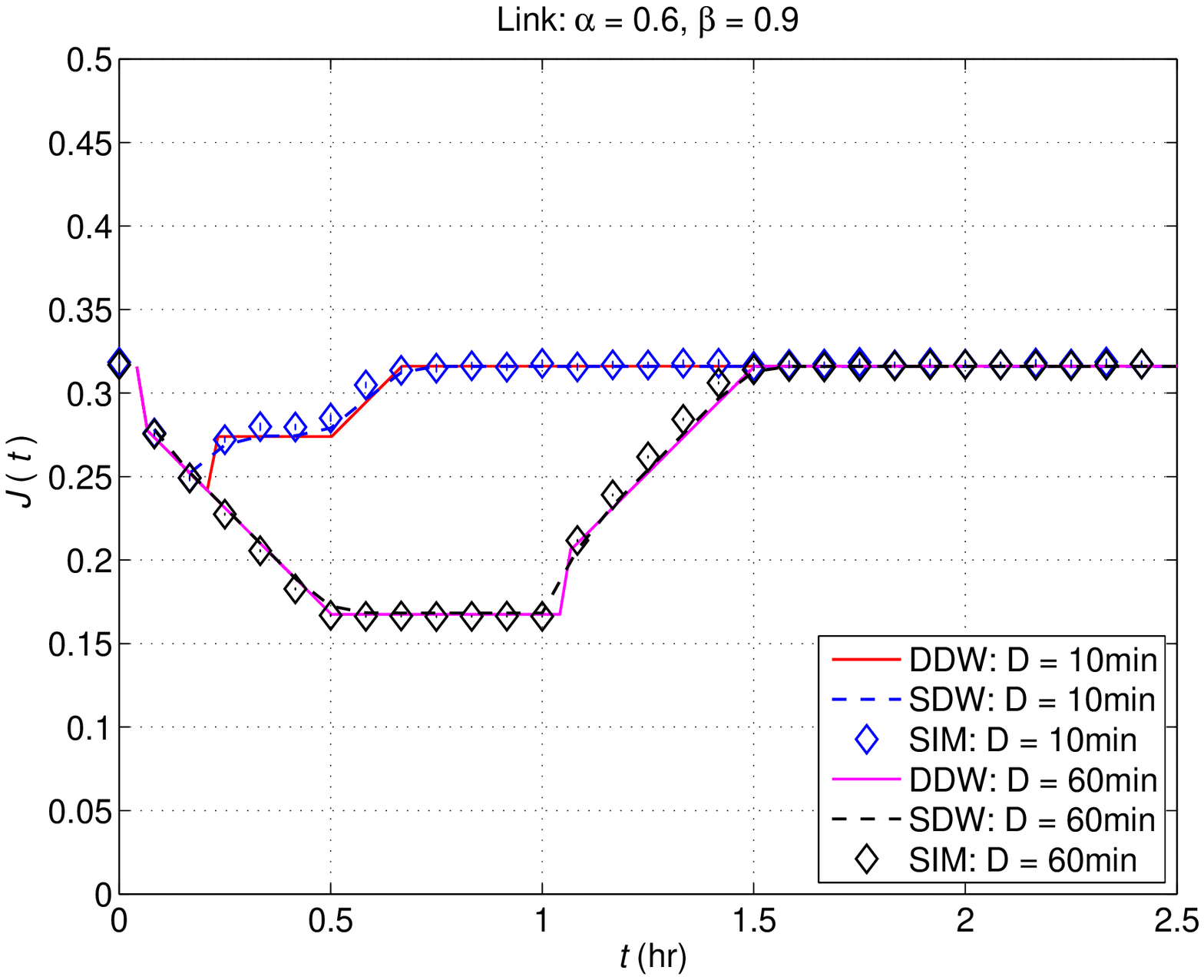}}
\def\dwmodelroutedensitycmp		{\includegraphics[scale=\oneup]{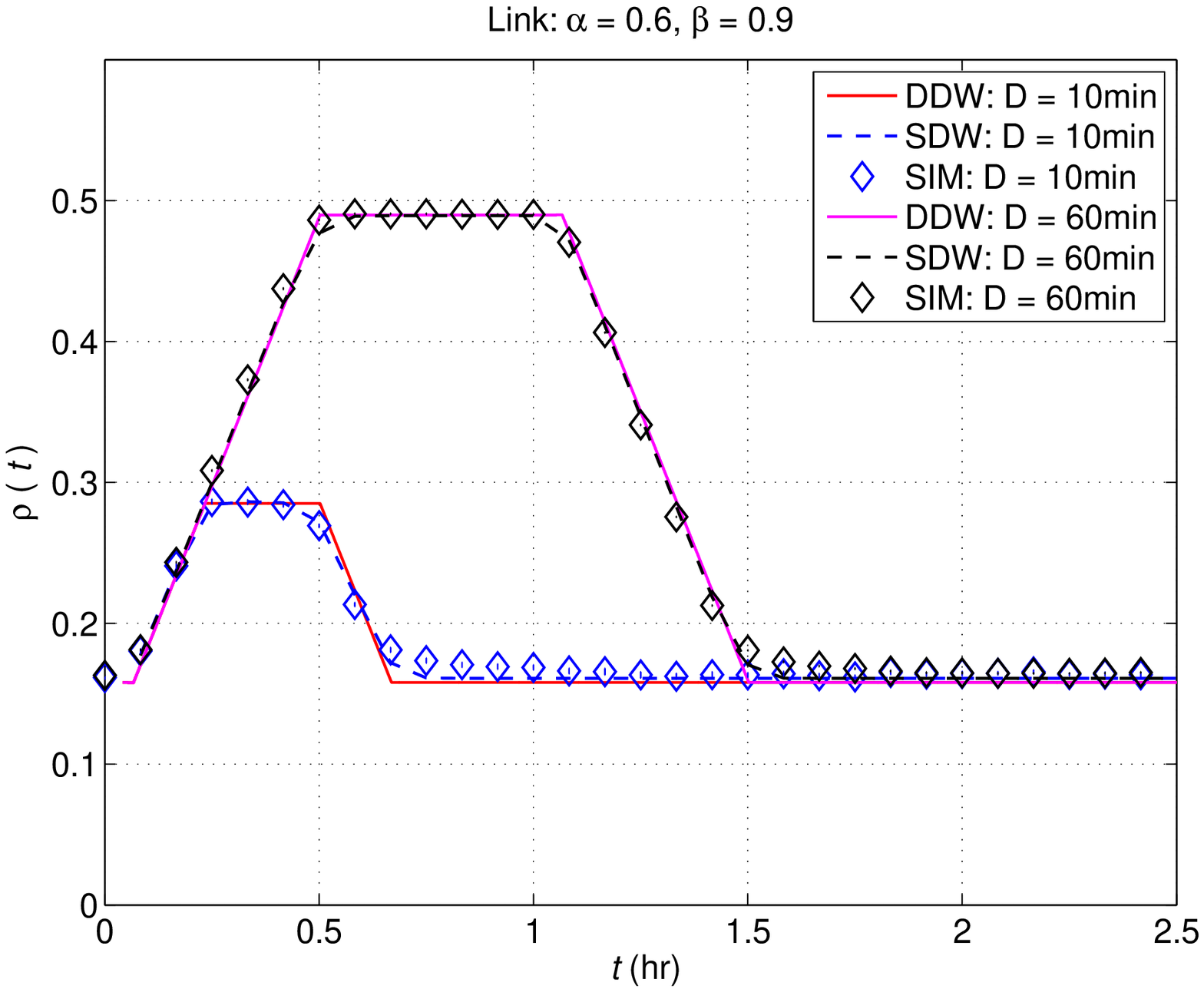}}
\def\dwmodelcelldensityMCL		{\includegraphics[scale=\oneup]{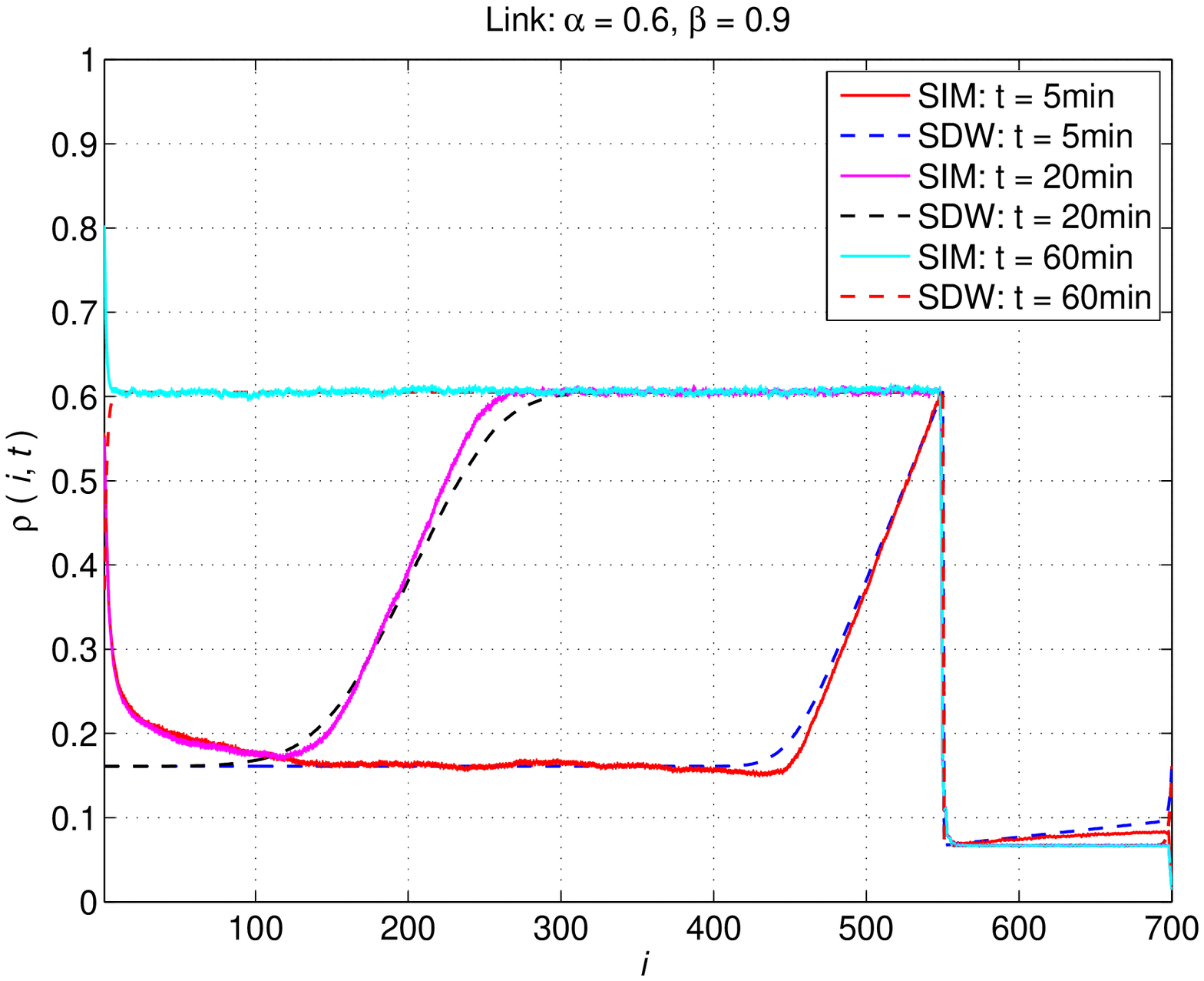}}
\def\dwmodelcelldensityMCR		{\includegraphics[scale=\oneup]{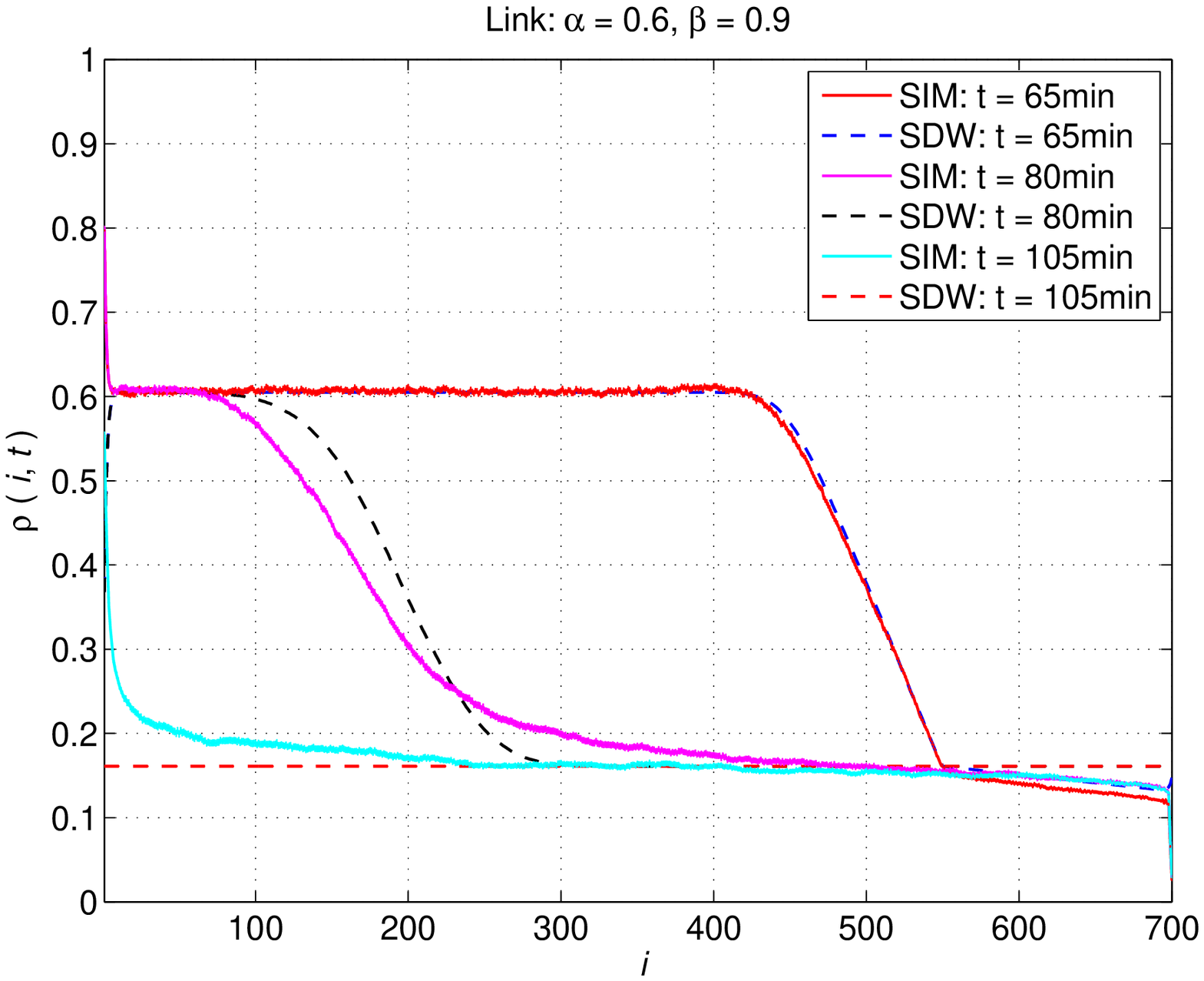}}
\def\dwmodelrouteflowH			{\includegraphics[scale=\oneup]{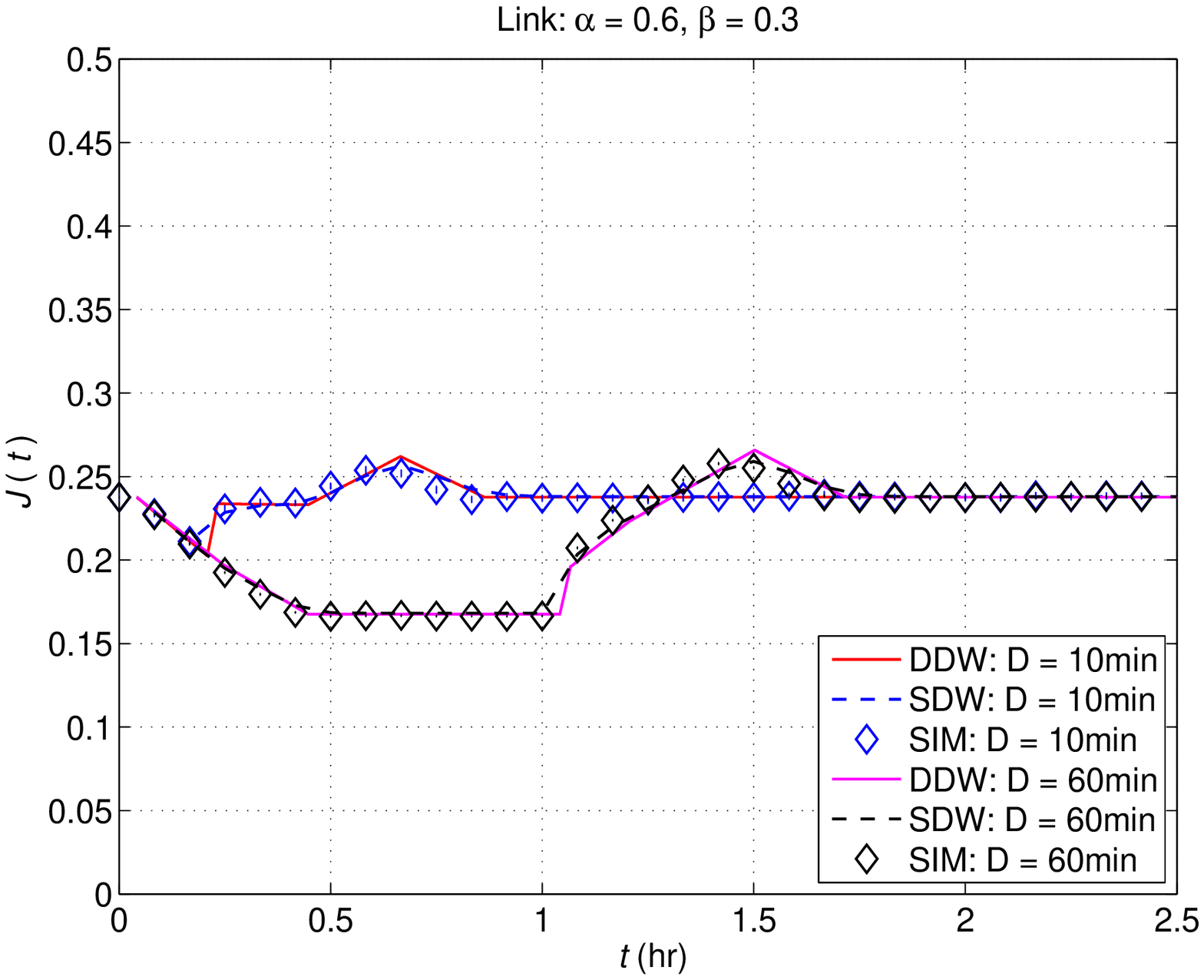}}
\def\dwmodelroutedensityH		{\includegraphics[scale=\oneup]{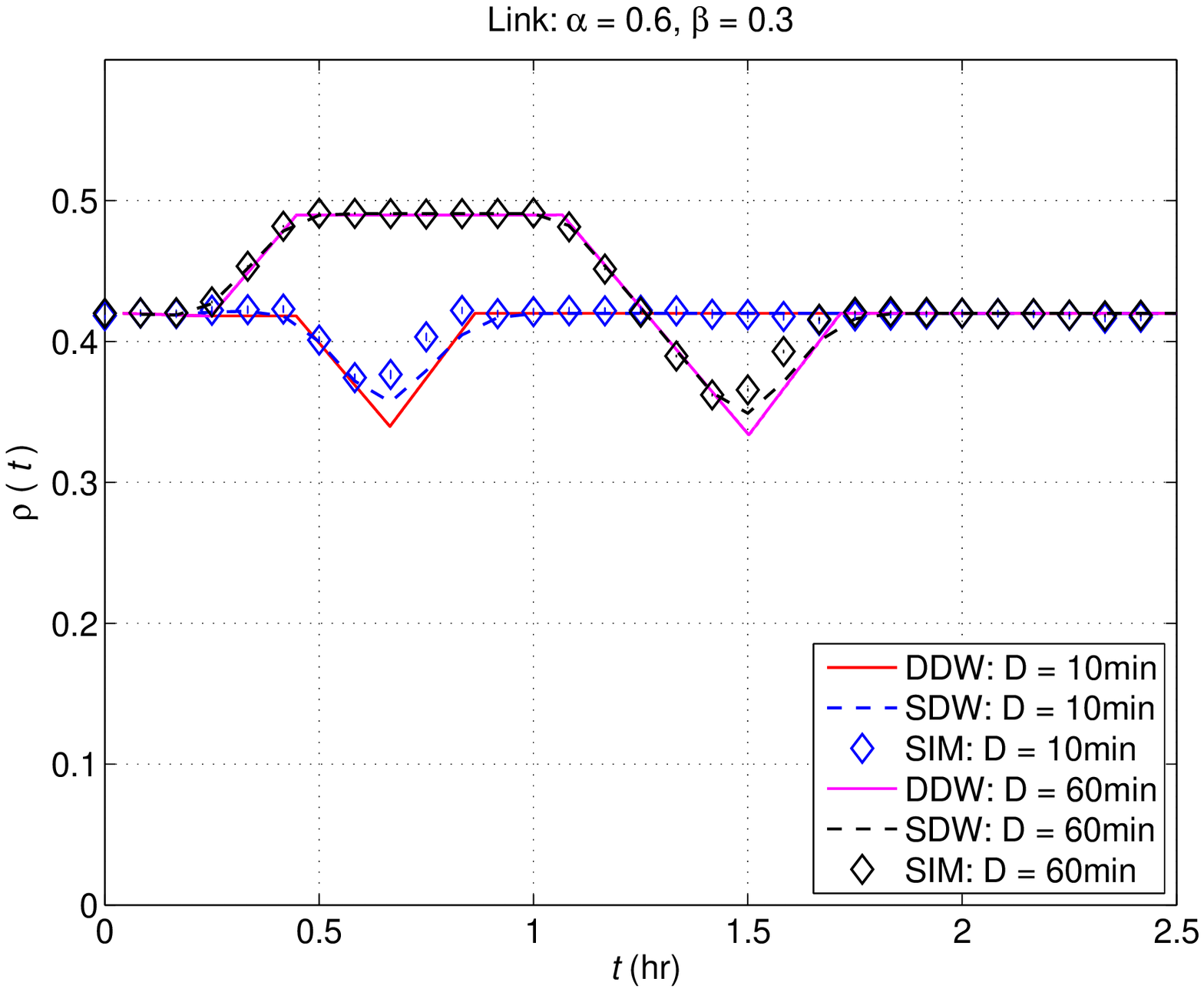}}
\def\dwmodelcelldensityHDL		{\includegraphics[scale=\oneup]{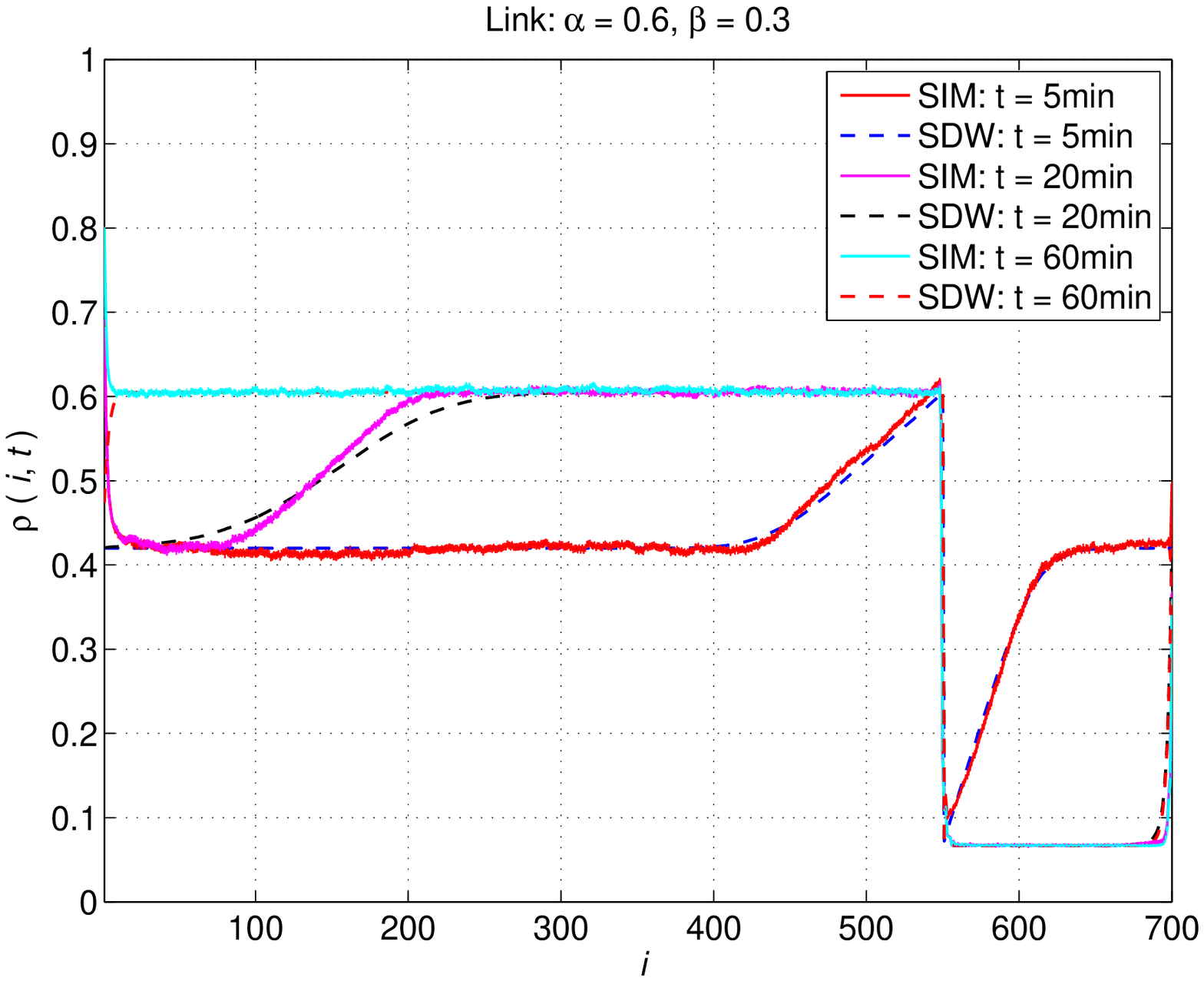}}
\def\dwmodelcelldensityHDR		{\includegraphics[scale=\oneup]{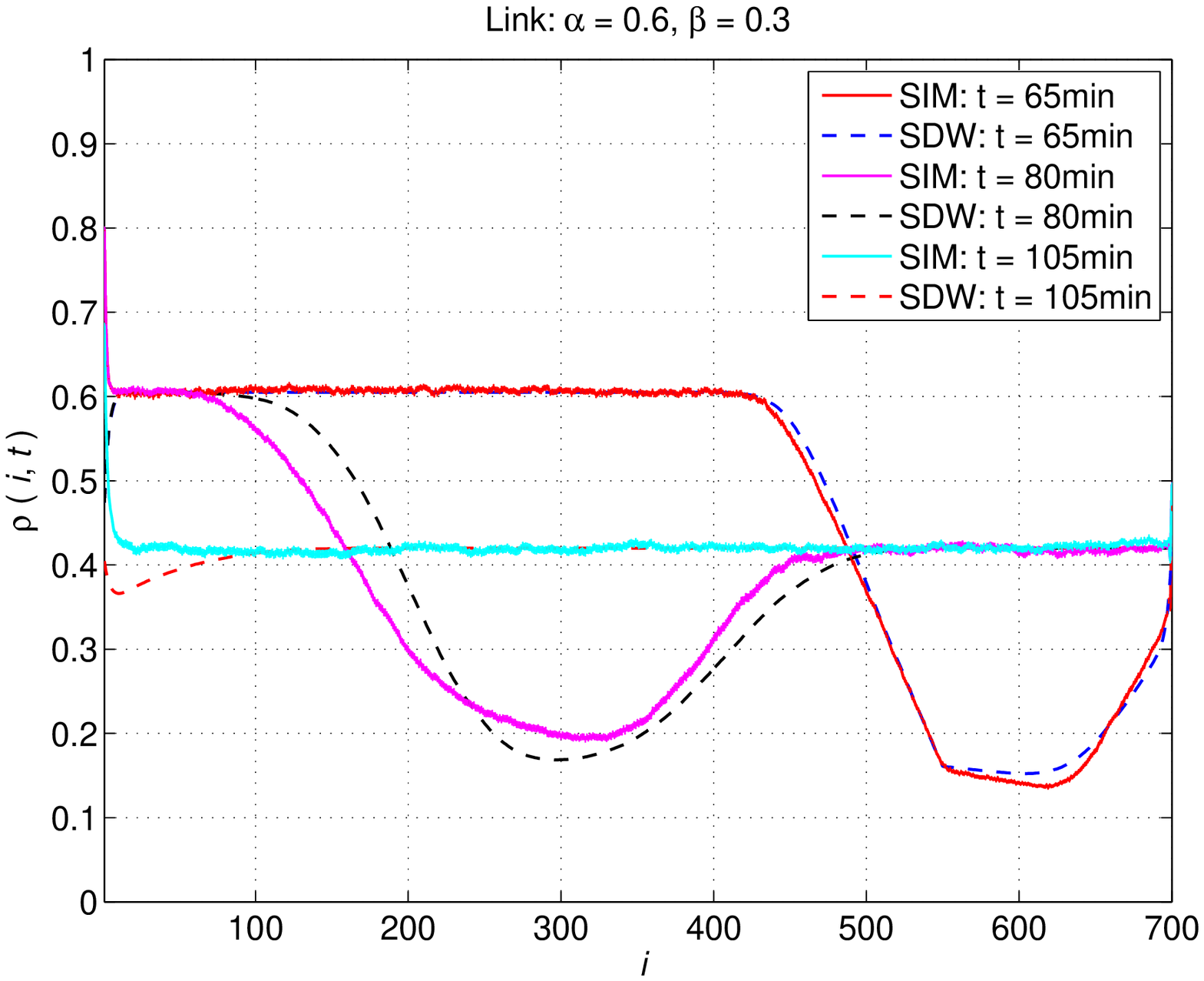}}
\def\LKeightflowAsixtyBninety	{\includegraphics[scale=\oneup]{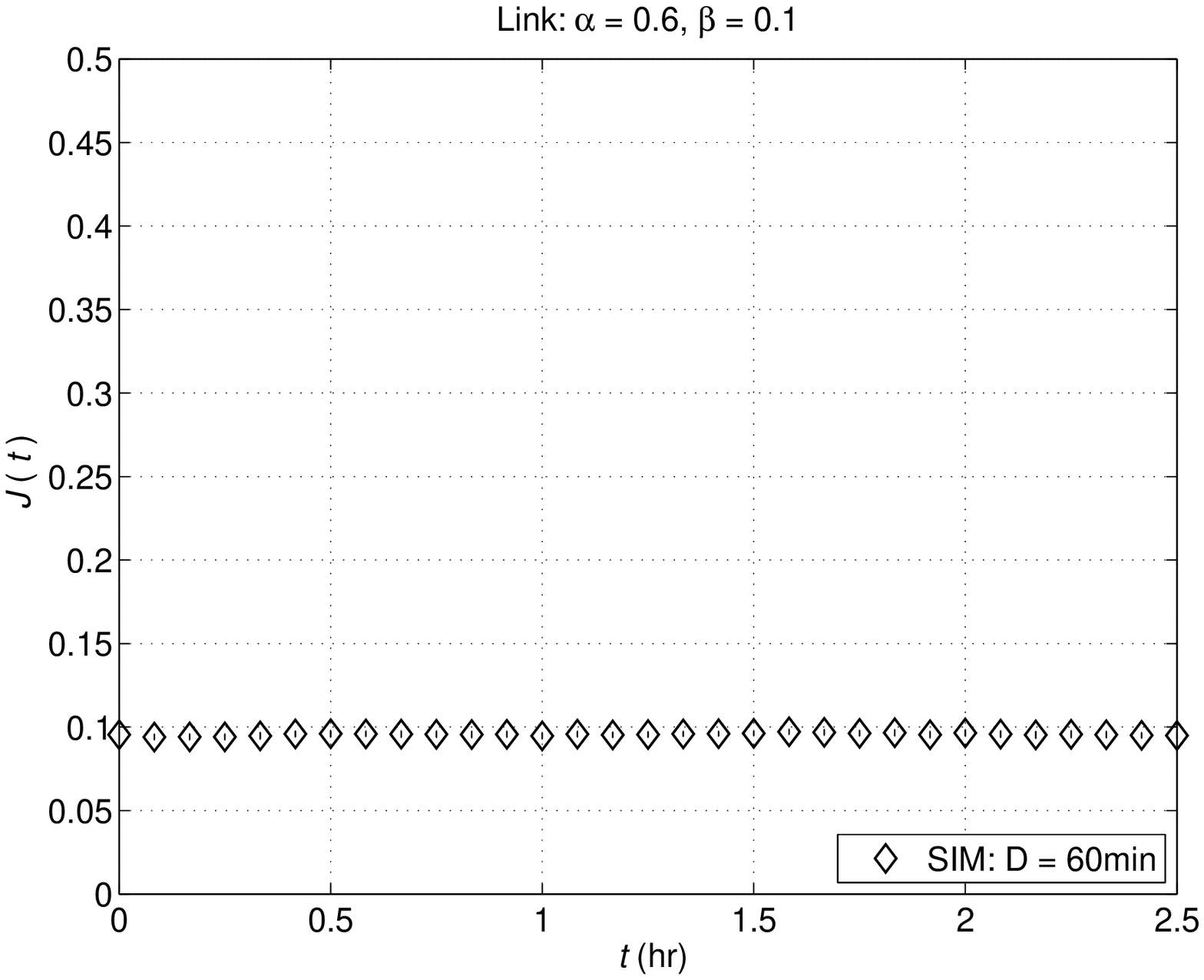}}
\def\LKeightdensityAsixtyBninety{\includegraphics[scale=\oneup]{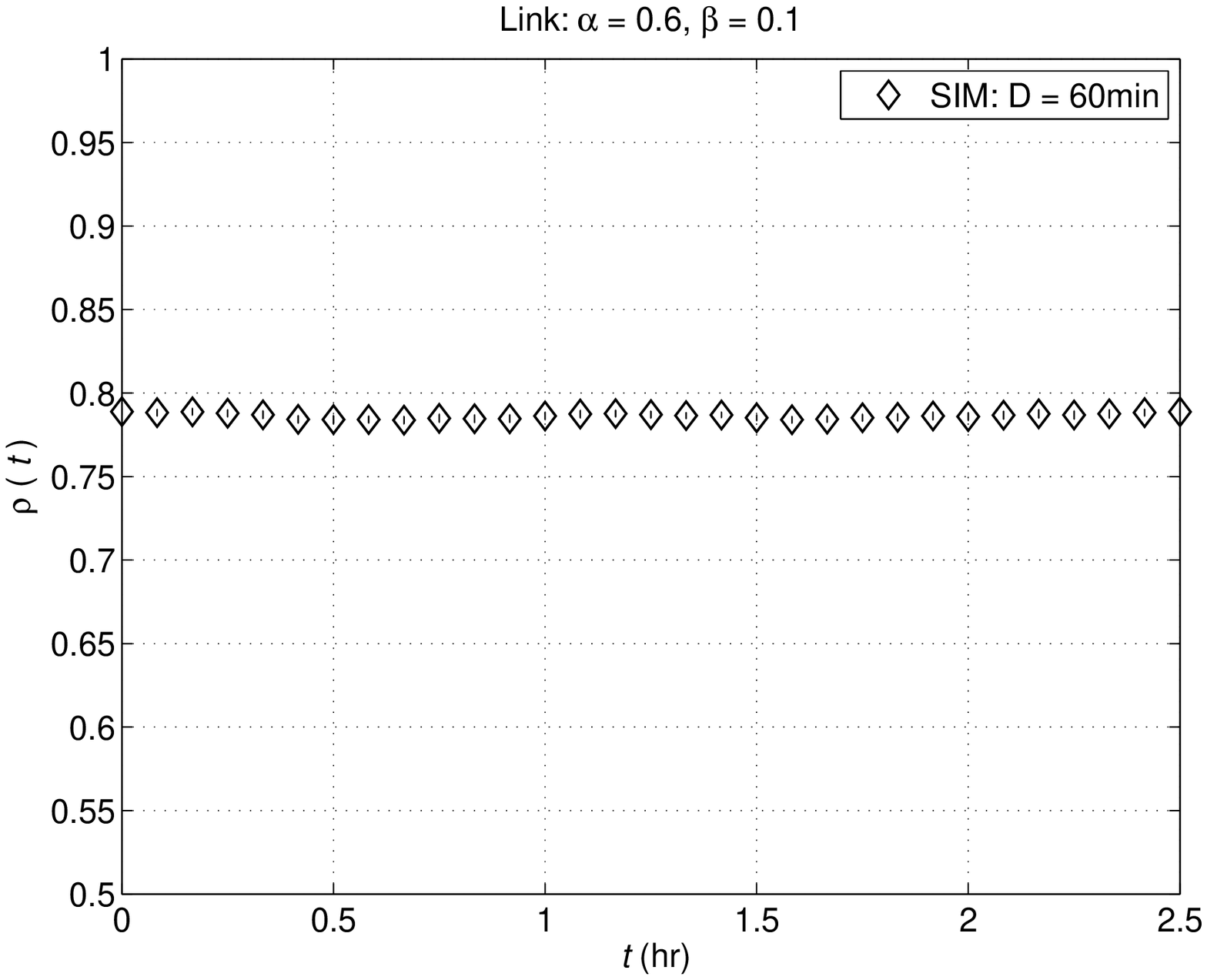}}

\def\network					{\includegraphics[scale=0.36]{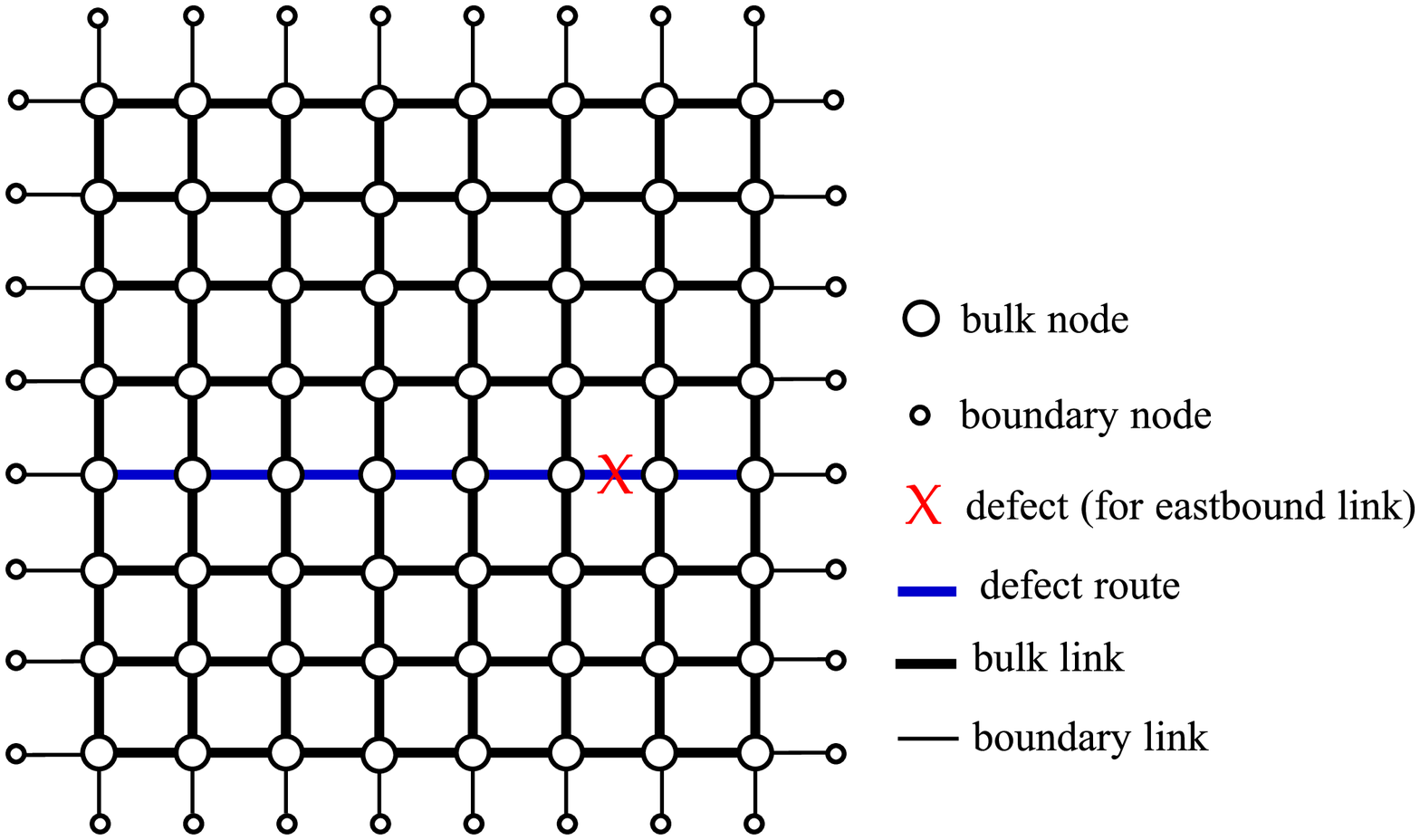}}
\def\phasea						{\includegraphics[scale=\smallest]{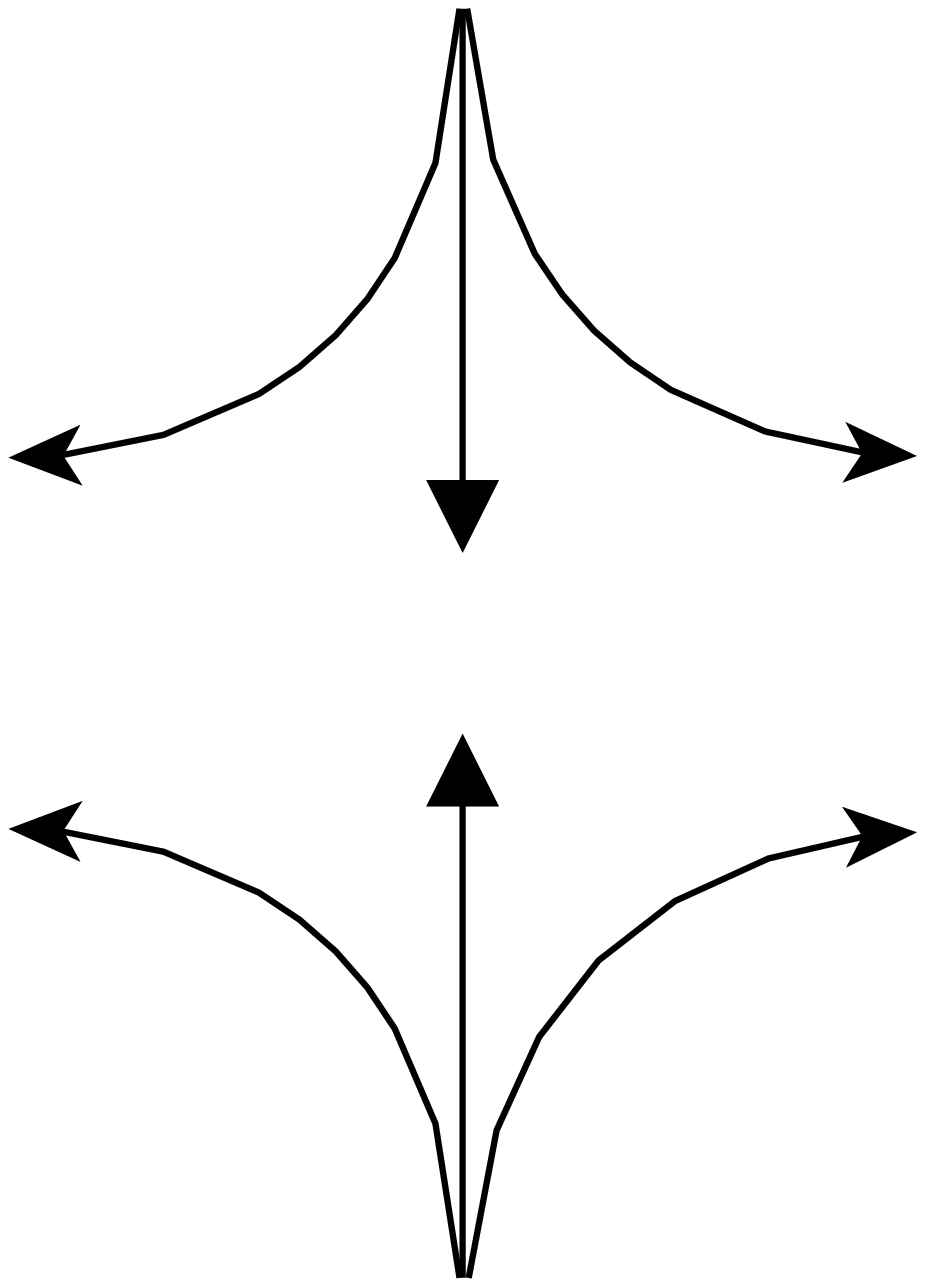}}
\def\phaseb						{\includegraphics[scale=\smallest, angle=90]{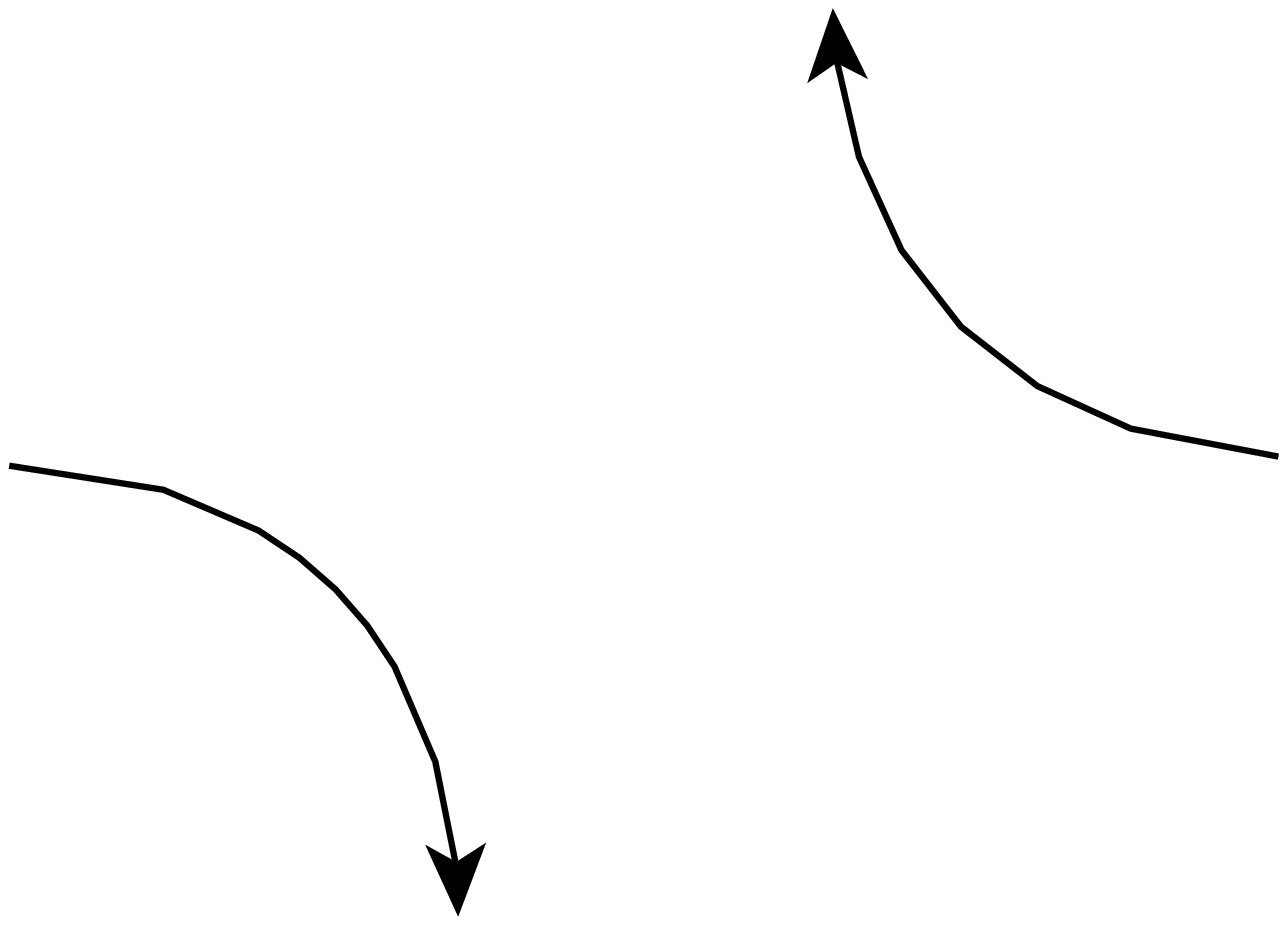}}
\def\phasec						{\includegraphics[scale=\smallest, angle=90]{fig10a.eps}}
\def\phased						{\includegraphics[scale=\smallest]{fig10b.eps}}
\def\CRfds						{\includegraphics[scale=\oneup]{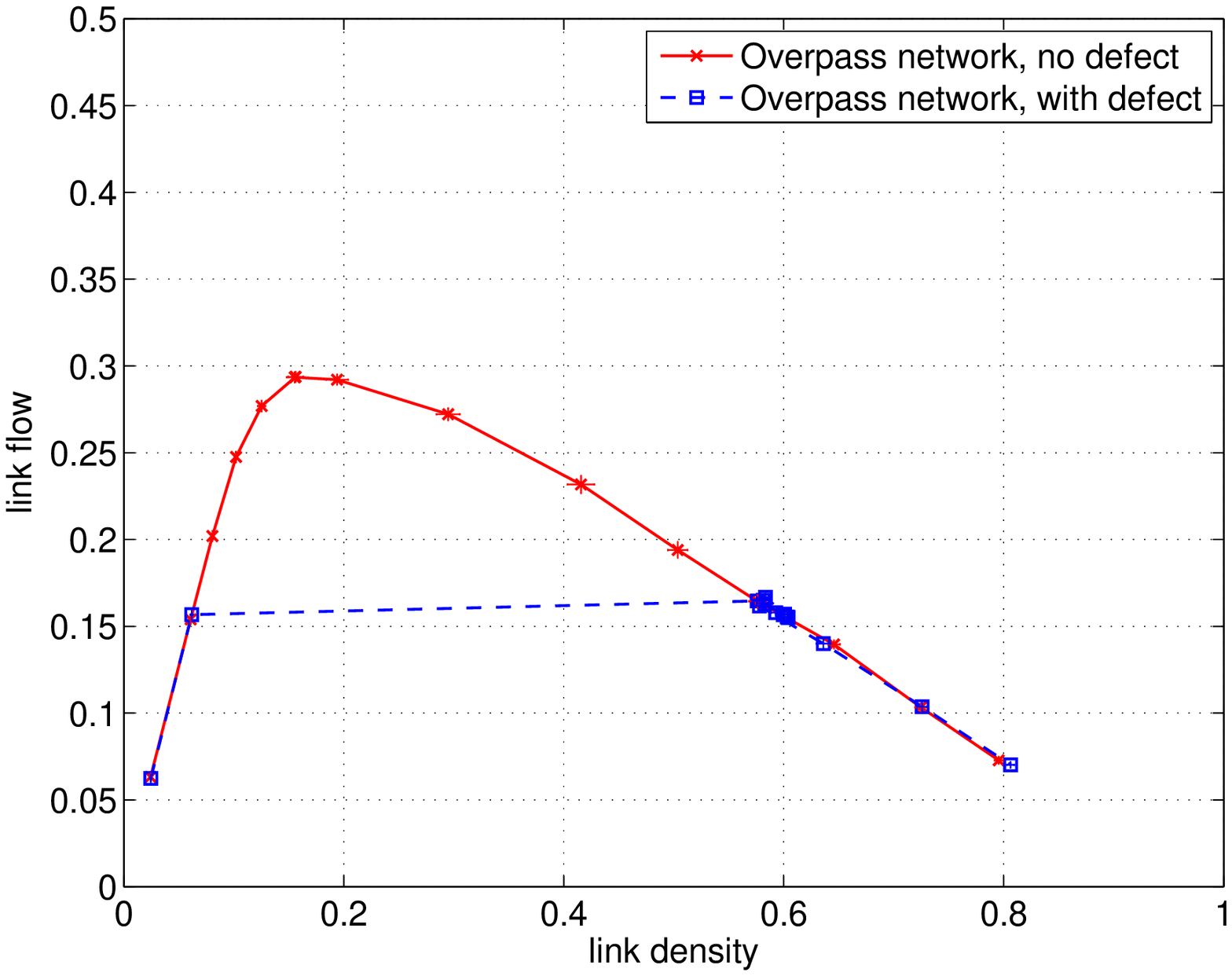}}

\def\dwmodelCRrouteflowLD		{\includegraphics[scale=\oneup]{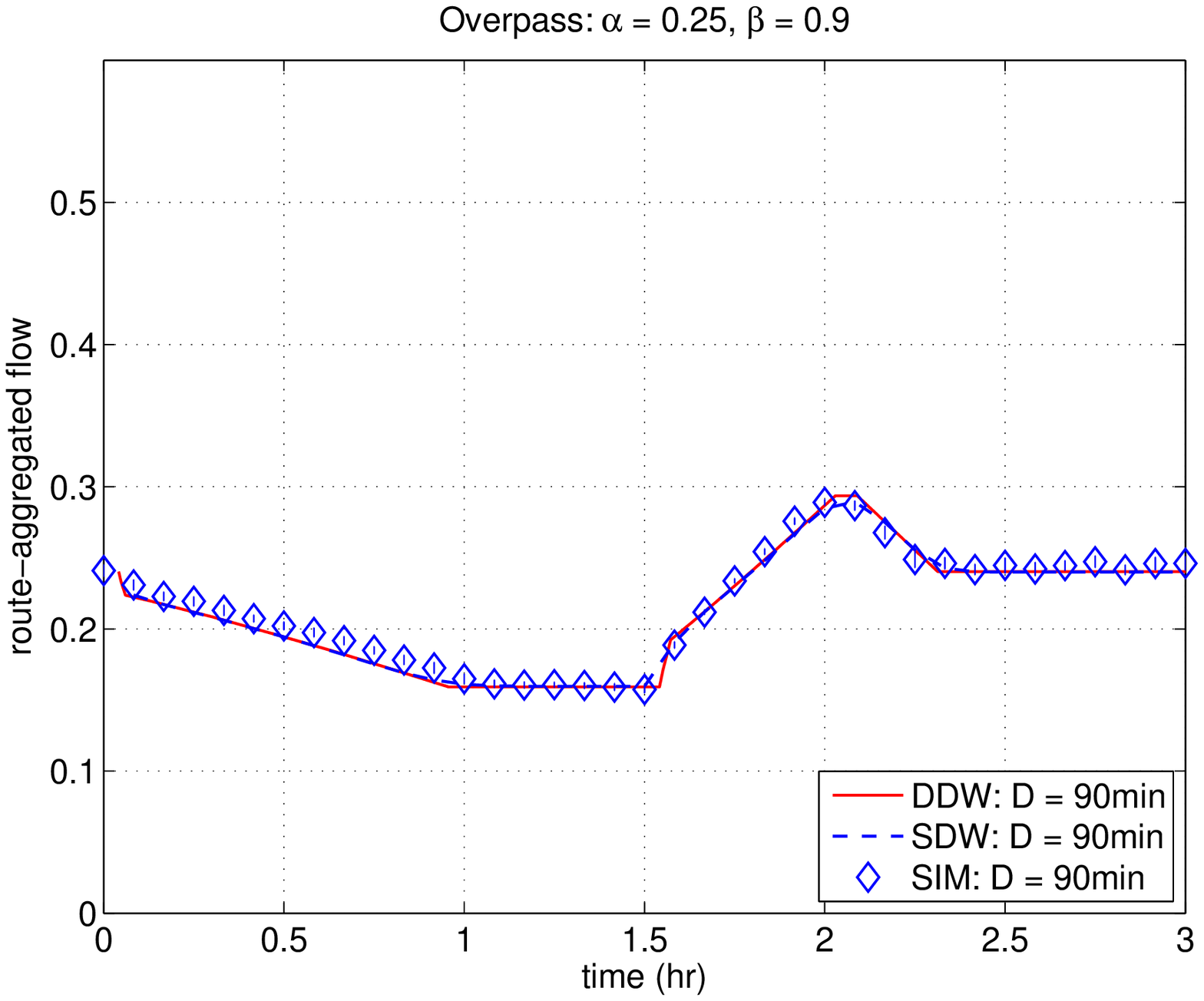}}
\def\dwmodelCRroutedensityLD	{\includegraphics[scale=\oneup]{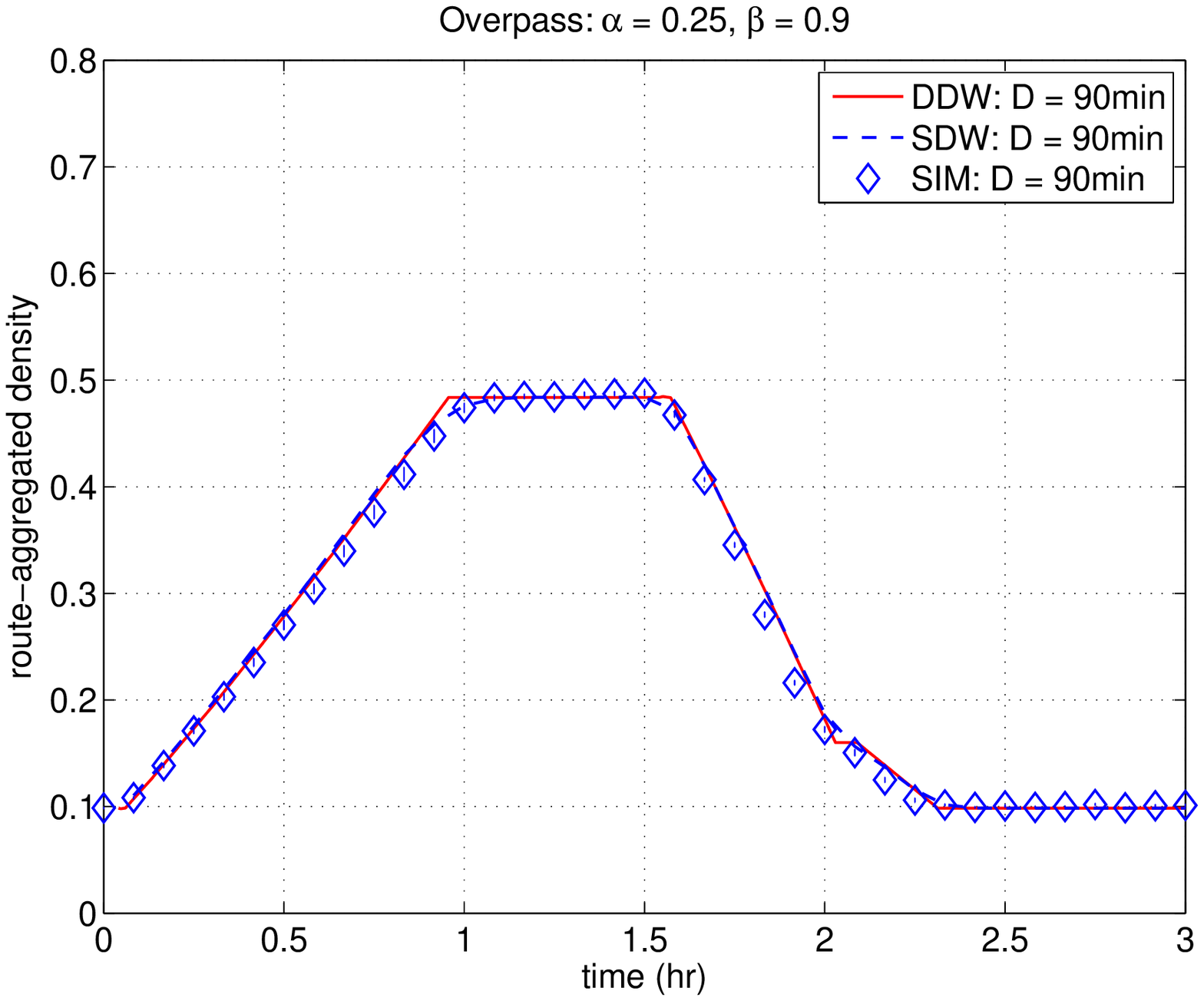}}
\def\dwmodelCRlinksflowLD		{\includegraphics[scale=\oneup]{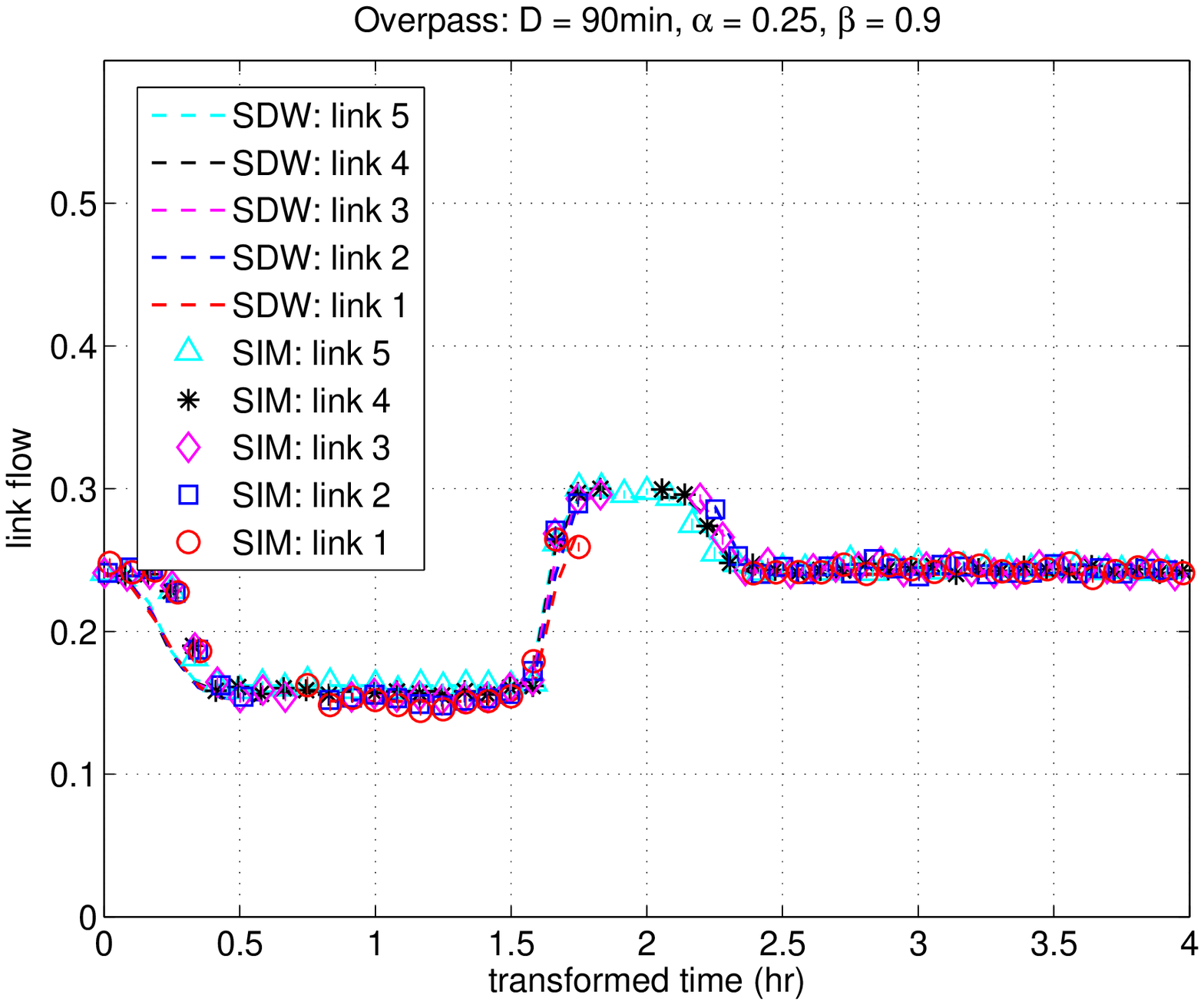}}
\def\dwmodelCRlinksdensityLD	{\includegraphics[scale=\oneup]{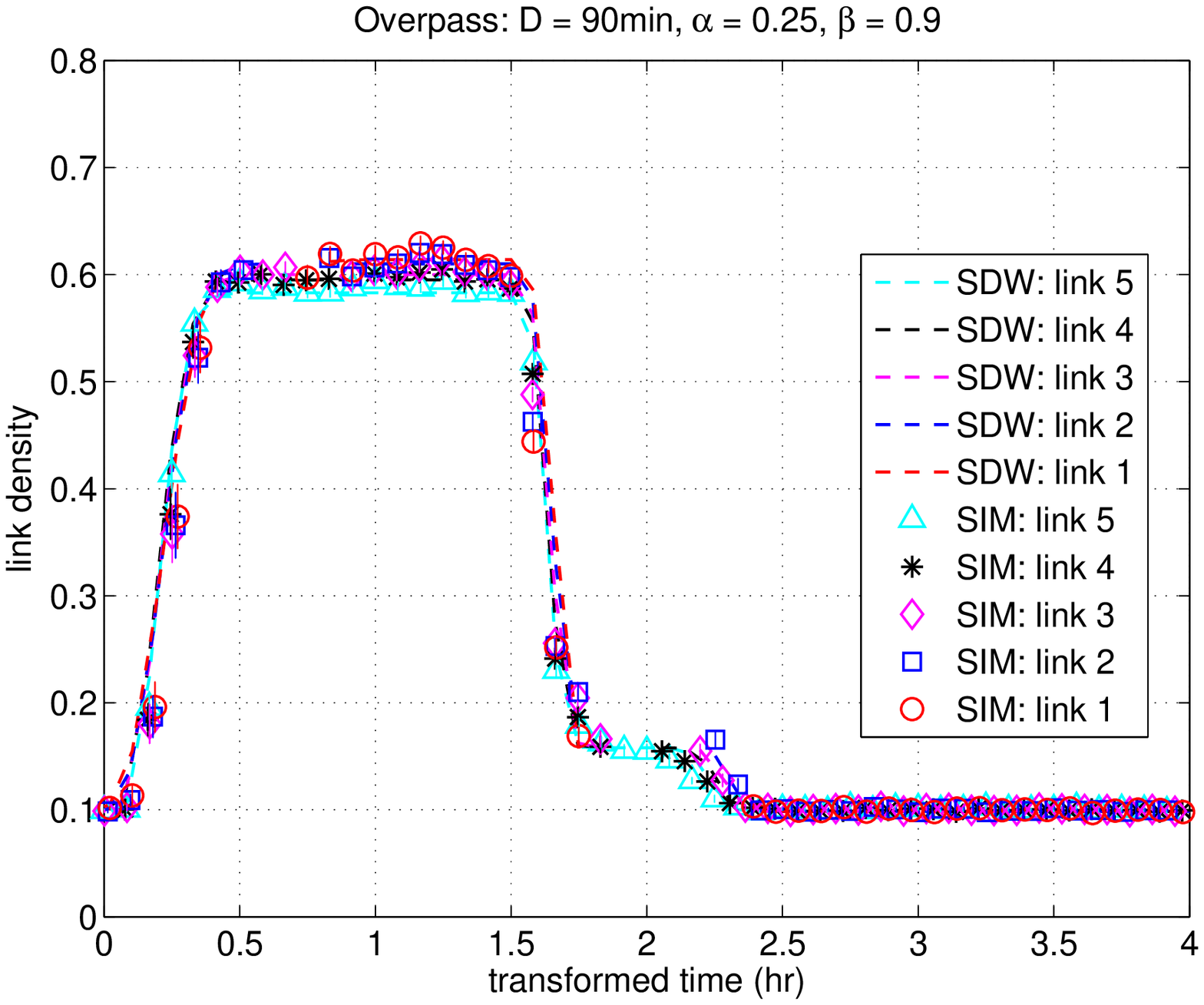}}

\def\dwmodelCRrouteflowMC		{\includegraphics[scale=\oneup]{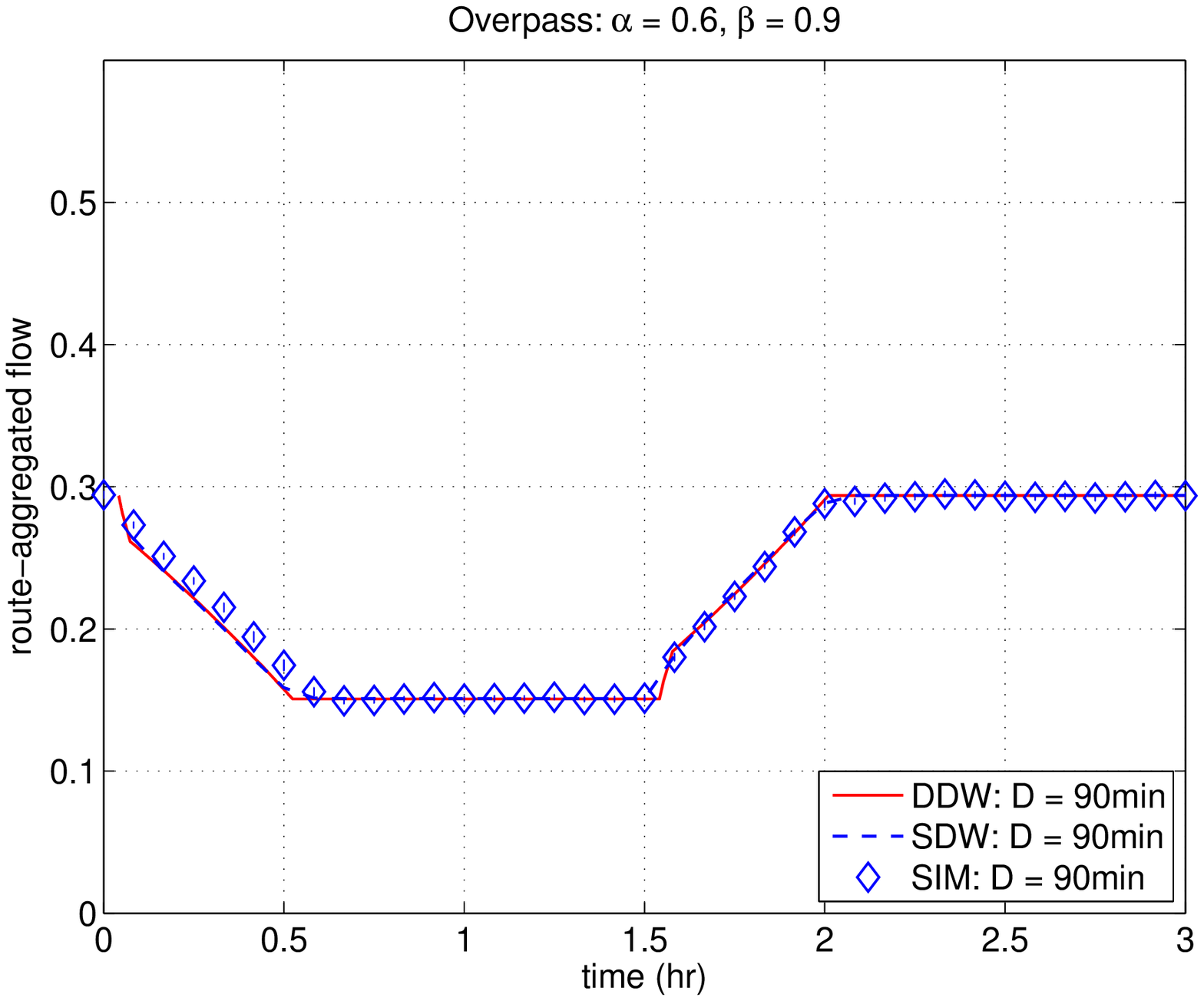}}
\def\dwmodelCRroutedensityMC	{\includegraphics[scale=\oneup]{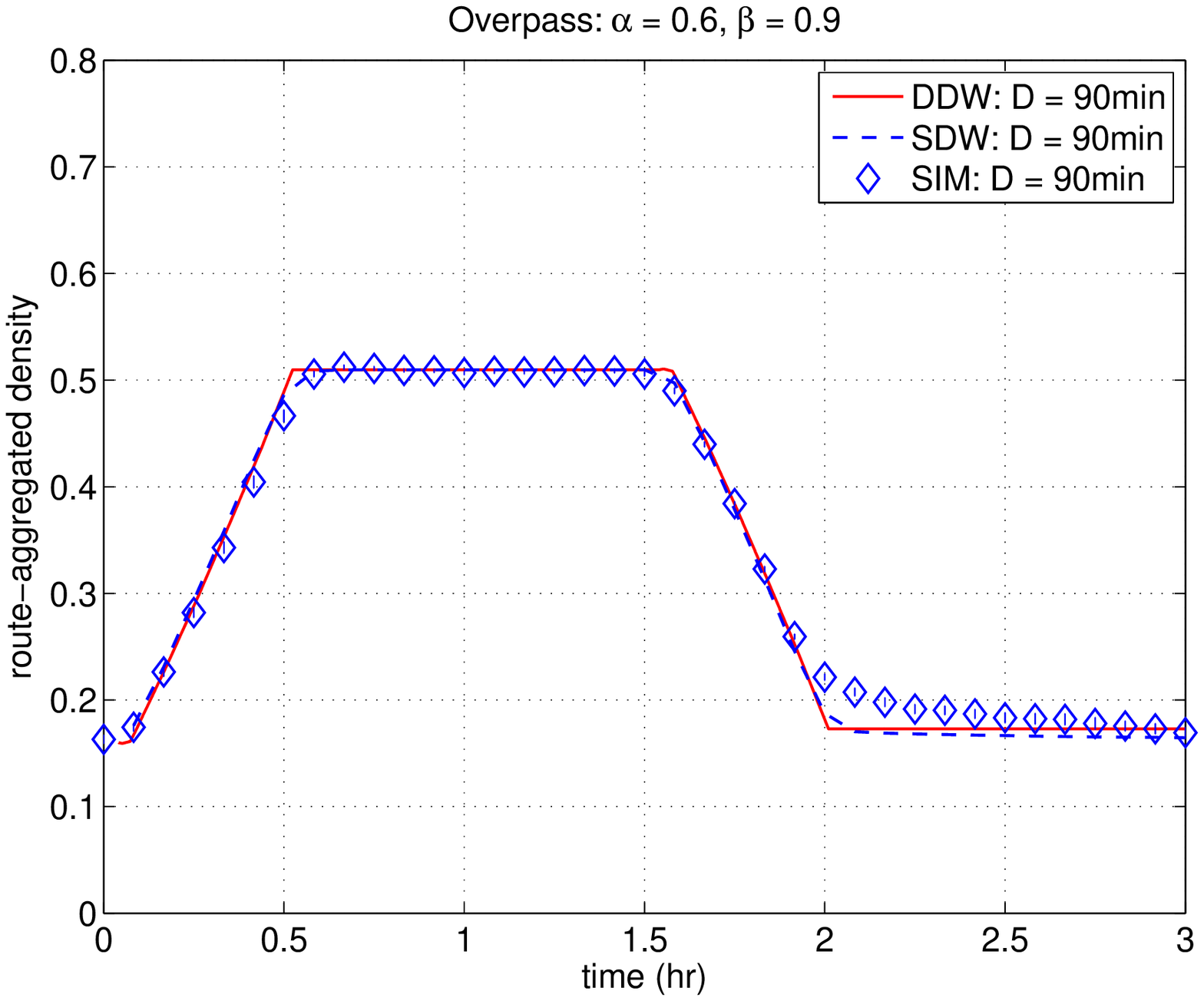}}
\def\dwmodelCRlinksflowMC		{\includegraphics[scale=\oneup]{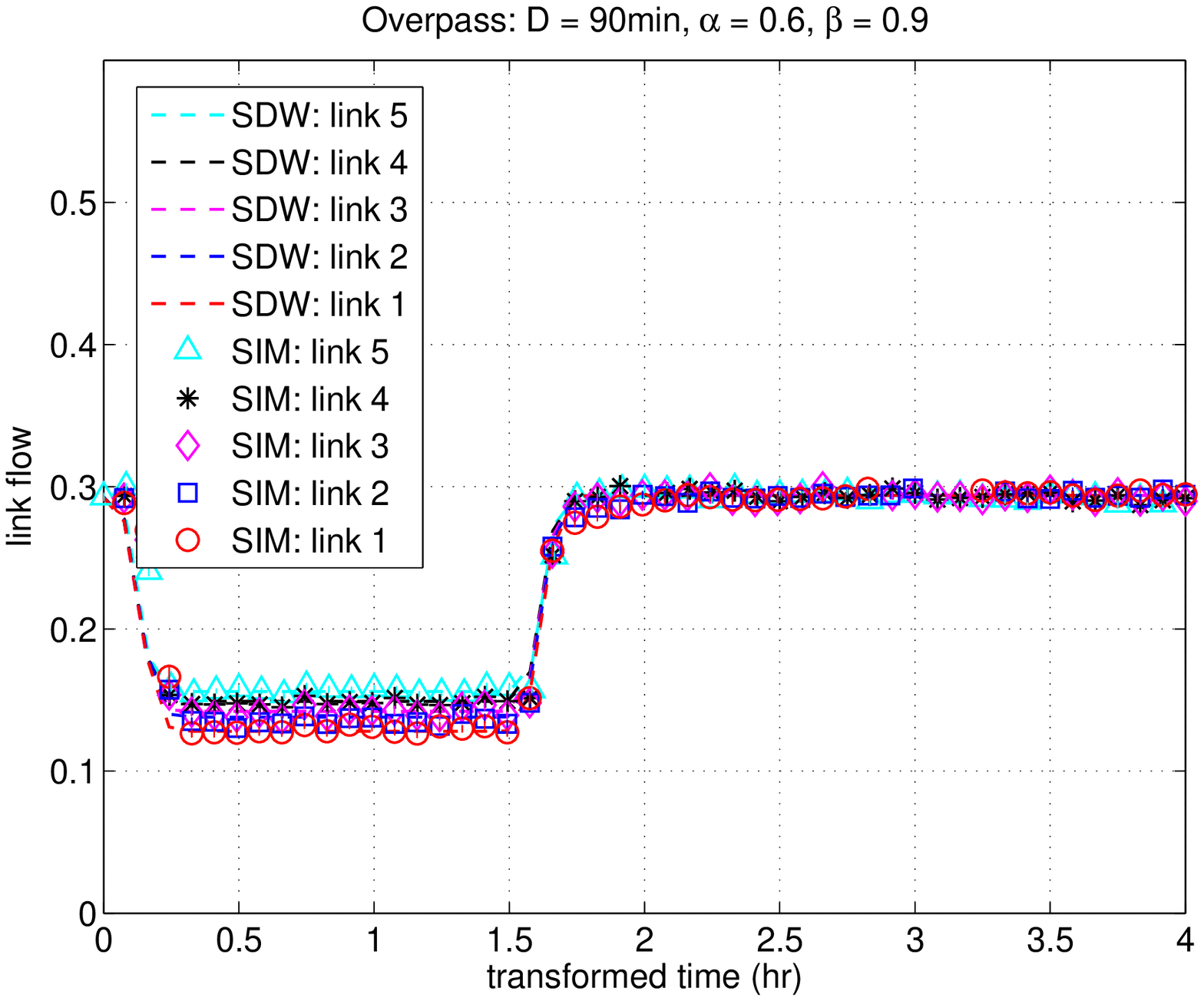}}
\def\dwmodelCRlinksdensityMC	{\includegraphics[scale=\oneup]{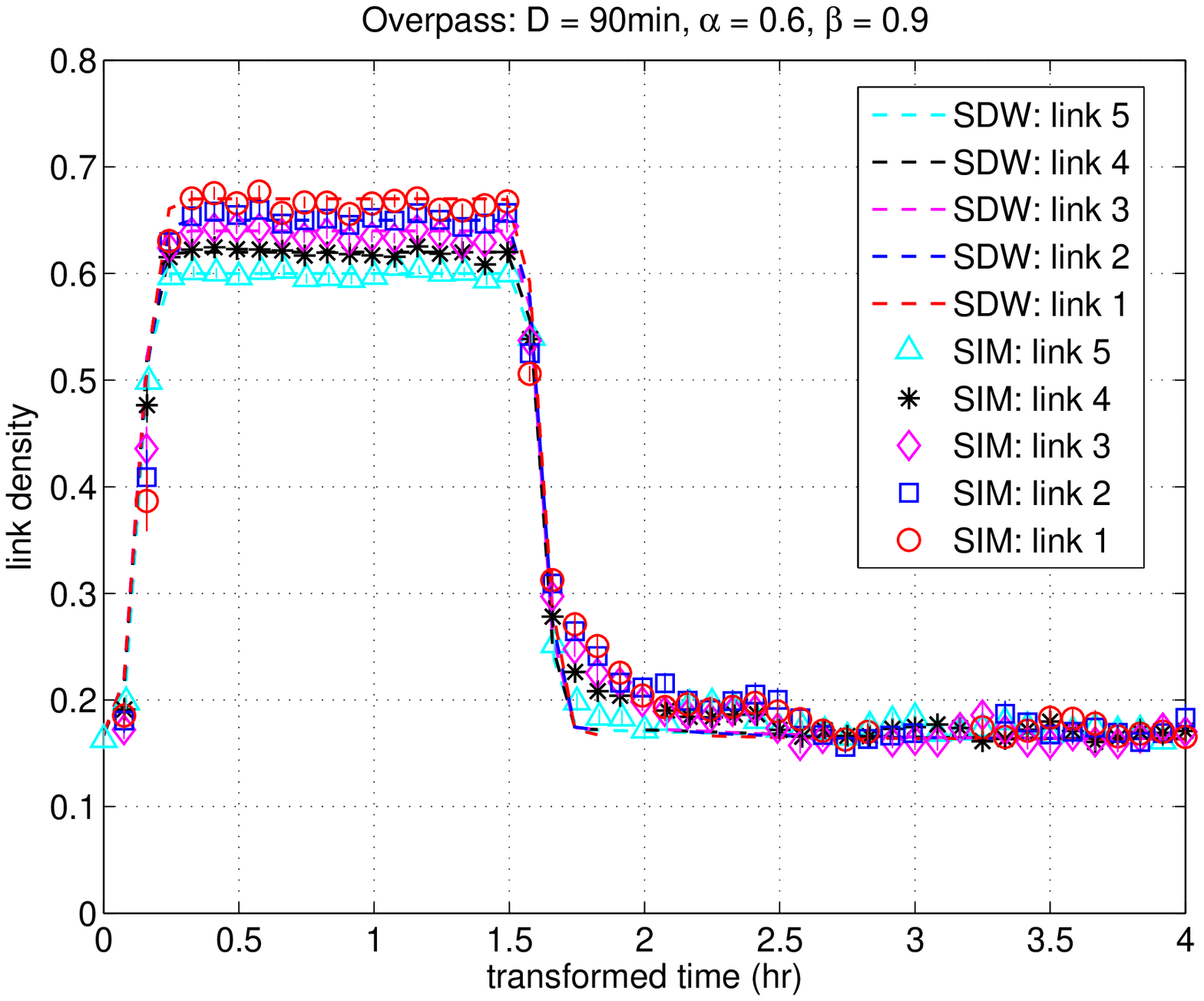}}

\def\dwmodelCRrouteflowHD		{\includegraphics[scale=\oneup]{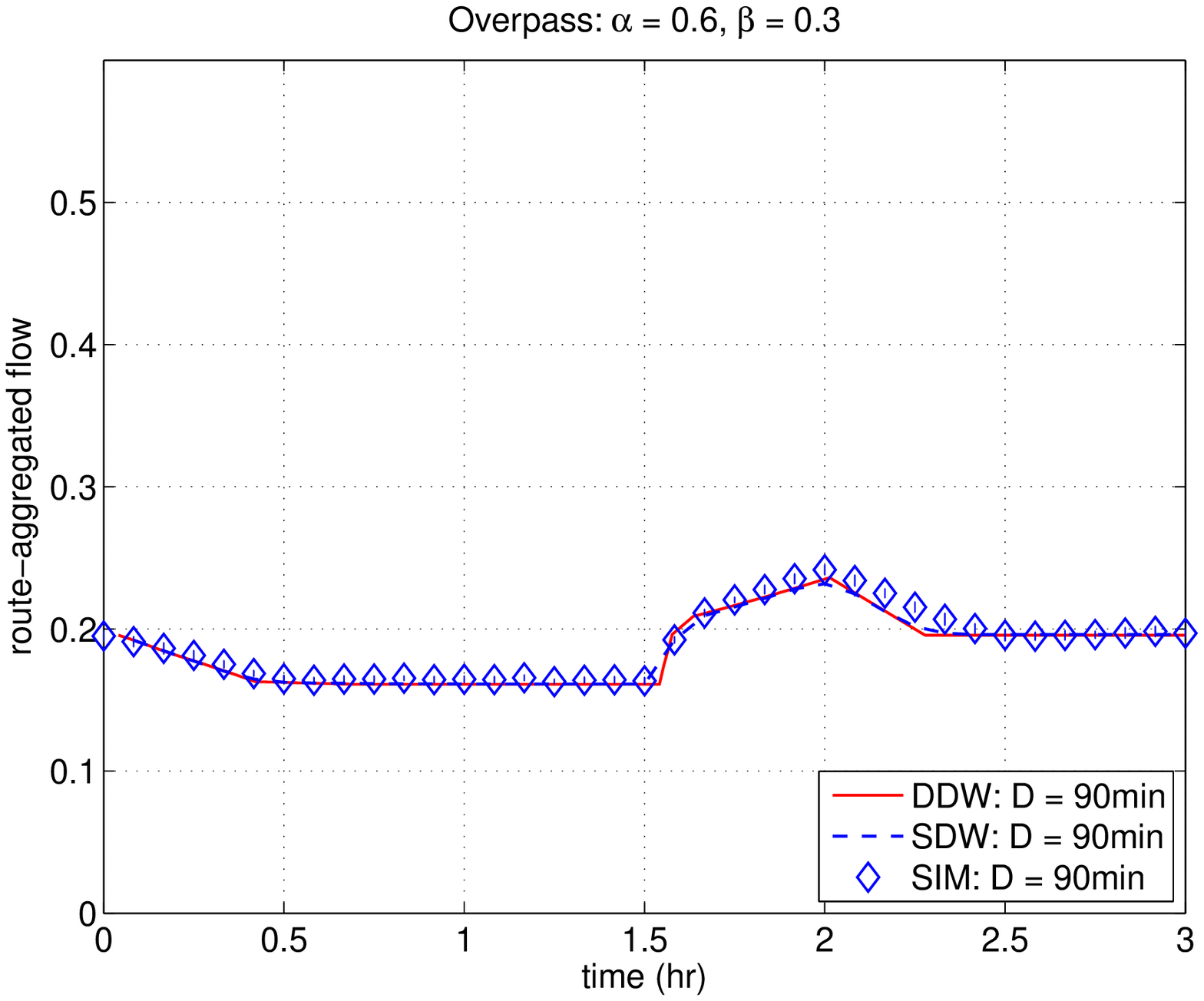}}
\def\dwmodelCRroutedensityHD	{\includegraphics[scale=\oneup]{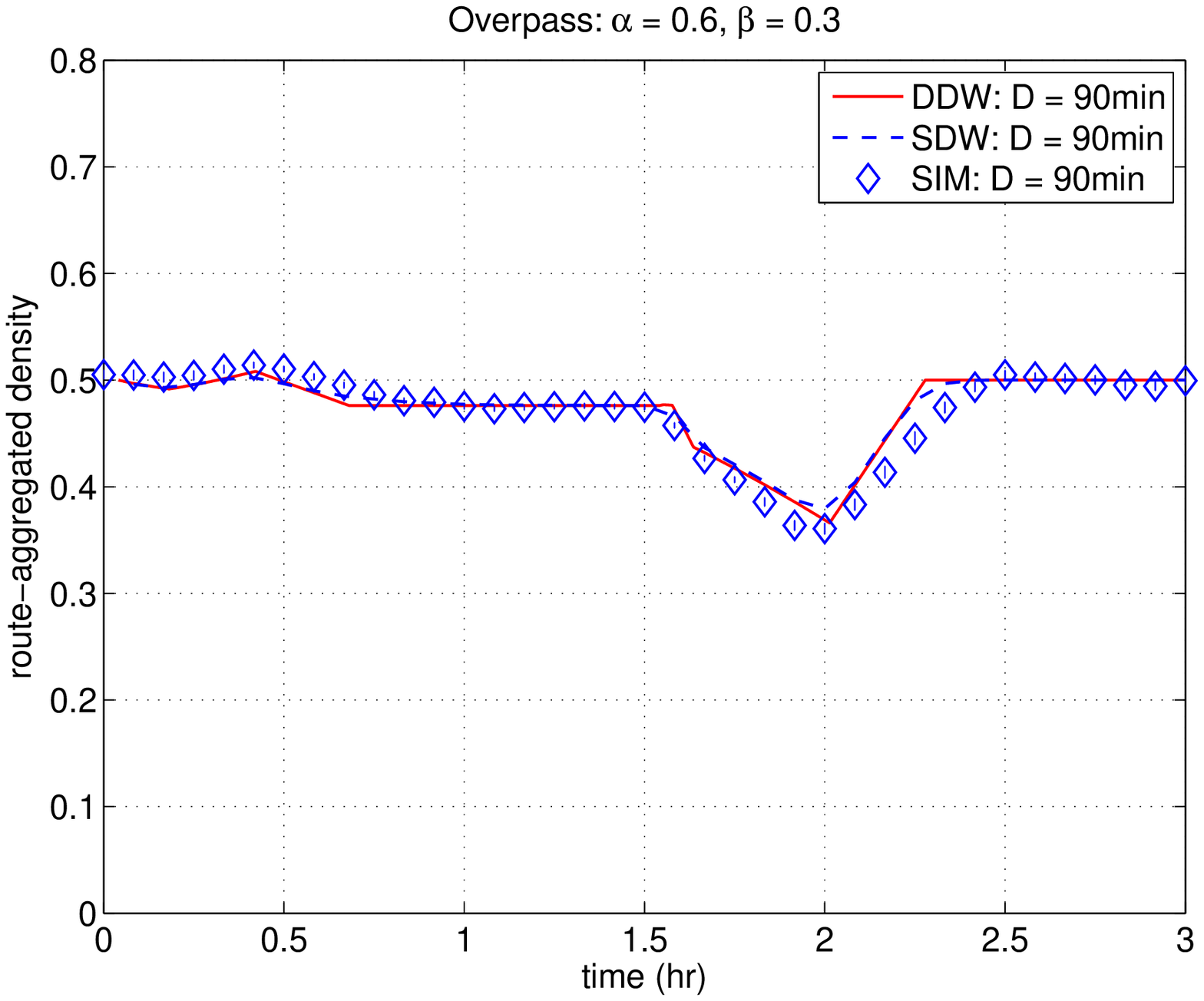}}
\def\dwmodelCRlinksflowHD		{\includegraphics[scale=\oneup]{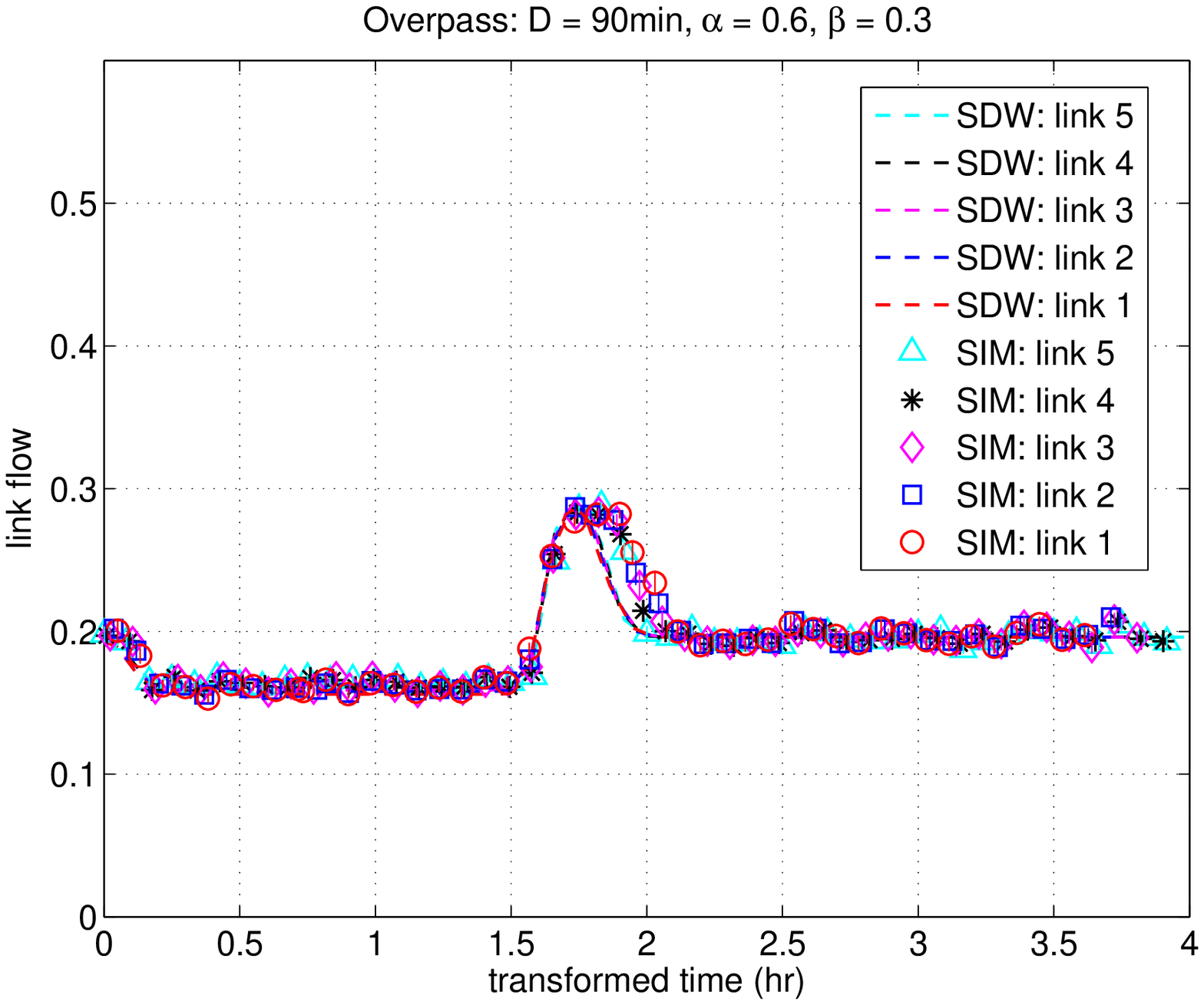}}
\def\dwmodelCRlinksdensityHD	{\includegraphics[scale=\oneup]{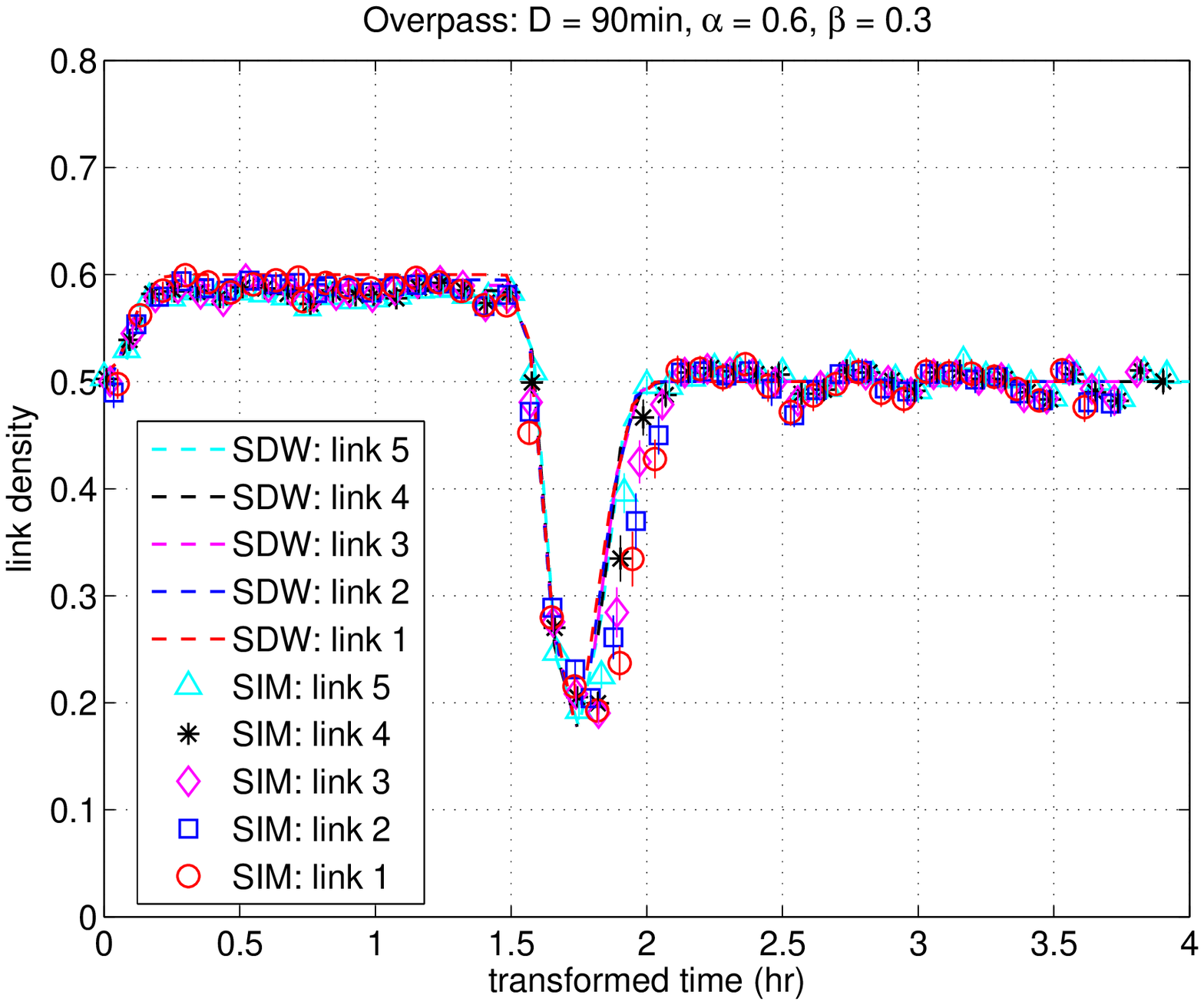}}

\def\ANflowmidblock				{\includegraphics[scale=\oneup]{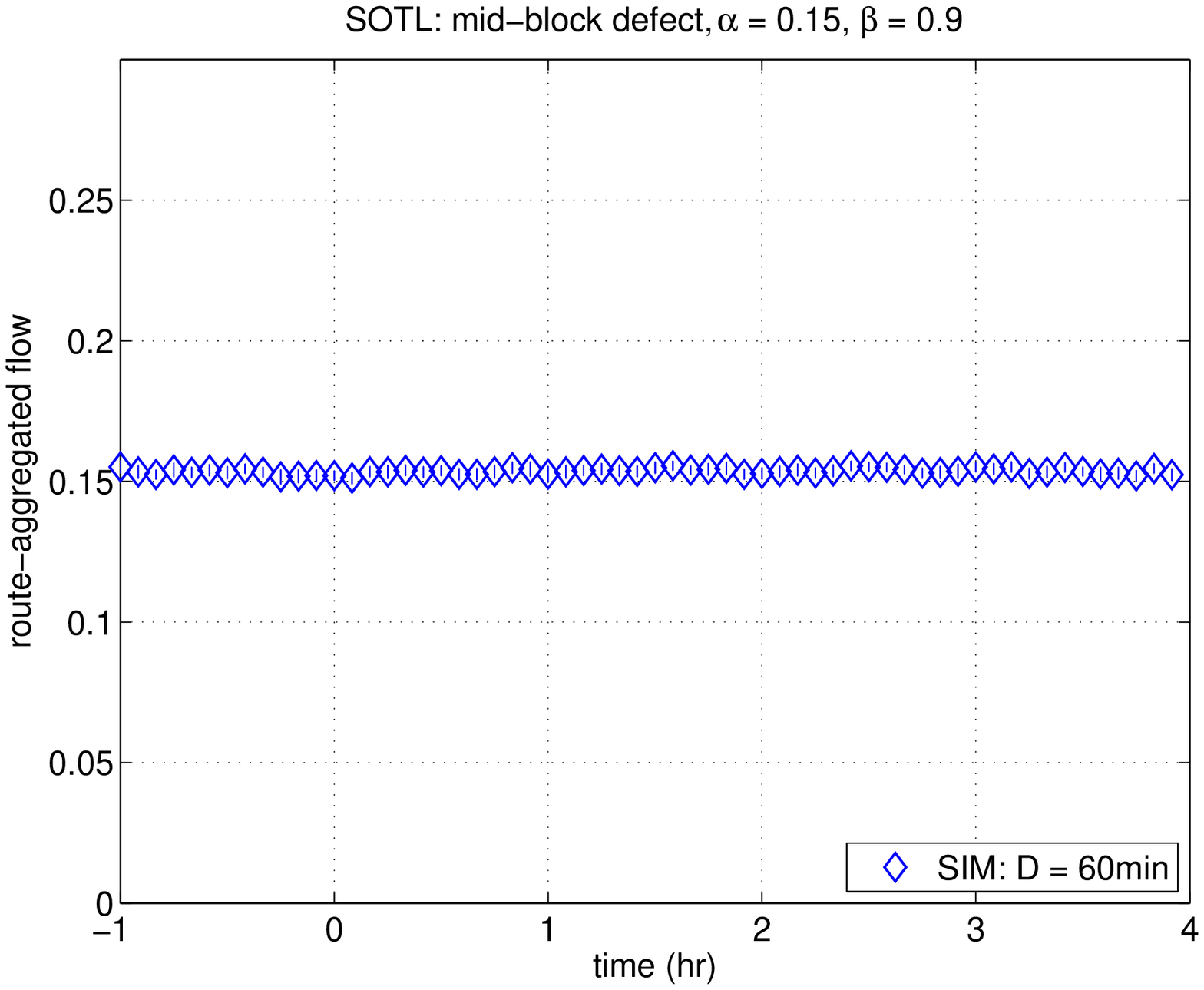}}
\def\ANdensitymidblock			{\includegraphics[scale=\oneup]{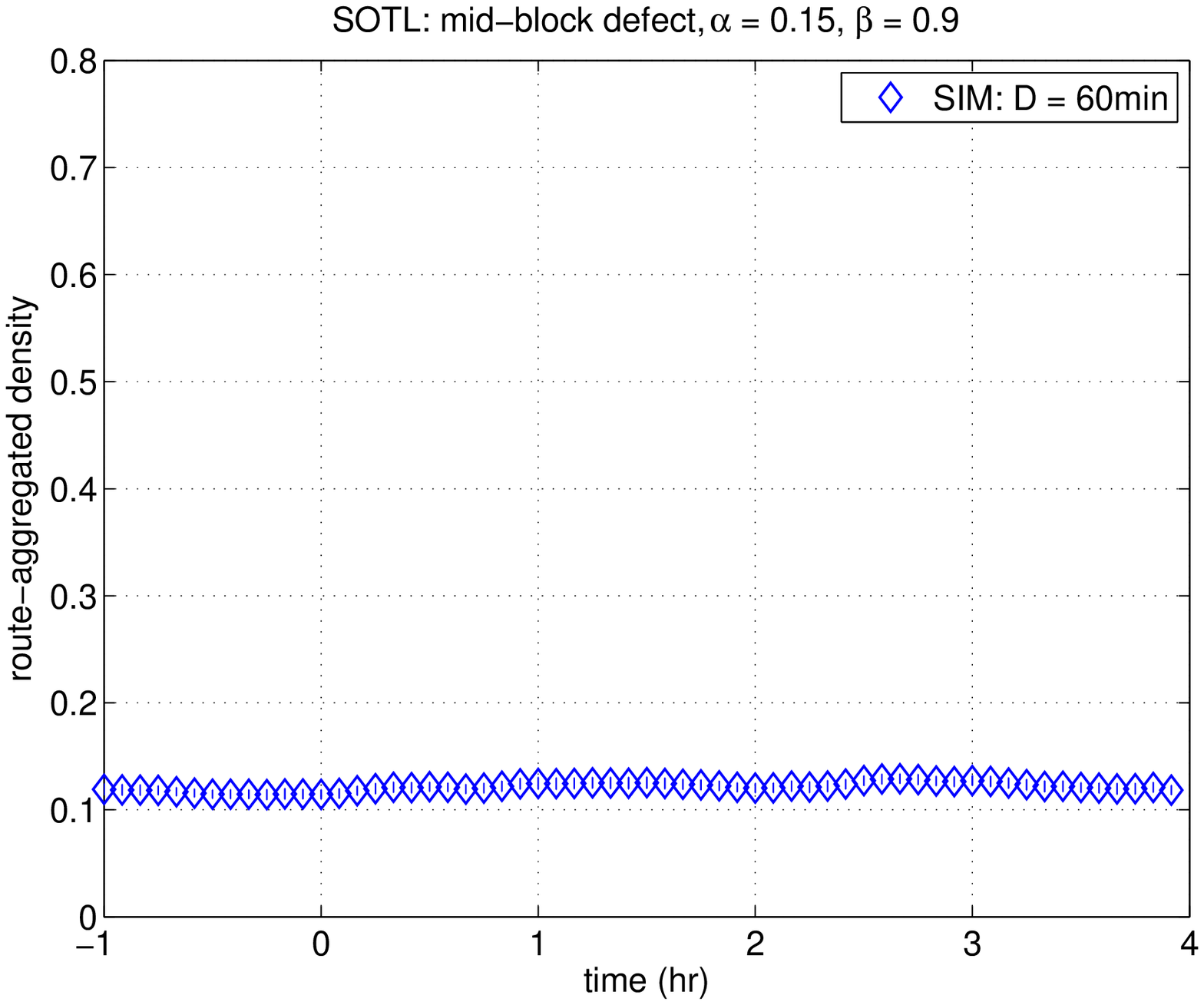}}
\def\ANlinkfdscmp				{\includegraphics[scale=\oneup]{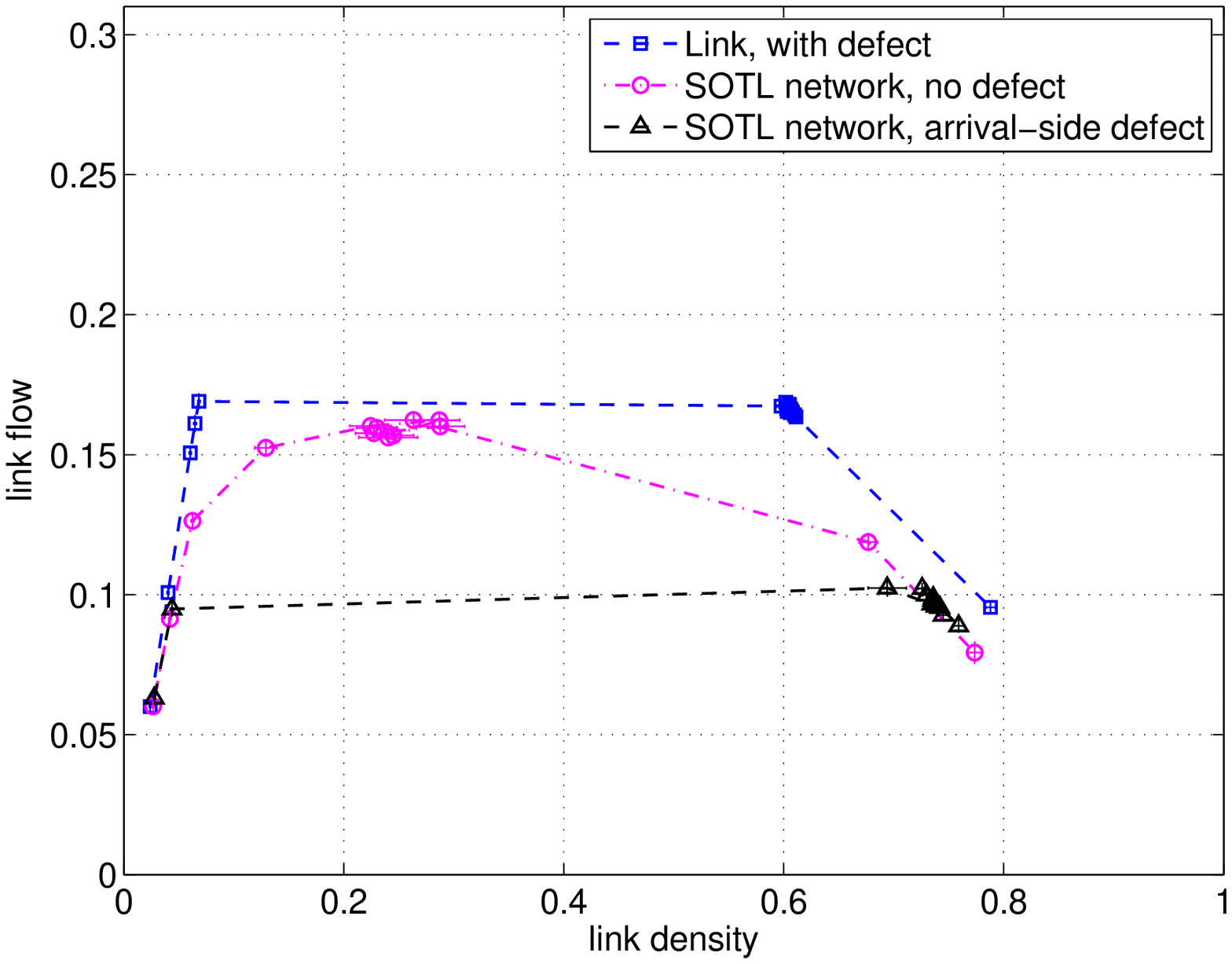}}
\def\ANlinkfds					{\includegraphics[scale=\oneup]{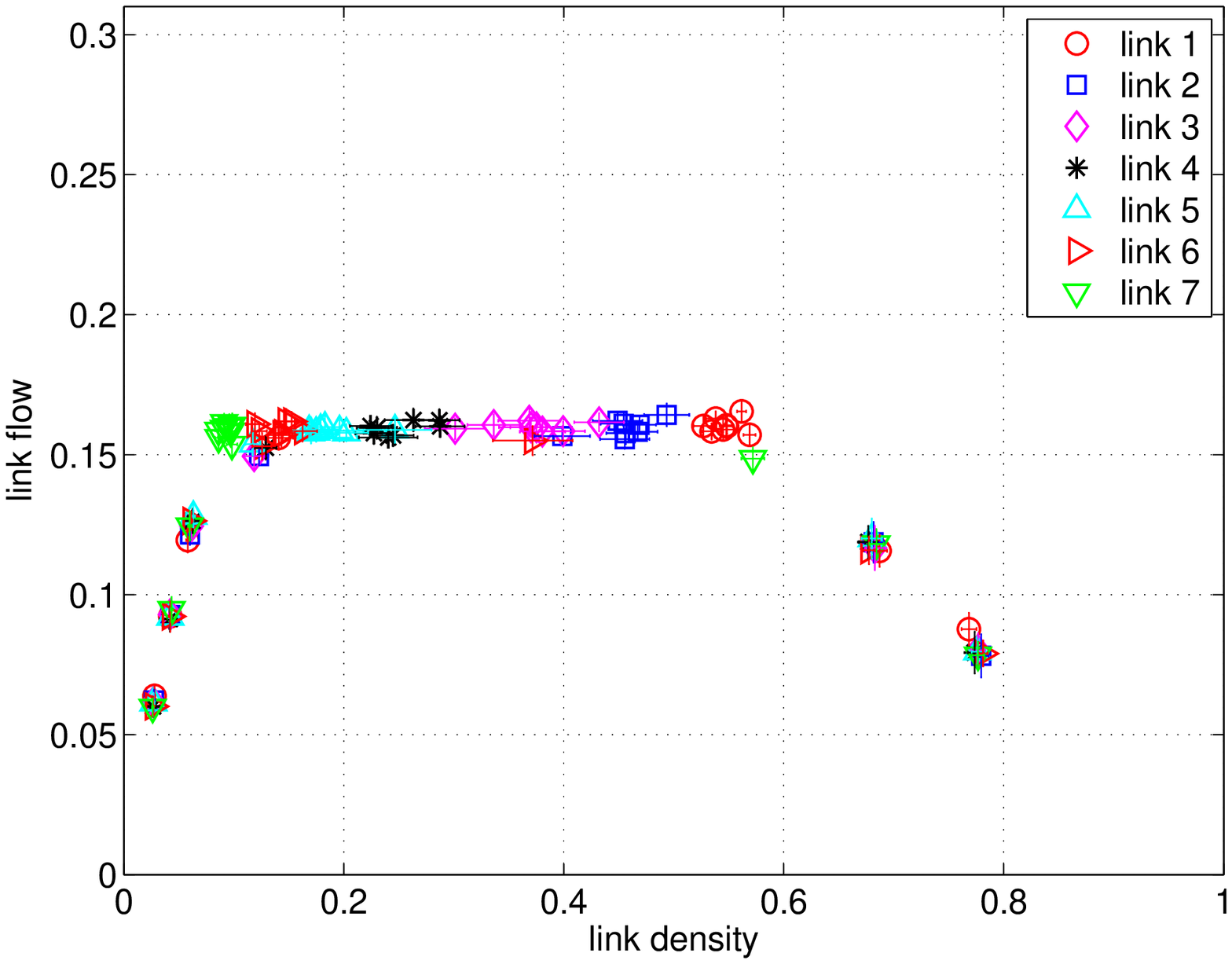}}

\def\dwmodelANrouteflowLD		{\includegraphics[scale=\oneup]{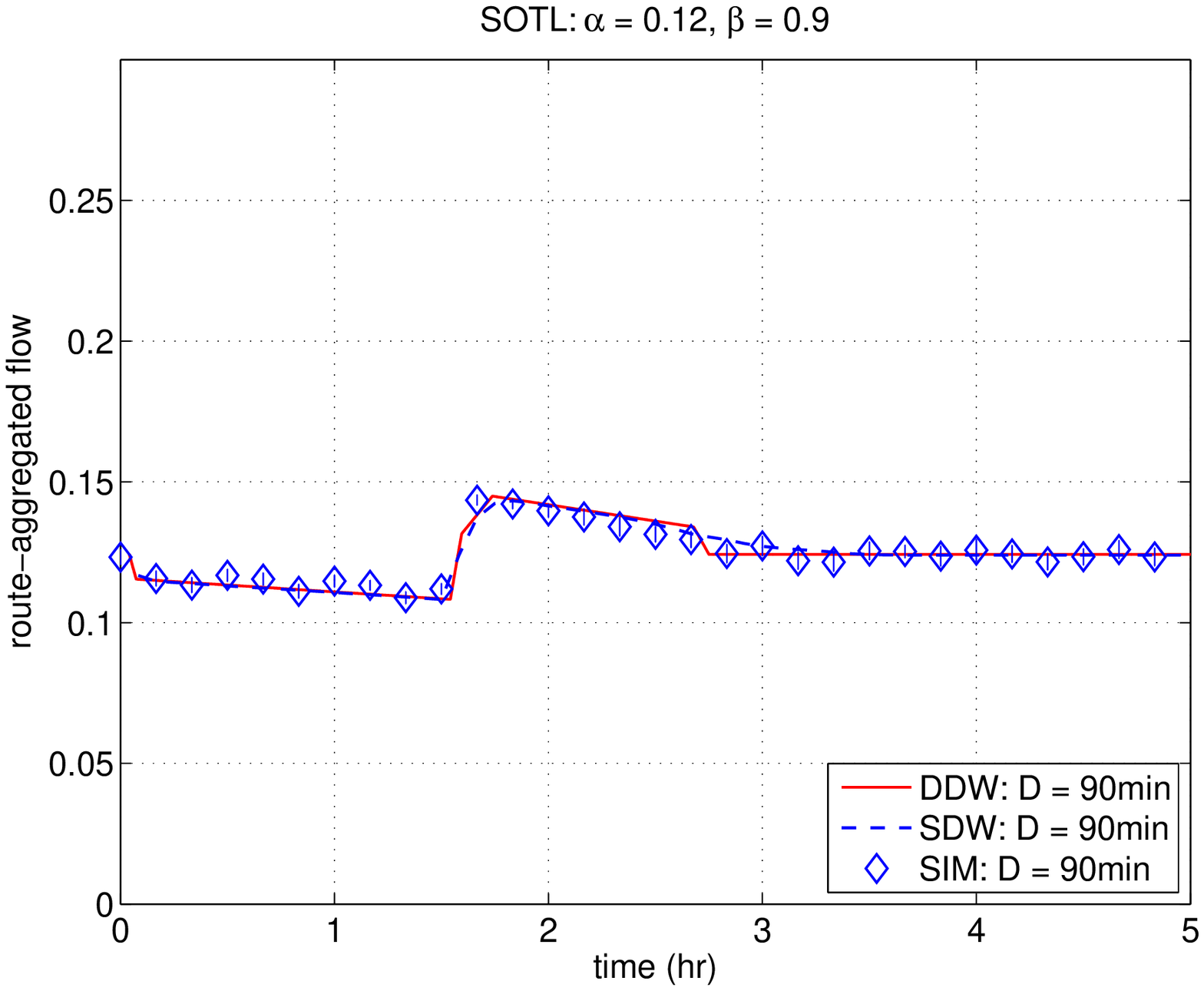}}
\def\dwmodelANroutedensityLD	{\includegraphics[scale=\oneup]{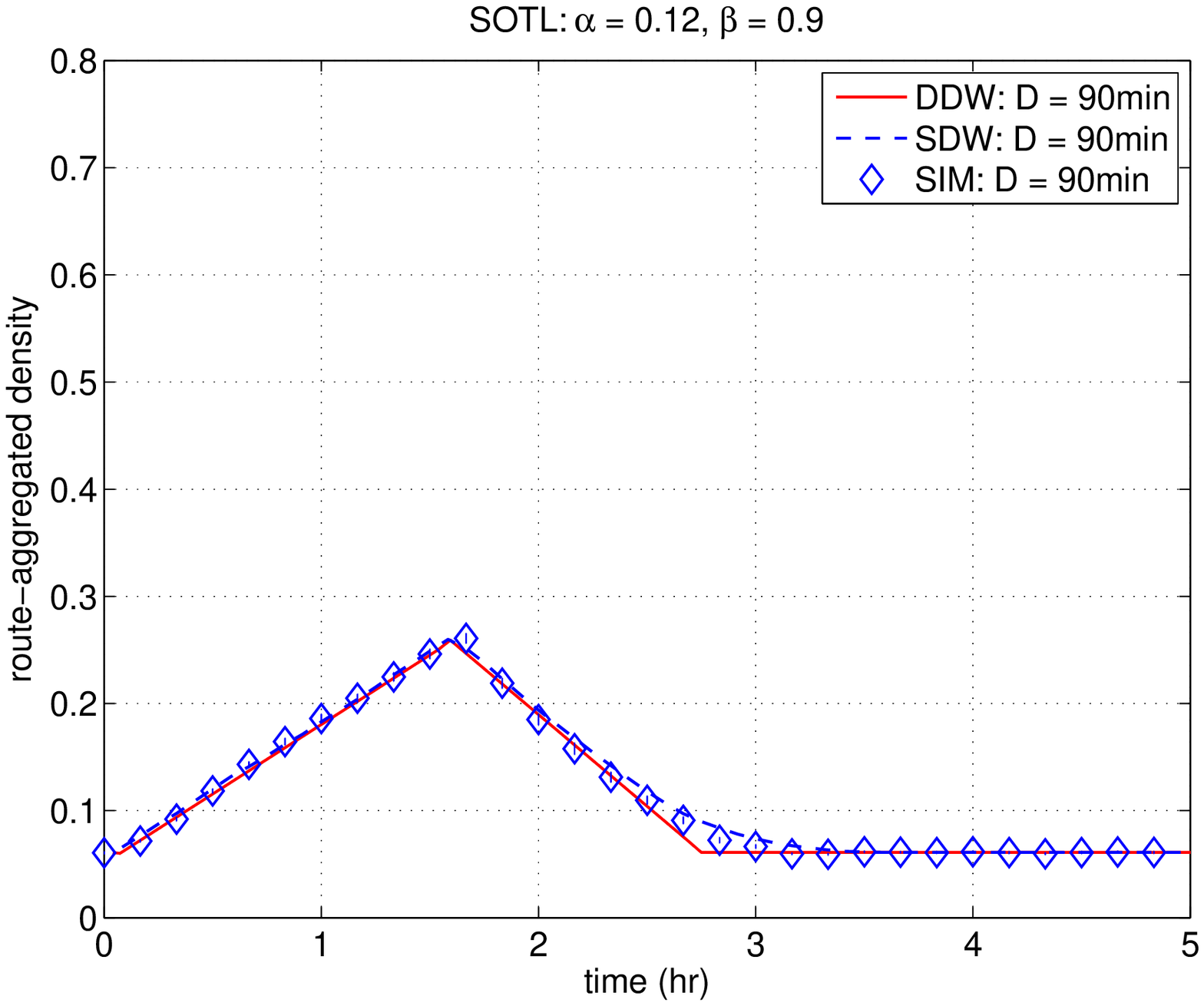}}
\def\dwmodelANrouteflowMC		{\includegraphics[scale=\oneup]{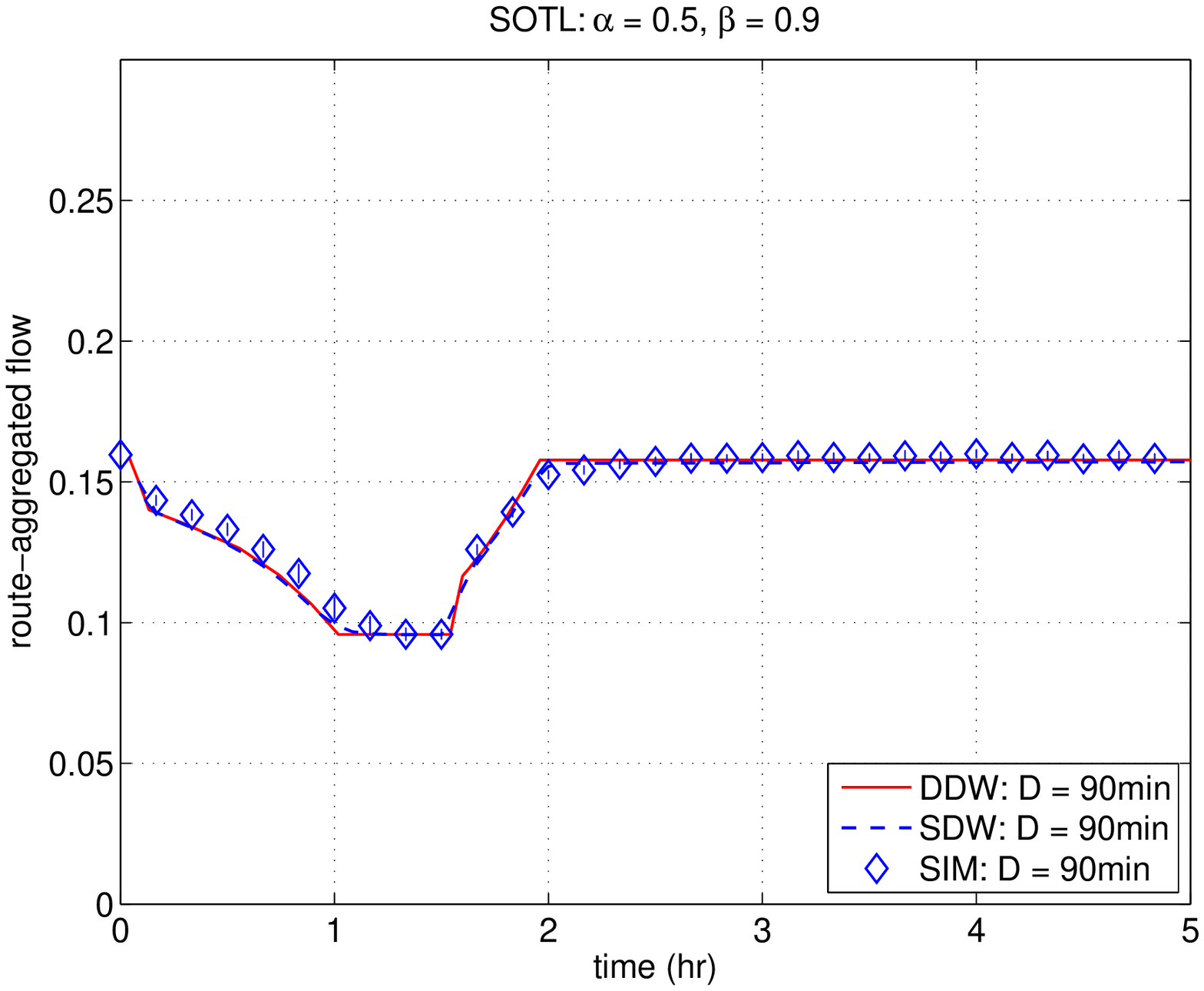}}
\def\dwmodelANroutedensityMC	{\includegraphics[scale=\oneup]{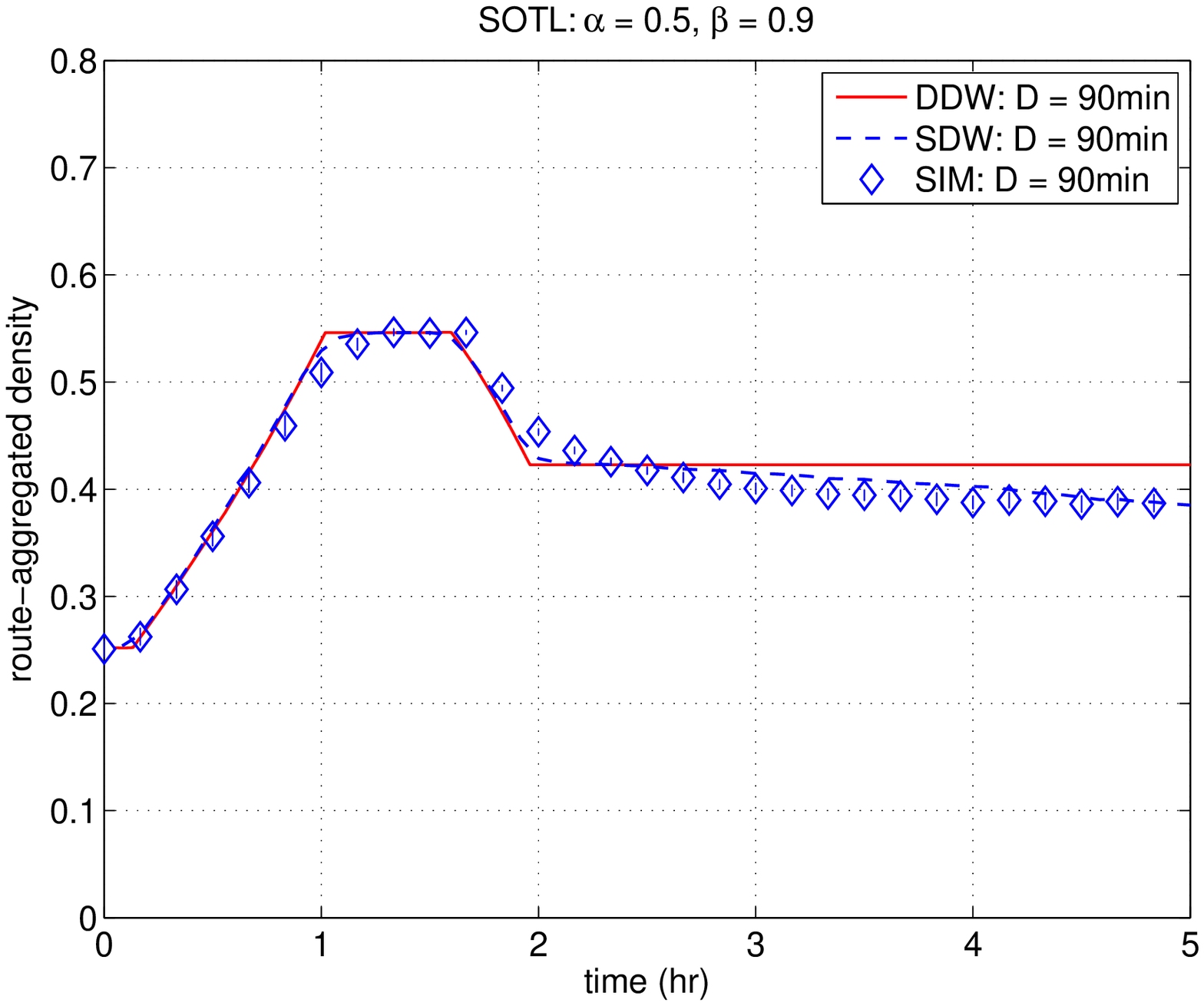}}
\def\dwmodelANrouteflowHD		{\includegraphics[scale=\oneup]{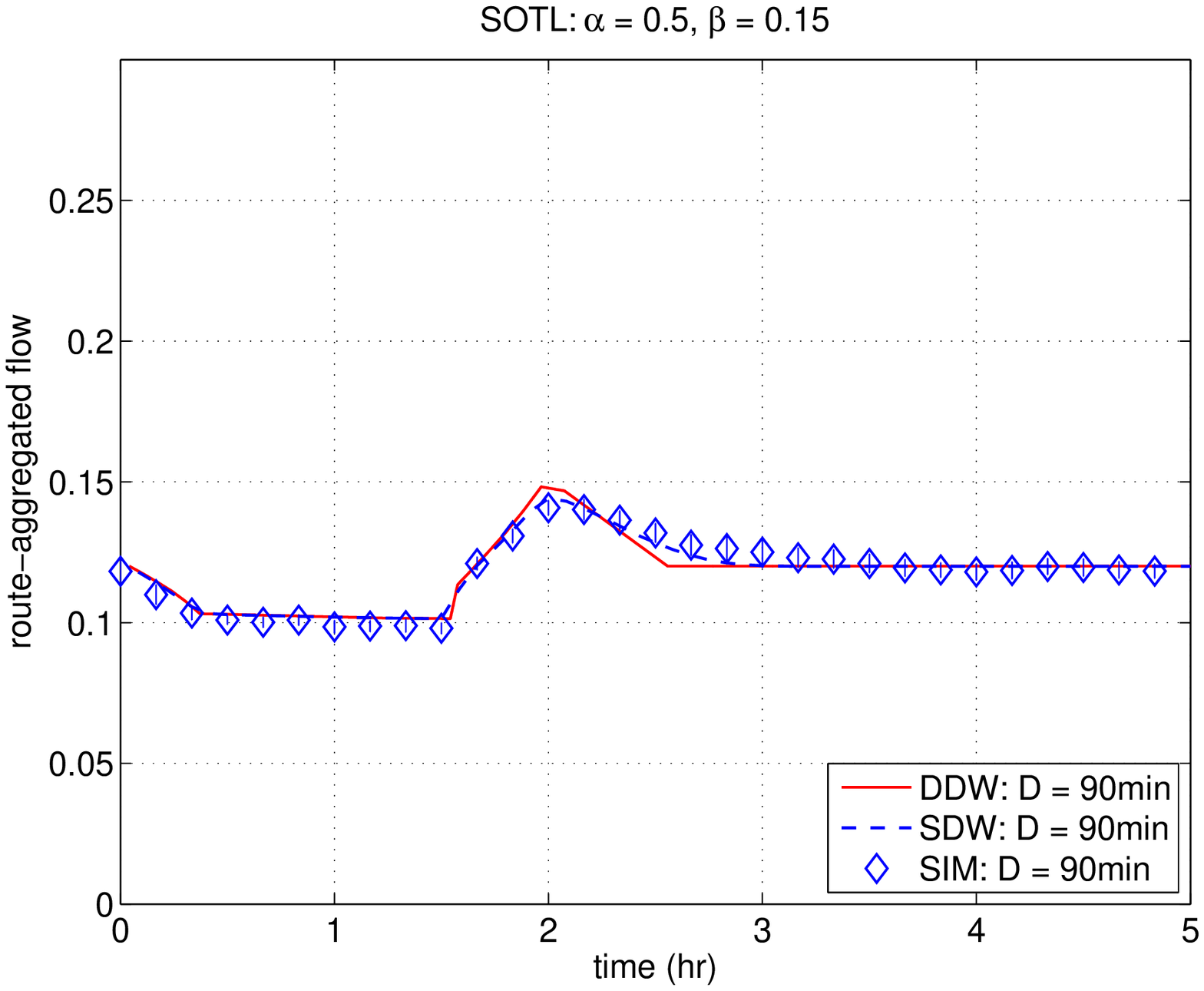}}
\def\dwmodelANroutedensityHD	{\includegraphics[scale=\oneup]{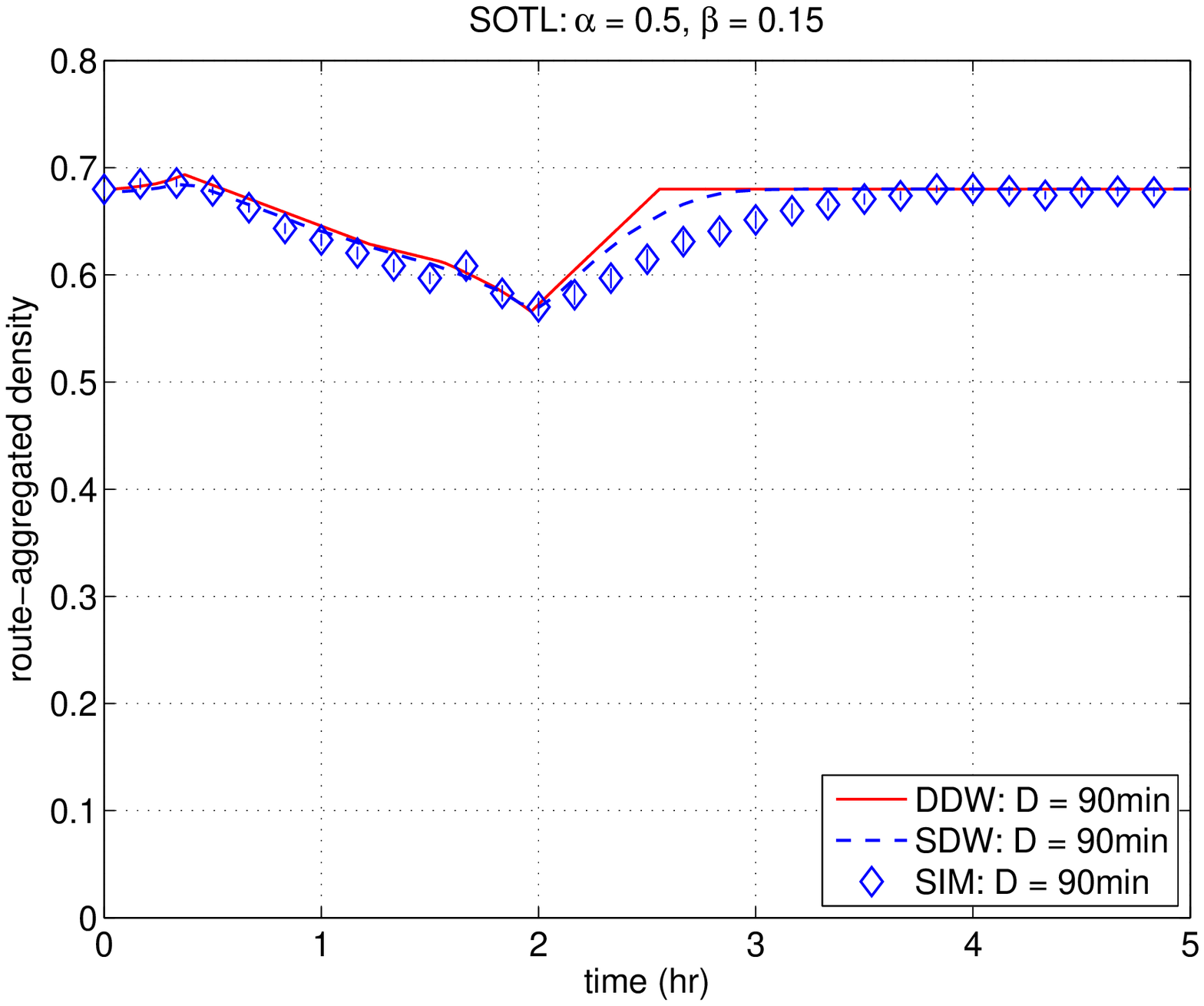}}

\def\dwmodelANrouteflowLDT			{\includegraphics[scale=\oneup]{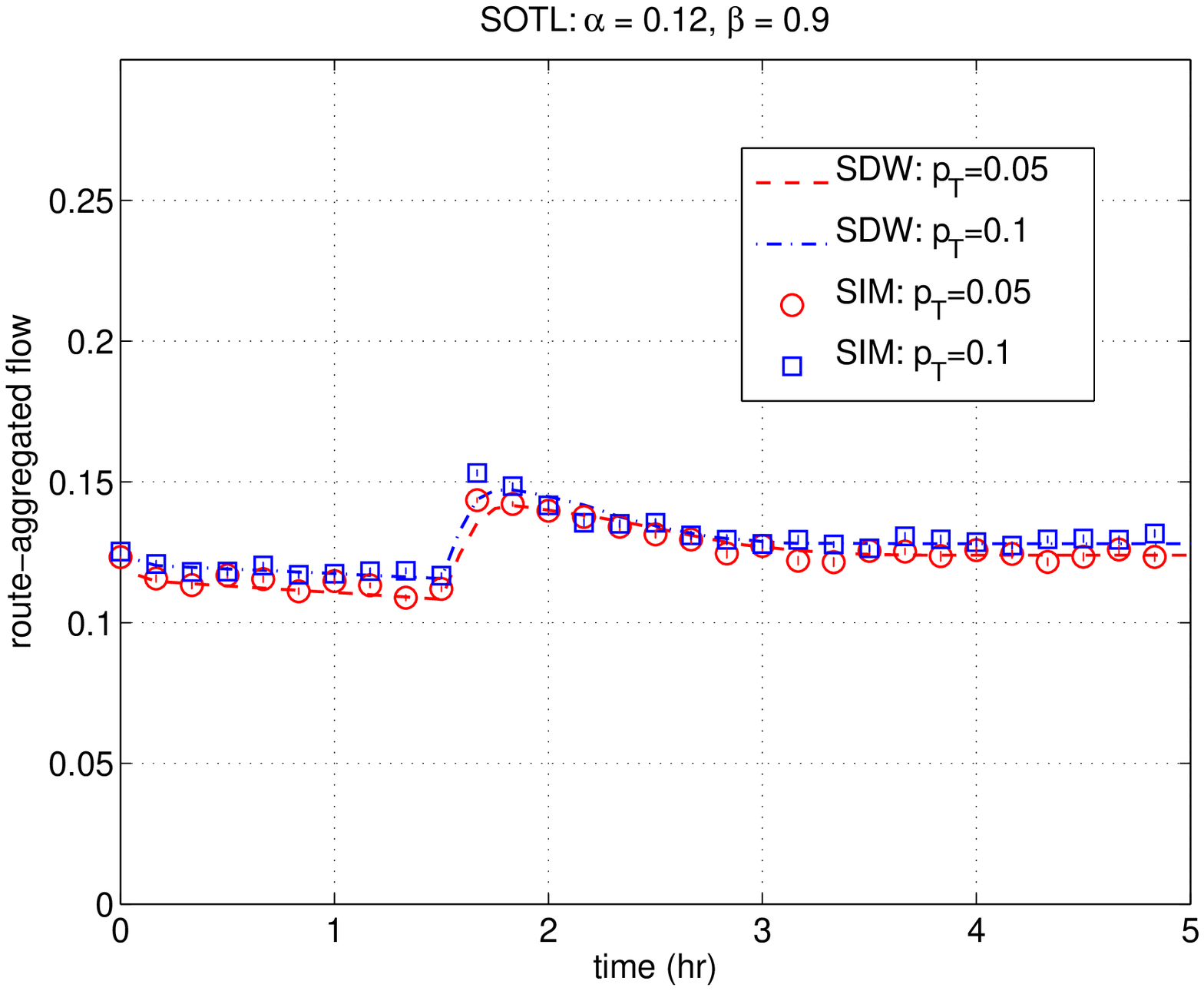}}
\def\dwmodelANroutedensityLDT		{\includegraphics[scale=\oneup]{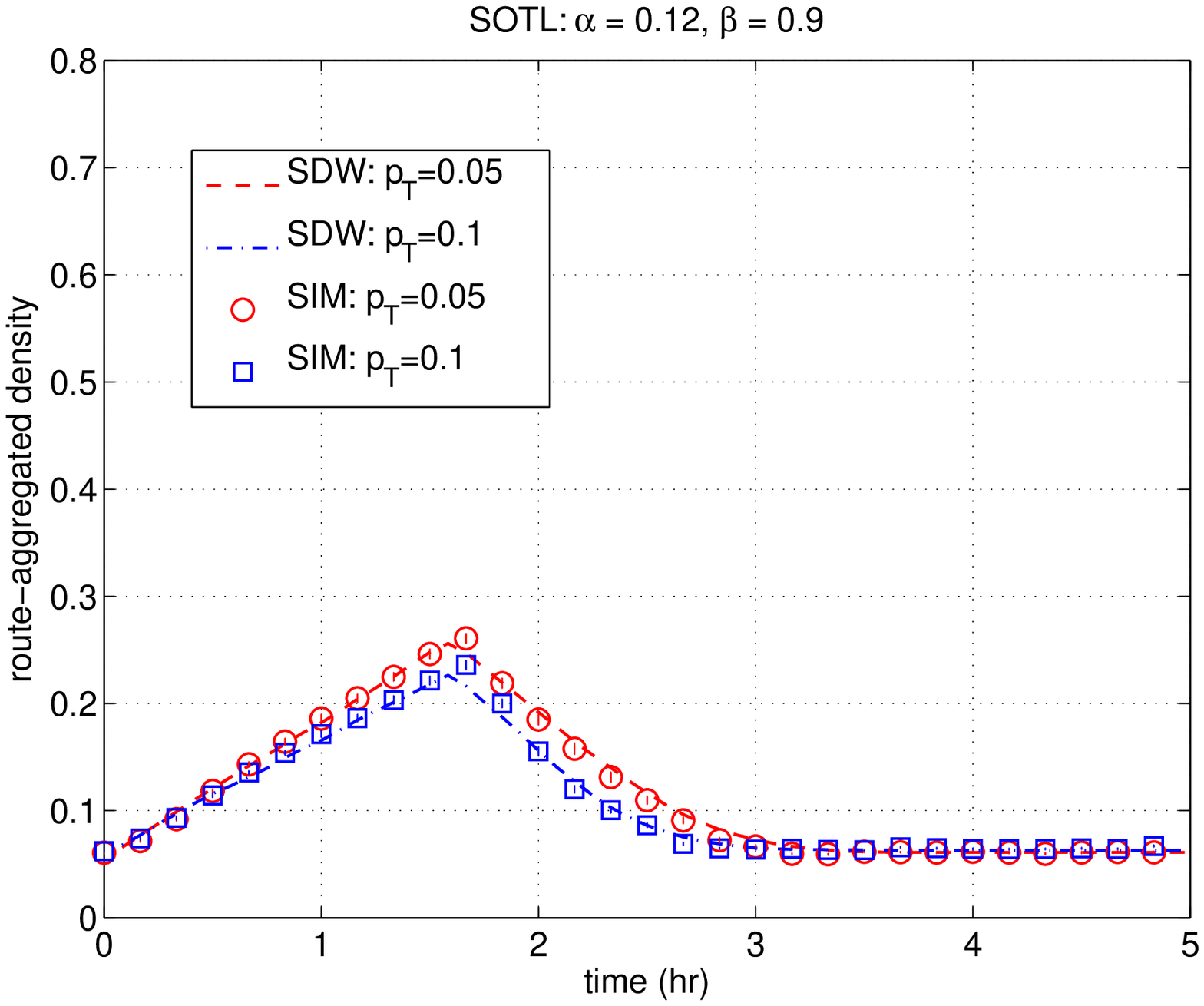}}
\def\dwmodelANrouteflowHDT			{\includegraphics[scale=\oneup]{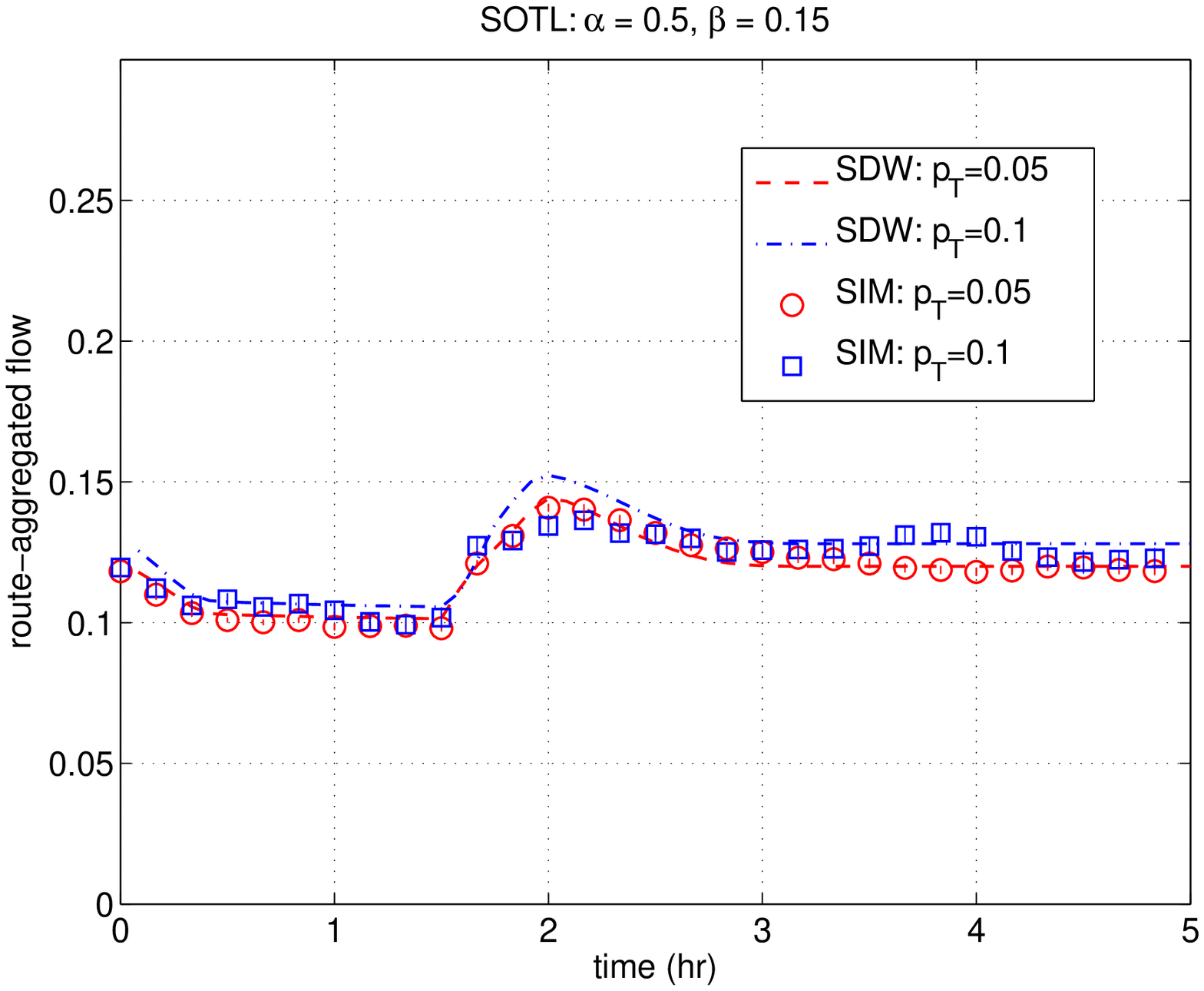}}
\def\dwmodelANroutedensityHDT		{\includegraphics[scale=\oneup]{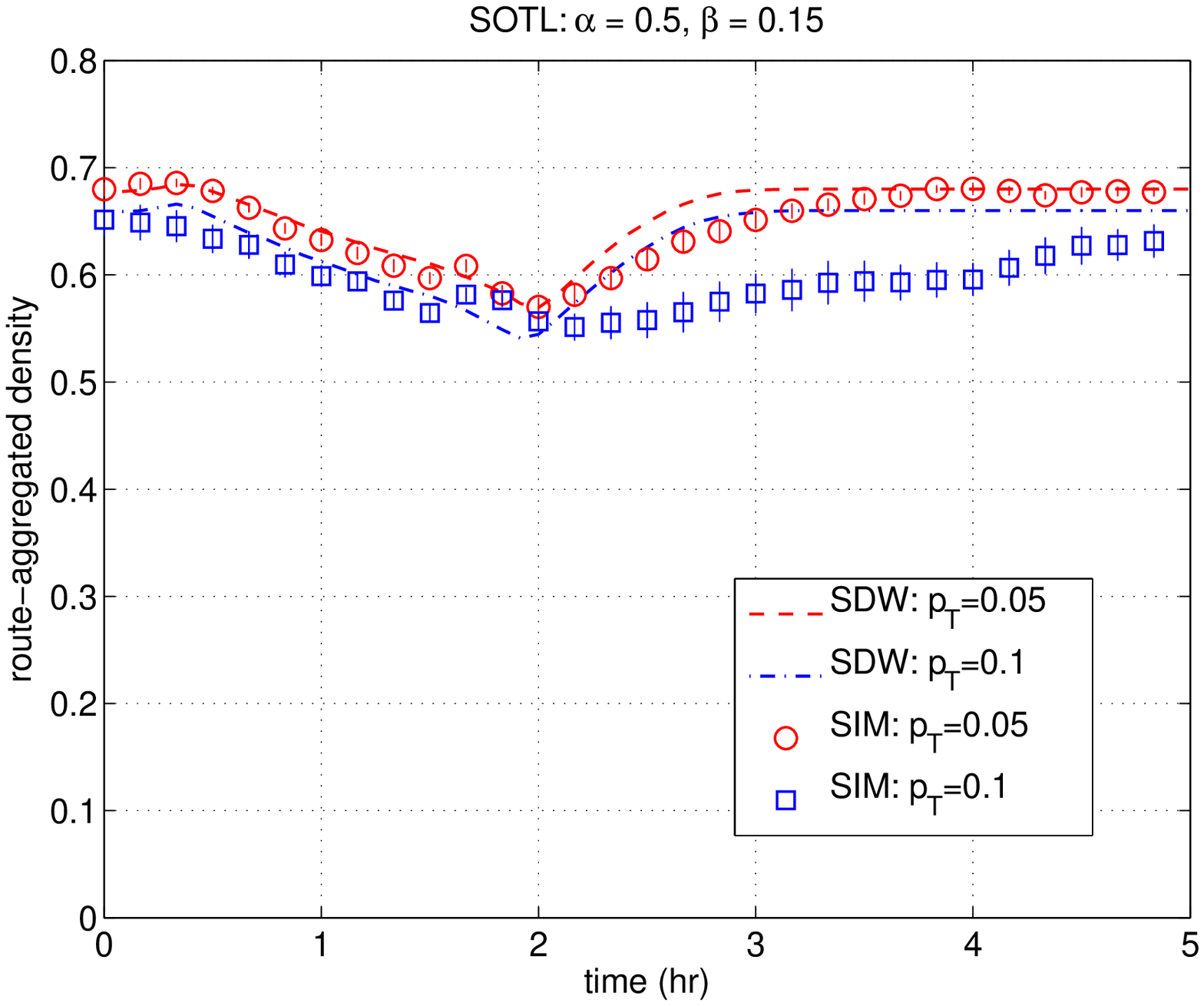}}

\begin{document}

\title{Traffic disruption and recovery in road networks}

{\author[monash]{Lele Zhang\corref{cor1}}
\ead{lele.zhang@monash.edu}
\author[melbourne]{Jan de Gier}
\ead{jdgier@unimelb.edu.au}}
\author[monash]{Timothy M. Garoni}
\ead{tim.garoni@monash.edu}

\cortext[cor1]{Corresponding author}
\address[monash]{School of Mathematical Sciences, Monash University, Clayton, Victoria~3800, Australia}
\address[melbourne]{Department of Mathematics and Statistics, University of Melbourne, Victoria~3010, Australia}

\begin{abstract}
We study the impact of disruptions on road networks, and the recovery process after the disruption is removed from the system.
Such disruptions could be caused by vehicle breakdown or illegal parking.
We analyze the transient behavior using domain wall theory, and compare these predictions with simulations of a stochastic cellular automaton model.
We find that the domain wall model can reproduce the time evolution of flow and density during the disruption and the recovery processes, for both one-dimensional systems and two-dimensional networks.
\end{abstract}
\maketitle

\section{Introduction}
\label{sec:intro}
The study of vehicular traffic has played an increasingly significant role in non-equilibrium statistical mechanics over recent years. 
A number of approaches, such as car following models, cellular automata and optimal velocity models have been applied to model traffic. 
The study of the impact of traffic bottlenecks, or {\em defects}, on network performance are of particular interest. 

Bottlenecks are a frequent cause of traffic congestion.
Many bottlenecks are {\em sitewise}, meaning the location of the bottleneck does not change in time.
Such bottlenecks can be further classified as either slowdown-regions or lane-reductions.
Ramps, slopes and bad weather act effectively as slowdown bottlenecks, whereas vehicle breakdown, illegal parking and road-work are the main causes of lane reduction. One could also view intersections with traffic lights as sitewise time-dependent bottlenecks.
The impact of slowdown bottlenecks for the asymmetric simple exclusion process (ASEP) was studied in \cite{JanowskyLebowitz92,JanowskyLebowitz94,Kolomeisky98}. 
Such studies have been extended to freeway networks; see for example \cite{Diedrich2000,HanauraNagataniTanaka07,IshibashiFukui01,KomadaMasukuraNagatani09,NagaiHanauraTanakaNagatani06,TanakaNagaiNagatani06}.

Compared to the extensive study of slowdown bottlenecks, there are few studies of lane reduction.
Recently, \cite{ZhangLiWangSunCui12} studied the traffic characteristics near a lane reduction bottleneck using the optimal velocity model.
In this model there is a merging section upstream of the bottleneck where vehicles slow down and make asymmetric lane changes in order to merge into the unblocked lane.
It was found that the road capacity increased as the length of the merging section decreased, and as the maximum speed in that section increased. In effect, the merging area acts as a slowdown bottleneck.
A similar scenario was studied in~\cite{JiaJiangWu03} using a two-lane cellular automaton model with a localized lane-reduction bottleneck.
It was found that the capacity at the bottleneck was slightly smaller than that of a single-lane road, and the density distribution in the merging area depended on its length.
A similar study was performed in~\cite{EbersbachSchneider04}.

It is of considerable theoretical and practical interest to study the impact of bottlenecks using macroscopic evolution equations, 
in order to obtain robust results which are independent of the specific microscopic details inherent in any particular model/system.
Domain wall (DW) theory \cite{KolomeiskySchutzKolomeiskyStraley98,SantenAppert02} is a phenomenological approach which has proved 
very successful in explaining both the stationary and transient behavior for ASEP, and is expected to be applicable to a rather general class particle transport systems.

The aim of the current work is to study the impact of lane-reduction defects on both one-dimensional and two-dimensional traffic networks. 
For the stationary state, we study the impact of such defects on both the phase diagram and the fundamental diagram.
In the transient regime, we discuss an extended domain wall model which allows multiple domain walls to exist in the system simultaneously.
We then compare the time evolutions of density and flow produced by the domain wall model with those produced by simulating a stochastic cellular automaton (CA) model.

For one-dimensional systems, the CA model we use consists essentially of two parallel NaSch models, with the addition of simple lane-changing rules.
For two-dimensional networks, we used the NetNaSch model \cite{deGierGaroniRojas11}. 

The remainder of this paper is organized as follows.
We study one-dimensional systems in Section~\ref{sec:1D}.
After giving a brief outline of the CA model and discussing the key features of the DW theory,
we study the phase diagram and fundamental diagram of the perturbed system.
We then use the domain wall model to study the response of the system to the imposition, and subsequent removal, of a defect (lane-reduction bottleneck), and then compare these results with 
those of our CA simulations.
In Section~\ref{sec:2D} we extend these studies to a two-dimensional network.
Finally, we conclude with a discussion in Section~\ref{sec:discussion}.

\section{One-Dimensional Traffic System}
\label{sec:1D}
In this section we study a one-dimensional traffic system of length $\mL$.
A localized defect (traffic disruption) is imposed on the system at location $\mx$, for duration $D$.
Prior to the imposition of the defect, the system is in a stationary state.
We consider two distinct models of this scenario, both of which are spatially and temporally discrete; 
a cellular automaton, based on the Nagel-Schreckenberg model, and a simple random walk model of domain walls.

\subsection{NaSch model with defects}
\label{ssec:defects}
The CA model studied here consists of two lanes oriented in the same direction.
The dynamics along each lane is governed by the Nagel-Schreckenberg (NaSch) model~\cite{NagelSchreckenberg92}, and additional rules governing lane-changing are imposed.
Each lane is discretized into cells, each of which can be either occupied by a vehicle or empty.
Each vehicle can move $0, 1, 2, \ldots,v_{\max}$ cells per time step, depending on local traffic conditions.
A random unit deceleration is applied with probability $\pnoise$.
We use open boundary conditions, and so the density in the network is not directly controlled.
On each lane, at each time step, vehicles are inserted into the system with rate $\alpha$.
The precise inflow mechanism used on each lane is that described in \cite{BarlovicHuisingaSchadschneiderSchreckenberg02}, to ensure the unperturbed system can reach the maximum flow regime.
A vehicle at the end of the system which is traveling with sufficient speed is allowed to exit with an output probability $\beta$.

Two alternative distinct types of lane changing rules were considered. The first type of lane changing corresponds to the {\em dynamic lane changing} rules described in~\cite{deGierGaroniRojas11}.
In essence, these rules allow a vehicle which could achieve a higher speed by making a lane change to change lanes provided no vehicle in the neighboring lane is forced to suddenly decelerate.
We shall refer to these lane changing rules as {\em symmetric}, since they apply symmetrically to left-to-right and right-to-left lane changes.

When the system is in stationarity, we impose a localized blockage in cell $\mx$ of the left lane.
A second set of lane changing rules are implemented to allow vehicles to navigate past the defect.
We refer to these rules as asymmetric lane changing, since they allow lane changes from left to right but not right to left.
In our model, vehicles are not aware of the defect until they arrive at the cell immediately upstream of it.
Vehicles in the left lane then change to the right lane whenever the adjacent cell is unoccupied, which may cause vehicles in the right lane to decelerate.
Once such vehicles pass the defect, they may change back to the left lane or not, according to the rules of symmetric lane changing.

In order to understand the effect on the system of the lane reduction we will focus on the density and flow.
The density of cell $i$ at time $t$ is simply the indicator for the event that cell $i$ is occupied at time $t$.
This quantity is a stochastic process, and we denote its expected value by $\rho(i,t)$.
The density of the cell containing the defect is not defined. For concreteness, we define the right-lane cell adjacent to the defect to belong to the downstream subsystem.
The density of the system $\rho(t)$ is then simply the arithmetic mean of $\rho(i,t)$ over all cells $1\le i\le L$.

Similarly, the flow from cell $i$ to cell $i+1$ along a given lane at time $t$ is simply the indicator for the event that a vehicle crosses the boundary between cells $i$ and $i+1$
during the $t$th time step. We denote its expected value by $J(i,t)$.
The flow per lane $J(t)$ is then simply the arithmetic mean of $J(i,t)$ over all cells $1\le i\le L$~\footnote{In practice, in our simulations we measured the flow only every 100 cells.}.
For simplicity, henceforth we shall take it as given that when we refer to density and flow we are referring to their expected values.
We will denote the corresponding stationary ($t\to\infty$) values of the system density and flow by $\rho$ and $J$ respectively.

\subsection{Domain wall model}
\label{ssec:1D_dwm}
The domain wall model has been shown to be capable of explaining the behavior of ASEP in both the stationary~\cite{KolomeiskySchutzKolomeiskyStraley98} and transient regimes~\cite{SantenAppert02}.
Domain walls are shocks separating two regions of different density. 
We shall assume that the width of such shocks is small compared to the macroscopic system size, so that to a good approximation we can consider the position of the domain wall to be a single point.

\begin{figure}[!t]
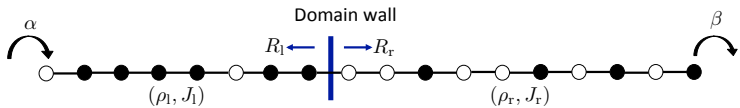

\centering
\domainwall
\caption{Illustration of domain wall motion in a one-dimensional system with $L=18$. Each circle corresponds to a cell. Solid circles are occupied while open circles are empty. States left and right to the domain wall are with density and flow $(\rho_{\rm l},J_{\rm l})$ and $(\rho_{\rm r},J_{\rm r})$ respectively.}
\label{fig:domain_wall}
\end{figure}

The motion of such a domain wall $W_{\rm l | r}$, separating states with density and flow $(\rho_{\rm l},J_{\rm l})$ to its left and $(\rho_{\rm r},J_{\rm r})$ to its right,
is described by a simple biased random walk with hopping rates
\be
R_{\rm l} = \frac{J_{\rm l}}{\rho_{\rm r}-\rho_{\rm l}},
\,\,\mathrm{ and }\,\,
R_{\rm r} = \frac{J_{\rm r}}{\rho_{\rm r}-\rho_{\rm l}},
\label{eqn:hoppingrates}
\ee
for moves to the left and right, respectively. See e.g. Fig.~\ref{fig:domain_wall}.
This random walk takes place on a finite linear chain of sites. 
For comparison with the cellular automaton described above, these sites can be considered as the boundaries between adjacent cells, and the cells themselves then correspond to links between adjacent sites. 
The possible positions of the domain walls are then constrained to the sites $0,1,\ldots,L$, while the vehicles themselves reside on cells $1,2,\ldots,L$.

The probability $\pr(i,t+1)$ of the wall $W_{\rm l | r}$ being at site $i$ at time $t+1$ satisfies the following simple biased diffusion,
\be
  \hspace{-0.0cm}
  \pr(i,t+1) = \left\{
  \begin{tabular}{ll}
    $R_{\rm r}\, \pr(i-1,t)+R_{\rm l}\, \pr(i+1,t)+[1-(R_{\rm r}+R_{\rm l})]\, \pr(i,t)$,&$0<i<\mL$\\
    $R_{\rm l}\, \pr(1,t)+(1-R_{\rm r})\, \pr(1,t)$, & $i=0$\\
    $R_{\rm r}\, \pr(\mL-1,t)+(1-R_{\rm l})\, \pr(\mL,t)$, & $i=L$.\\
    \end{tabular}\right.
  \label{eqn:wallmotion_mid}
\ee
The corresponding drift velocity is
\be
V_{{\rm l | r}}=\frac{J_{\rm l}-J_{\rm r}}{\rho_{\rm l}-\rho_{\rm r}}.
\label{eqn:wallspeed}
\ee

In order to model the effect of the imposition and later removal of the traffic disruption, we will need to allow for the possibility of multiple domain walls coexisting in the system simultaneously.
We assume that the dynamics of each wall is independent of all the other walls, unless two walls choose to move to the same site. In such a case the two walls immediately merge to form a single wall.
The number of walls present in the system at time $t$ is therefore random and not conserved.

As illustrated by the explicit form of the hopping rates in~(\ref{eqn:hoppingrates}), each domain wall is associated with a specific density gradient between the density on its left and the density on its right.
These densities are assumed to be bulk quantities that are homogeneous between consecutive domain walls.
For a given cell $i$, let $\varrho(i,t)$ denote the value of $\rho_{\rm l}$ corresponding to the domain wall immediately to the right of $i$ at time $t$.
Clearly $\varrho(i,t)$ is a well-defined stochastic process induced by the dynamics of the domain walls, and
\be
\rho(i,t) = \EE\,\varrho(i,t).
\label{eqn:celldensity}
\ee

As a simple deterministic approximation to the stochastic dynamics described above, the position of a domain wall at time $t+1$ can be approximated via the recurrence
\be
P(t+1) = \max\{1,\min\{P(t) + \lfloor V_{\rm l | r} \rfloor, \mL\}\}.
\label{eqn:dw_speed_position}
\ee
For multiple walls, similar recurrences are again run independently unless multiple walls occupy the same site. Merging of shocks for ASEP was studied in \cite{Belitsky2002,Ferrari2000}. In this paper, for simplicity, we assume that the walls are immediately merged into a single wall. 
We note that in this approximation the system density is simply a linear function of the drift speeds.
We will refer to this deterministic approximation as the {\em deterministic domain wall} (\ddw) model.
When the distinction between the domain wall model and this deterministic approximation requires emphasis we shall refer to the domain wall model as the {\em stochastic domain wall} model.

\subsection{Stationary state}
\label{ssec:1D_stationary}
We now discuss the effect of a lane reduction on the stationary state. In particular, we study the impact on the fundamental diagram (FD).

The behavior of an ASEP in the presence of a localized slowdown defect has been studied previously~\cite{JanowskyLebowitz92,Kolomeisky98}.
In \cite{Kolomeisky98}, a continuous-time ASEP with open boundary conditions was studied, in which the hopping rate of a fixed cell was rescaled by $q < 1$.
A mean-field approximation showed that the perturbed FD is the same as the unperturbed FD when $\rho< q/(q+1)$ or $\rho> 1/(q+1)$, however the subsystems upstream and downstream
of the defect were unable to achieve densities in the range $(q/(q+1),1/(q+1))$.
At system density $1/2$, the system achieved a maximum flow of $q/(q+1)^2$, compared with $1/4$ for the unperturbed system.

\begin{figure}[!t]
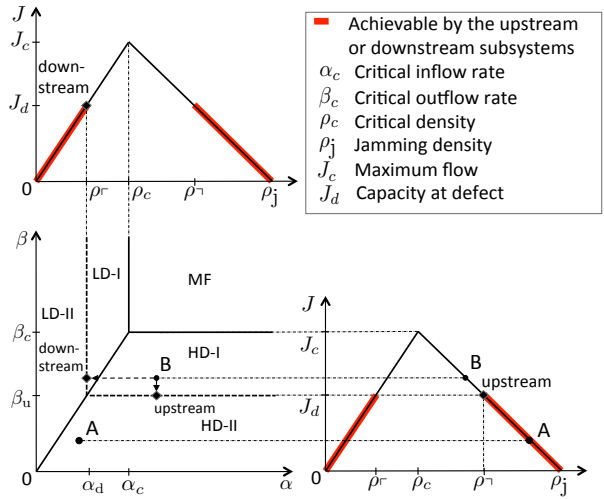

\centering
\phasediagram
\caption{Phase diagram and fundamental diagram of the two-lane system with a defect based on the domain wall model.
In the phase diagram $\alphac$ and $\betac$ are critical rates that separate the maximum-flow phase from the high-density and low-density phases.
$\alpha_{\rm d}$ and $\beta_{\rm u}$ are respectively the maximum inflow and outflow rates for the downstream and the upstream subsystems, in the presence of the defect.
If the initial state is A, then the defect does not have an impact on the system.
If the initial state is B, the upstream section becomes more congested and the downstream section becomes free-flow after the system reaches the perturbed stationary state.
}
\label{fig:phase_diagram}
\end{figure}

Our simulations show qualitatively similar behavior for the two-lane NaSch model with a lane-reduction described in Section~\ref{ssec:defects}.
In order to describe what happens in this case, we partition the system into two subsystems, upstream and downstream of the defect. Both subsystems exclude the defect. 
We make the following physically reasonable assumptions for a domain wall interpretation of the perturbed system. Relevant notation is given in Fig.~\ref{fig:phase_diagram}.
\begin{enumerate}
\item\label{FD assumption} Neglecting finite-size effects, the upstream and the downstream subsystems should have the same FD as the unperturbed system.
\item When the defect is introduced, the capacity at the defect site reduces to $\Jd<\Jc$.
\item The maximum outflow rate for the upstream section reduces to $\beta_{\rm u}<\beta_{\rm c}$. This is because for the upstream section the defect is downstream and thus limits its outflow rate.
\item The maximum inflow rate for the downstream section reduces to $\alpha_{\rm d}<\alpha_{\rm c}$.
\end{enumerate}

The phase diagram based on those assumptions for the system with defect is shown in Fig.~\ref{fig:phase_diagram}.
If the unperturbed system is in either sub-phase LD-II ($\alpha \leq \alpha_{\rm {d}}$) or HD-II ($\beta \leq \beta_{\rm u}$), 
then the presence of the defect does not affect the state of either the upstream or downstream subsystem.
Otherwise, phase separation occurs. 
For example, suppose that the unperturbed system is in state B in Fig.~\ref{fig:phase_diagram}.
Achieving state B requires an inflow rate in the downstream subsystem higher than $\alpha_{\rm {d}}$ and an outflow rate of the upstream subsystem higher than $\beta_{\rm u}$.
Such rates cannot be achieved after introducing the defect.
The decrease in the outflow rate results in congestion in the upstream section, whereas the decrease in the inflow rate results in free-flow in the downstream section.

In terms of the FD, if the initial flow $\Jo$ is smaller than the capacity at the defect site $\Jd$, implying that the system is in either in the LD-II or HD-II regimes, 
then neither the flow nor the density are affected.
Otherwise, the flow of the perturbed system decreases to $\Jd$.
Let $\rhodl$ and $\rhodr$ be the unique densities satisfying $J(\rhodl)=\Jd=J(\rhodr)$ with respect to the unperturbed FD, with $\rhodl<\rhoc<\rhodr$.
In the presence of the defect, the upstream region will be congested and the downstream region will be in free-flow, so 
assumption~(\ref{FD assumption}) implies that the the upstream region has density $\rhodr$ while the downstream region has density $\rhodl$.
Assumption~(\ref{FD assumption}) also implies that neither the upstream nor the downstream system can achieve densities in the range $(\rhodl, \rhodr)$, nor flows in the range $(\Jd,\Jc]$.

\begin{figure}[!t]
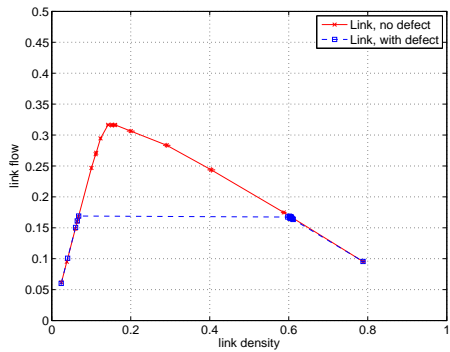

\centering
\LKfds
\caption{Fundamental diagram for the subsystem upstream of the defect in the one-dimensional system from simulations.
  The solid curve corresponds to the unperturbed FD, whilst the dashed curve illustrates the unachievable region in the perturbed FD.
  Error bars corresponding to one standard deviation are shown.}
\label{fig:1d_fds}
\end{figure}

Fig.~\ref{fig:1d_fds} confirms this picture for the two-lane NaSch model with lane reduction.
The figure shows the FDs of the upstream section in the unperturbed and the perturbed stationary states.
We observe that all states $(\alpha,\beta)$ corresponding to the density range $(\rhodl, \rhodr)$ for the unperturbed system collapse to the point $(\rhodr,\Jd)$ in the perturbed system.
The FDs of the downstream section are essentially the same, except that all states $(\alpha,\beta)$ corresponding to densities in the range $\rho\in(\rhodl, \rhodr)$ 
collapse to $(\rhodl,\Jd)$ on the FD of the perturbed system.

The capacity for the perturbed system is approximately half of that for the unperturbed system.
The aggregated flow of the two-lane link in the presence of the defect, which is $0.33$ ($=2J_{\rm d}$), is slightly higher than the capacity of a single-lane NaSch model with the same parameters, which we estimate to be $\approx0.295$.
Previous studies~\cite{JiaJiangWu03} found that the capacity of a two-lane system with a lane-reduction defect would be slightly less than the single-lane capacity.
The likely explanation is that the size of the defect in our simulations is localized to a single cell, and hence it has lower impact on the system.

In summary, based on the impact of the defect, we divide the FD into 5 segments:
\begin{enumerate}[label=(\alph*)]
\item\label{seg a} the range $0\leq \rho\leq \rhodl$. The defect has no impact on the FD.
\item\label{seg b} the free-flow regime between segment~\ref{seg a} and capacity. Defects have a significant impact on the system.
\item\label{seg c} the capacity regime, where $J=\Jc$. 
  For one-dimensional systems, this corresponds to a single value of the density $\rho=\rhoc$, but it may correspond to an interval of densities for two-dimensional networks.
\item\label{seg d} the high-density regime between segments~\ref{seg c} and~\ref{seg e}. Defects have moderate impact on the system.
\item\label{seg e} the range $\rhodr \le \rho \le \rho_{\rm j}$. The defect has no impact on the FD.
\end{enumerate}

Finally, we note that to construct the unperturbed FD that we used as input to the DW theory we performed a simple linear interpolation of the simulated FD shown in Fig.~\ref{fig:1d_fds}.

\subsection{Transient behavior}
\label{ssec:1D_transient}
In this subsection we introduce a domain wall model to interpret the system's transient behavior in the disruption and recovery processes.
After the unperturbed system reaches an initial stationary state, we impose the defect for duration $D$.
Depending on the value of $D$, the system may or may not have reached the perturbed stationary state when the defect is removed.

Given that the defect does not affect the state of the system if the state of the unperturbed system corresponds to segments~\ref{seg a} or~\ref{seg e},
unless stated otherwise, we assume that the initial system is in a state corresponding to segments~\ref{seg b}, \ref{seg c} or~\ref{seg d}.
The perturbed system is dynamically partitioned into several regions of different densities by domain walls.
Within each region, density and flow are assumed to be homogeneous.

\subsubsection{Disruption process}
We assume a defect is inserted at link $\mx$ (cell $\mx$ in the CA model) at time $t=0$, and we study the transient behavior of the resulting process.
In this disruption process, the defect cell effectively acts as boundary.
The domain walls in the upstream and downstream subsystems do not interact with each other.
We can therefore discuss them separately.

\begin{figure}[!t]
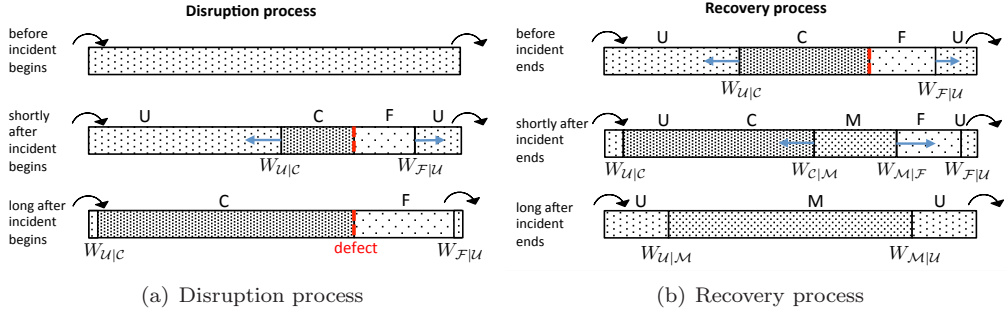

\centering
\subfigure[Disruption process\label{subfig:disruption_demo}] {\LKregiondemo}\subfigure[Recovery process\label{subfig:recovery_demo}]{\LKrecoveryB}
\caption{Defect disruption and recovery processes. Arrows show the expected direction of motion. The motion (and existence) of $W_{\mU|\mM}$ and $W_{\mM|\mU}$ depends on the initial traffic state.}
\label{fig:disruption_recovery_demo}
\end{figure}

During the disruption process, let $\mU$ denote the region which is still in the unperturbed state.
In the upstream subsystem, introducing the defect induces a local high-density region, denoted by $\mC$.
It follows that a domain wall $W_{\mU|\mC}$ forms separating $\mU$ and $\mC$, which drifts upstream, 
see Fig.~\ref{subfig:disruption_demo}.
Conversely, in the downstream subsystem, a low-density region $\mF$ is formed.
Between $\mF$ and $\mU$ a domain wall $W_{\mF|\mU}$ forms, which drifts downstream.
The wall $W_{\mU|\mC}$ is confined to sites $0\le i <\mx$, whereas the wall $W_{\mF|\mU}$ is confined to sites $\mx \leq i \le L$, so no merging of walls is possible during the disruption process.
The system has reached the perturbed stationary state when both $W_{\mU|\mC}$ and $W_{\mF|\mU}$ have arrived at the boundaries, where they then remain localized.

By definition, $\rho_\mU=\rho_{\rm o}$ and $J_\mU=\Jo$, where $\rho_{\rm o}$ and $\Jo$ are the density and flow in the unperturbed stationary state.
We assume that domains $\mC$ and $\mF$ are in the perturbed stationary state for the upstream and the downstream subsystems respectively.
In the perturbed stationary state, the flows through the upstream and downstream subsystems should be equal and identical to the capacity at the defect $\Jd$.
Therefore, $\rho_\mC=\rhodr$, $\rho_\mF=\rhodl$, and $J_\mC=J_\mF=\Jd$. 
In practice, the values of the parameters $\rhodl$, $\rhodr$ and $\Jd$ were obtained numerically from the simulated stationary fundamental diagram.

From the behavior of the walls $W_{\mU|\mC}$ and $W_{\mF|\mU}$ one can obtain the transient behavior of the density profile $\rho(i,t)$, as described in Section~\ref{ssec:1D_dwm}.

\subsubsection{Recovery process}
\label{recovery process}
Compared with the disruption process, the recovery process is somewhat more complicated, especially if the defect is removed before the perturbed stationary state is reached.
Once the defect is removed, there is no mechanism to maintain the density jump at the defect. A domain $\mM$ emerges in the neighborhood of the defect.
Simulations show that $\mM$ is in the maximum flow state. A simple argument explains this observation.

Consider a system with a unimodal fundamental diagram. 
Suppose two copies of the system are concatenated, so that the inflow of system $\mB$ is the outflow of system $\mA$, and that these systems are in states with densities $\rho_{\mA}$ and $\rho_{\mB}$.
The inflow into system $\mB$ will be a function $\alpha_\mB=f(\rho_\mA)$ of the density of system $\mA$, and it is reasonable to assume $f$ is non-decreasing.
In the particular case that $\rho_\mA=\rho_\mB=\rhoc$, the entire concatenated system will be in maximum flow, implying $f(\rhoc)\ge\alphac$.
In the general case, the inequality $f(\rhoc)\ge\alphac$ together with the fact that $f$ is non-decreasing implies that whenever $\rho_\mA\ge\rhoc$ we have $\alpha_\mB=f(\rho_\mA) \ge f(\rhoc) \ge \alphac$.
Analogous arguments imply that $\beta_\mA = g(\rho_\mB) \geq g(\rhoc) \ge \betac$ whenever $\rho_\mB\le\rhoc$, with $g$ some non-increasing function.
In summary then, $\rho_\mA\ge\rhoc$ implies $\alpha_\mB\ge\alphac$ while $\rho_\mB\le\rhoc$ implies $\beta_\mA\ge\betac$.

Since $\rho_\mC>\rhoc$, applying the above argument to the regions $\mC$ and $\mM$ implies $\alpha_\mM\ge\alphac$.
Likewise, since $\rho_\mF<\rhoc$, applying the above argument to the regions $\mM$ and $\mF$ implies $\beta_\mM \geq\betac$. 
It follows that the region $\mM$ is at maximum flow.

The region $\mM$ expands via the motion of two domain walls, $W_{\mC|\mM}$ and $W_{\mM|\mF}$, which, according to (\ref{eqn:wallspeed}),
respectively drift upstream and downstream. Fig.~\ref{subfig:recovery_demo} illustrates the recovery process.
For simplicity, in the domain wall model we will forbid $W_{\mC|\mM}$ and $W_{\mM|\mF}$ to hop over the location (site $\mx$) of the (now removed) defect.
This implies that the recovery processes in the downstream and the upstream sections do not interfere with each other
before $W_{\mC|\mM}$ ($W_{\mM|\mF}$) meets $W_{\mU|\mC}$ ($W_{\mF|\mU}$).
When $W_{\mU|\mC}$ and $W_{\mC|\mM}$ meet, they merge to form a new wall $W_{\mU|\mM}$, provided $\mU\neq\mM$.
Depending on the state of $\mU$, the wall $W_{\mU|\mM}$ will remain localized at the in-boundary or drift downstream.
Analogous arguments apply to $W_{\mM|\mF}$ and $W_{\mF|\mU}$.
The recovery process is complete when $W_{\mU|\mM}$ and $W_{\mM|\mU}$ merge and the system returns to the initial state.

We can summarize how the recovery process depends on the initial conditions. Assuming that the system recovers from the perturbed stationary state:
\begin{itemize}
\item Maximum-flow case: If the unperturbed system is at maximum flow, then the system recovers after merges between $W_{\mU|\mC}$ and $W_{\mC|\mM}$, and between $W_{\mM|\mF}$ and $W_{\mF|\mU}$.
\item Low-density case:  If the unperturbed system is in low density,  then the domain wall $W_{\mU|\mM}$ will drift downstream and merge with $W_{\mM|\mU}$, which remains localized at the out-boundary.
\item High-density case: If the unperturbed system is in high density, then the domain wall $W_{\mM|\mU}$ will drift upstream   and merge with $W_{\mU|\mM}$, which remains localized at the in-boundary.
\end{itemize}

\subsection{Simulations}
\label{ssec:1D_sim}
We now test the above domain-wall model by comparing against simulations of the cellular automata model described in Section~\ref{ssec:defects}. The simulation parameters are given in Subsection~\ref{ssec:para_nasch}.

\subsubsection{Parameters for NaSch model}\label{ssec:para_nasch}
In our simulations, we used $v_{\max} = 3$, $\pnoise=0.5$, $\mL=700$ and $\mx = 550$. 
To set our time scale we make the usual assumption~\cite{SchadschneiderChowdhuryNishinariBook11} that the length of a cell is 7.5m, corresponding to the typical space occupied by a vehicle in a jam.
When measuring the instantaneous densities and flows we bin the data into bins of size 5 minutes.
While such binning is not really necessary for the one-dimensional system studied in the present section,
it is necessary to get meaningful results for the networks with traffic signals studied in Section~\ref{sec:2D}, so for consistency we apply the same binning procedure in both cases.

For the purposes of studying the transient behavior of the disruption and recovery processes it is sufficient to focus on one value of $L$, since provided $L$ is sufficiently large so that finite-size effects
are negligible. In the case of these processes, varying $L$ simply results in a trivial rescaling of the time scales arising in the transient processes, without making any qualitative changes.

For each distinct choice of boundary conditions we performed between $30$ and $300$ independent simulations.

\subsubsection{Moderately low density}
\begin{figure}[!t]
\centering
\subfigure[System flow $J(t)$\label{subfig:1D_dwm_jroute_ld}]{\dwmodelrouteflowLD}
\subfigure[System density $\rho(t)$\label{subfig:1D_dwm_droute_ld}]{\dwmodelroutedensityLD}\\
\subfigure[Density profile $\rho(i,t)$: disruption process\label{subfig:1D_dwm_dcell_dis_ld}]{\dwmodelcelldensityLDL}
\subfigure[Density profile $\rho(i,t)$: recovery process\label{subfig:1D_dwm_dcell_rec_ld}]{\dwmodelcelldensityLDR}
\caption{Comparison of transient time evolution of flow and density derived from the domain wall model (\ddw\ and \sdw) and the CA simulations (SIM) of the one-dimensional system 
initially in a moderately low density unperturbed stationary state with $\alpha=0.27$, $\beta=0.9$.
The curves marked SDW correspond to the (stochastic) domain wall model and those marked DDW correspond to its deterministic approximation.
Figs.~\subref{subfig:1D_dwm_jroute_ld} and~\subref{subfig:1D_dwm_droute_ld} show the system flow $J(t)$ and density $\rho(t)$ vs time $t$ with defect durations of $D=10,60$min.
Figs.~\subref{subfig:1D_dwm_dcell_dis_ld} and~\subref{subfig:1D_dwm_dcell_rec_ld} show a comparison of density profiles $\rho(i,t)$ vs position $i$ at various fixed $t$ values
in the disruption and recovery processes with $D=60$min. 
Error bars corresponding to one standard deviation are shown.}
\label{fig:1D_dwm_dj_ld}
\end{figure}

Fig.~\ref{fig:1D_dwm_dj_ld} compares the results derived from the domain wall model with the numerical results from the simulation of the CA model with $\alpha=0.27$, $\beta=0.9$.
With respect to $J(t)$ and $\rho(t)$ illustrated in~Figs.~\ref{subfig:1D_dwm_jroute_ld} and~\ref{subfig:1D_dwm_droute_ld}, 
the predictions of the DW model are numerically indistinguishable (within two error bars) from the simulated CA results.
The results derived from the domain wall model (\sdw) and its deterministic approximation (\ddw) also differ very little.

For $D=60$min, the system has relaxed to the perturbed stationary state by the time the defect is removed.
Neglecting boundary effects, in the perturbed stationary state the density of the subsystem upstream of the defect is $\rhodr$ and that of the downstream subsystem is $\rhodl$.
Therefore, the route-aggregated density should be
\be
\rho= \frac{\rhodr \mx + \rhodl (\mL-\mx)}{\mL}\approx 0.49,
\label{eqn:rho_u}
\ee
using $\rhodl\approx 0.07$ and $\rhodr\approx 0.61$ from Fig.~\ref{fig:1d_fds}.
This is illustrated by the density plateau from hours 0.8 to 1 in Fig.~\ref{subfig:1D_dwm_droute_ld}.

The qualitative features of the $J(t)$ and $\rho(t)$ curves are readily understood by considering the deterministic approximation to the DW model.
Consider the $D=60$ case.
When the incident begins, domain walls $W_{\mU|\mC}$ and $W_{\mF|\mU}$ start drifting upstream and downstream respectively, leading to a sharp drop in $J(t)$.
The wall $W_{\mF|\mU}$ arrives at the out-boundary at time $t=(\mL-\mx)/|V_{\mF|\mU}|\approx 0.02$hr, which causes the rate of decrease of $J(t)$ to lessen.
The flow $J(t)$ then continues to decrease, at a reduced rate, until $W_{\mU|\mC}$ arrives at the in-boundary at time $t=\mx/|V_{\mU|\mC}|\approx 0.8$hr.
From this time onwards, until the defect is removed at time $t=1$hr, the system is in the perturbed stationary state.
Once the defect is removed, the expansion of domain $\mM$ results in a very sharp increase in $J(t)$ until $W_{\mM|\mF}$ reaches the out-boundary at time $t=D+(\mL-\mx)/|V_{\mM|\mF}|\approx 1.03$hr.
The flow $J(t)$ then continues to increase, at a reduced rate, until $W_{\mC|\mM}$ arrives at the in-boundary at time $t=t_{\rm c}=D+\mx/|V_{\mC|\mM}|\approx 1.5$hr, at which time $J(t)=\Jc$.
Since the inflow from the boundary is insufficient to maintain the system at maximum flow, the wall $W_{\mU|\mM}$ then begins to drift downstream, causing $J(t)$ to decrease.
Finally, $W_{\mU|\mM}$ reaches the out-boundary at time $t=t_{\rm c} + L/|V_{\mU|\mM}|\approx 1.7$hr, at which time the system has recovered, and is once again in the unperturbed stationary state.

The above reasoning describes in detail how the DDW curve for $J(t)$ was calculated in Fig.~\ref{subfig:1D_dwm_jroute_ld} for $D=60$min. 
Analogous reasoning is applied to calculate all the DDW curves we present; the details will be omitted henceforth.
We note that in the DDW argument given above, one should consider the walls $W_{\mU|\mC}$ and $W_{\mF|\mU}$ to remain pinned to the boundaries after they arrive, 
so that the later arrivals of $W_{\mC|\mM}$ and $W_{\mM|\mF}$ at the boundaries result in merges with $W_{\mU|\mC}$ and $W_{\mF|\mU}$.
For the stochastic DW model however, the walls $W_{\mU|\mC}$ and $W_{\mF|\mU}$ fluctuate near the boundaries, rather than being pinned there, and so
the merges with $W_{\mC|\mM}$ and $W_{\mM|\mF}$ in fact occur in the bulk in the stochastic case.

For the shorter disruption duration $D=10$min, the system has not yet reached the perturbed stationary state when the defect is removed.
The domain wall $W_{\mU|\mC}$ typically merges with $W_{\mC|\mM}$ before $W_{\mC|\mM}$ reaches the in-boundary.
It therefore typically does not occur that the whole system is in maximum flow.
As a result, we observe that the maximum flow during the recovery process is less than $\Jc$.

Figs.~\ref{subfig:1D_dwm_dcell_dis_ld} and~\ref{subfig:1D_dwm_dcell_rec_ld} 
show a comparison of density profiles $\rho(i,t)$ vs position $i$ at various fixed $t$ values in the disruption and recovery processes with $D=60$min. 
We see that the domain wall model not only provides good approximations of the system observables $J(t)$ and $\rho(t)$ but also of the local density profiles $\rho(i,t)$.

The agreement is not perfect however. Near the beginning of the recovery process, at say $t=65$min as illustrated in Fig.~\ref{subfig:1D_dwm_dcell_rec_ld}, the 
DW and CA curves essentially overlap each other. By $t=80$min however the curves have started to separate.
This is a consequence of the shock in the CA model gradually spreading out and becoming non-localized.
Since both $W_{\mC|\mM}$ and $W_{\mM|\mF}$ are downward shocks, i.e. they satisfy $\rho_{\rm l}>\rho_{\rm r}$, this behavior is to be expected~\cite{SchadschneiderChowdhuryNishinariBook11}. 
The width of the downward shock grows as the excess mass disperses away from the center of mass.
However, for moderately-sized systems the shocks do not smear out completely before arriving at the boundary, 
and so despite this instability the domain wall model still produces good approximations of the transient behavior.
We note that in physical units, the length of the system under study is over 5km, which is a realistic length scale in an urban road network.

\subsubsection{Maximum flow}
\label{sssec:1D_mf}
We now study the case when the unperturbed state is at capacity.
Figs.~\ref{subfig:1D_dwm_jroute_mc} and~\ref{subfig:1D_dwm_droute_mc} compare the results for $J(t)$ and $\rho(t)$ obtained via the domain wall model and the CA simulation with $\alpha=0.6$ and $\beta=0.9$.
Again we find that the domain wall results are in an excellent agreement with the CA simulations.
Because the inflow rate is higher than in the low density case shown in Fig.~\ref{fig:1D_dwm_dj_ld}, the congestion at the upstream side of the defect grows faster,
and thus it takes less time for the system to relax to the perturbed stationary state.
\begin{figure}[!t]
\centering
\subfigure[System flow $J(t)$\label{subfig:1D_dwm_jroute_mc}]{\dwmodelrouteflowcmp}
\subfigure[System density $\rho(t)$\label{subfig:1D_dwm_droute_mc}]{\dwmodelroutedensitycmp}
\subfigure[Density profile $\rho(i,t)$: disruption process\label{subfig:1D_dwm_dcell_dis_mc}]{\dwmodelcelldensityMCL}
\subfigure[Density profile $\rho(i,t)$: recovery process\label{subfig:1D_dwm_dcell_rec_mc}]{\dwmodelcelldensityMCR}
\caption{Comparison of transient time evolution of flow and density derived from the domain wall model (\ddw\ and \sdw) and the CA simulations (SIM) of the one-dimensional system 
initially in a maximum-flow unperturbed stationary state with $\alpha=0.6$ and $\beta=0.9$.
The curves marked SDW correspond to the (stochastic) domain wall model and those marked DDW correspond to its deterministic approximation.
Figs.~\subref{subfig:1D_dwm_jroute_mc} and~\subref{subfig:1D_dwm_droute_mc} show the system flow $J(t)$ and density $\rho(t)$ vs time $t$ with defect durations of $D=10,60$min.
Figs.~\subref{subfig:1D_dwm_dcell_dis_mc} and~\subref{subfig:1D_dwm_dcell_rec_mc} show a comparison of density profiles $\rho(i,t)$ vs position $i$ at various fixed $t$ values
in the disruption and recovery processes with $D=60$min.
Error bars corresponding to one standard deviation are shown.}
\label{fig:1D_dwm_dj_mc}
\end{figure}

As for the low-density case, for $D=10$min, the system has not yet reached the perturbed stationary state when the defect is removed.
The short plateau on the $J(t)$ curve in Fig.~\ref{subfig:1D_dwm_jroute_mc} for $0.2\lesssim t\lesssim 0.5$ warrants some discussion.
After the defect is removed, the downstream end of the defect-induced congestion starts to dissipate whilst the upstream end continues to grow. 
The speed of the congestion growth at the upstream end is approximately the same as that of the dissipation at the downstream end. 
Therefore, the size of the congested region stays approximately constant while it drifts upstream.
This leads to the short plateau in the flow.
At the same time, the effective inflow and outflow rates of the system remain unchanged, which results in the short plateau observed in $\rho(t)$ in Fig.~\ref{subfig:1D_dwm_droute_mc}.

Figs.~\ref{subfig:1D_dwm_dcell_dis_mc} and~\ref{subfig:1D_dwm_dcell_rec_mc}
show a comparison of density profiles $\rho(i,t)$ vs position $i$ at various fixed $t$ values in the disruption and recovery processes with $D=60$min.
The behavior is qualitatively similar to that observed for low density, and we again see that the domain wall model generally provides a good approximation.
We note that the discrepancy between the DW and CA curves at time $t=105$min in Fig.~\ref{subfig:1D_dwm_dcell_rec_mc} is due simply to boundary effects, rather than being related to the dissipation of shocks.
By $t=105$min the system has completely recovered, for both the domain wall model and CA simulations. 
However, while the DW model assumes a perfectly flat density profile, the CA simulations display a noticeable finite-size effect at the boundaries.

\subsubsection{Moderately high density}
\begin{figure}[!t]
\centering
\subfigure[System flow $J(t)$\label{subfig:1D_dwm_jroute_hd}]{\dwmodelrouteflowH}
\subfigure[System density $\rho(t)$\label{subfig:1D_dwm_droute_hd}]{\dwmodelroutedensityH}
\subfigure[Density profile $\rho(i,t)$: disruption process\label{subfig:1D_dwm_dcell_dis_hd}]{\dwmodelcelldensityHDL}
\subfigure[Density profile $\rho(i,t)$: recovery process\label{subfig:1D_dwm_dcell_rec_hd}]{\dwmodelcelldensityHDR}
\caption{Comparison of transient time evolution of flow and density derived from the domain wall model (\ddw\ and \sdw) and the CA simulations (SIM)
initially in a moderately high density unperturbed stationary state with $\alpha=0.6$ and $\beta=0.3$.
The curves marked SDW correspond to the (stochastic) domain wall model and those marked DDW correspond to its deterministic approximation.
Figs.~\subref{subfig:1D_dwm_jroute_hd} and~\subref{subfig:1D_dwm_droute_hd} show the system flow $J(t)$ and density $\rho(t)$ vs time $t$ with defect duration of $D=60$min.
Figs.~\subref{subfig:1D_dwm_dcell_dis_hd} and~\subref{subfig:1D_dwm_dcell_rec_hd} show a comparison of density profiles $\rho(i,t)$ vs position $i$ at various fixed $t$ values
in the disruption and recovery processes also with $D=60$min.
Error bars corresponding to one standard deviation are shown.}
\label{fig:1D_dwm_dj_hd}
\end{figure}

Fig.~\ref{fig:1D_dwm_dj_hd} shows results for a system initially in a moderately-high density unperturbed stationary state.
As observed for the previous two cases, the results produced by the domain wall model are generally in a good agreement with the simulation results.

We observe from Fig.~\ref{subfig:1D_dwm_droute_hd} that there exists a delay in the density variation for around 15 minutes after the defect is inserted.
The continuity equation for the system density $\rho(t)$ states that $\d \rho / \d t = J_{\rm in} - J_{\rm out}$, where $J_{\rm in}$ and $J_{\rm out}$ denote the boundary inflow and outflow, respectively.
In the deterministic approximation, $J_{\rm in}$ will not change until $W_{\mU|\mC}$ reaches the in-boundary, and $J_{\rm out}$ will not change until $W_{\mF|\mU}$ reaches the out-boundary.
It follows that $\rho(t)$ will vary only when a domain wall reaches the boundary, as observed.
In principle, this initial plateau also occurs in the low density and the maximum flow cases,
however because the travel velocity of the wall $W_{\mF|\mU}$ is much higher in these cases the resulting delay is extremely short, and therefore difficult to observe.

As noted in Section~\ref{sssec:1D_mf}, 
similar arguments also explain the short density plateau observed for $0.2\lesssim t\lesssim 0.5$ for the maximum flow case with $D = 10$min in Fig.~\ref{subfig:1D_dwm_droute_mc}.
Quite generally, the deterministic approximation of the domain wall model predicts that the slope of $\rho(t)$ can change only when a domain wall reaches the boundary.

\subsubsection{Very low and high densities}
\begin{figure}[!t]
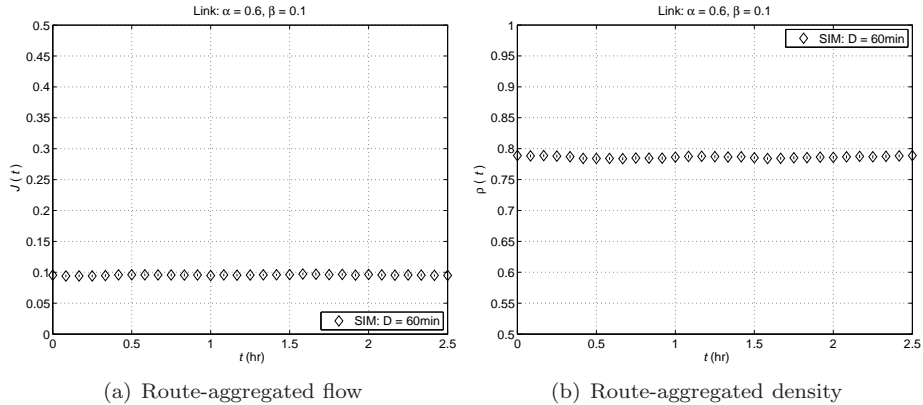

\centering
\subfigure[Route-aggregated flow]{\LKeightflowAsixtyBninety} \subfigure[Route-aggregated density]{\LKeightdensityAsixtyBninety}
\caption{
Transient time evolution of flow and density from the CA simulations (SIM), initially in a very high density unperturbed stationary state with $\alpha=0.6,\ \beta=0.1$. 
Figs.~\subref{subfig:1D_dwm_jroute_hd} and~\subref{subfig:1D_dwm_droute_hd} show the system flow $J(t)$ and density $\rho(t)$ vs time $t$ with defect duration of $D=60$min.
Error bars corresponding to one standard deviation are shown.}
\label{fig:1D_sim_dj_hd}
\end{figure}

Finally, we discuss the cases of very low and very high initial density.
Fig.~\ref{fig:1D_sim_dj_hd} shows $J(t)$ and $\rho(t)$ for the very high density case of $\alpha=0.6$ and $\beta=0.1$.
The unperturbed traffic flow is smaller than $J_d$ and thus the presence of the defect does not cause any change to either the flow or density.
Similar behavior is observed for the case of very low initial density.
In these cases it is the boundaries, not the defect, which act as the dominant bottleneck.

\section{Two-Dimensional Network}
\label{sec:2D}
In this section we study the effect of introducing a lane-reduction in a single link within a two-dimensional network. 
We will again compare the results obtained by simulating a cellular automaton model with the predictions of domain wall theory.
\subsection{Network model}
\label{ssec:netnasch}
\subsubsection{NetNaSch model}
The CA model we employ was introduced in~\cite{deGierGaroniRojas11}. 
This model essentially uses the NaSch model to define the motion of vehicles along individual lanes, and provides a simple set of rules for gluing such lanes together to form a network.
These rules are sufficiently general that they are easily applied to model networks of arbitrary topology; the rules can be applied to any directed multigraph of lanes.
We refer to this model as the {\em NetNaSch} model. 
In addition to the NaSch dynamics along lanes, the NetNaSch model includes rules for lane changes between adjacent lanes, rules for making turning decisions at intersections, and appropriate rules for 
determining how vehicles traverse intersections. 
A variety of possibilities exist for the latter rules, and the NetNaSch model was originally designed to allow easy comparison of different traffic signal systems, 
via appropriate choices of the rules governing intersections. In this section we shall consider two possible choices for these rules, which we outline in Section~\ref{sssec:signals}.

To mimic origin-destination behavior, the NetNaSch model demands that each vehicle makes a random decision about which link it wants to turn into at the approaching intersection according to some turning probability. This turning decision is made when the vehicle first enters a link and should influence 
its lane changing decision as it travels along the link. In order to guarantee the robustness of the model, vehicles are allowed to adaptively change their turning decisions when faced with very high levels of congestion. Specifically, suppose that a vehicle is queued at an intersection due to spillback on the link onto which it wishes to turn. In this instance, the NetNaSch model allows the vehicle to remake its random turning decision.

As already mentioned, 
the {\em symmetric} lane changing rules discussed in Section~\ref{ssec:defects} coincide with the {\em dynamic} lane changing rules introduced for the NetNaSch model in~\cite{deGierGaroniRojas11}. 
The NetNaSch model includes an additional set of lane changing rules, referred to as {\em topological} lane changing in~\cite{deGierGaroniRojas11}, which govern the behavior of vehicles wishing to 
change lanes to facilitate their decisions of which link to turn into at the upcoming intersection. 
For the NetNaSch model, we shall refer to the combination of dynamic and topological lane changes described in~\cite{deGierGaroniRojas11} as {\em symmetric} lane changing. 
As for the one-dimensional case described in Section~\ref{ssec:defects}, in the presence of a lane reduction defect on a given link in the network, 
we augment these symmetric lane changing rules with additional asymmetric lane changing rules, allowing vehicles to navigate around the defect.
The asymmetric lane changing rules we use for the NetNaSch model are identical to those described in Section~\ref{ssec:defects} for the one-dimensional case.

\subsubsection{Network parameters}
\begin{figure}[!t]
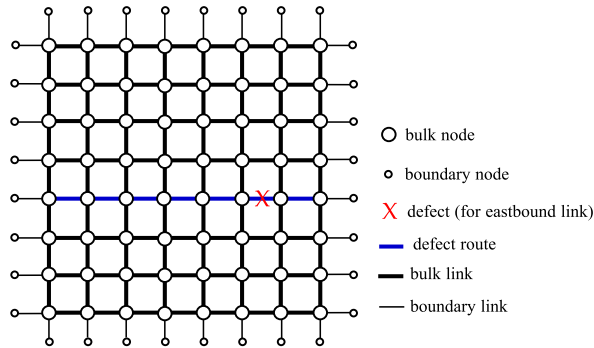

\centering
{\network}
\caption{An 8 by 8 square-lattice network with a defect on an eastbound link. Each edge between adjacent nodes corresponds to two oppositely-oriented directed links.
}
\label{fig:network}
\end{figure}
We simulated a regular $8\times 8$ square grid, as shown in Fig.~\ref{fig:network}.
Each adjacent pair of nodes is in fact connected by two oppositely-oriented directed links.
In turn, each such link consists of two main lanes, plus an additional right-turning lane~\footnote{As in Australia, vehicles drive on the left side of the road in our CA model.}.
Vehicles are inserted with an input probability $\alpha$ at the upstream end of each {\em boundary in-link}.
At the downstream end of each {\em boundary out-link}, a vehicle that wishes to leave the system is allowed to exit with an output probability $\beta$.
The same value of the inflow rate $\alpha$ is applied to all in-links, and the same outflow rate $\beta$ to all out-links.
The length of each bulk link was set to 100 cells, and each turning lane to 12 cells. Using a cell length of 7.5m, these lengths correspond to typical values for an arterial network.
The boundary links simply act as input/output buffers in the NetNaSch model, the length of each boundary link was set to 50 cells. 
The boundary links are not considered part of the network for the purposes of measuring observables.

Each link was assigned the same turning probability $p_{\rm T}$ for left and right turns, implying a probability $1-2p_{\rm T}$ for continuing straight ahead.
For the majority of our simulations, $p_{\rm T}$ was set to $0.05$.
For comparison, in Section~\ref{ssec:turns} we discuss simulations using $p_{\rm T} = 0.1$.

We refer to the sequence of the seven adjacent east-bound links in the fifth row as the {\em defect route}.
When the network has reached stationarity, a defect is inserted into a single cell of the left lane of the sixth link in the defect route.
See Fig.~\ref{fig:network}.
We considered placing the defect either mid-block, or at the upstream end of the link.
Our focus will be understanding the transient behavior of the flow and density along the defect route.
We note that with the parameter values chosen, the length of the defect route is the same as the length of the one-dimensional system studied in Section~\ref{sec:1D}.
The positions of the defects are also comparable in the two cases.

On each link $l$ along the defect route we measured the density $\rho_l(t)$ and flow $J_l(t)$. The flow $J_l(t)$ was measured at a single point close to the start of the link $l$.
From the link observables we then defined the corresponding {\em route aggregated} observables, $\rho(t)$ and $J(t)$, 
to be the arithmetic means of the link variables, over all links in the defect route.
For each distinct choice of traffic signal systems and boundary conditions, we performed between $30$ and $300$ independent simulations.

\subsubsection{Signal systems}
\label{sssec:signals}
\begin{figure}[!t]
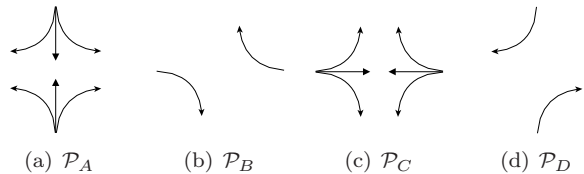

\centering
\subfigure[$\mP_A$]{\phasea}\quad\subfigure[$\mP_B$]{\phaseb}\quad\subfigure[$\mP_C$]{\phasec}\quad\subfigure[$\mP_D$]{\phased}
\caption{Phases used at each node of the simulated networks. Only SOTL uses the turning phases, $\mP_B$ and $\mP_D$.}
\label{fig:phases}
\end{figure}
We study two possible mechanisms governing how and when vehicles can traverse intersections.
As an example of an adaptive traffic signal system, we study a variation of the `self-organizing' traffic lights (SOTL) model studied in~\cite{deGierGaroniRojas11}.
Such traffic signal systems have been the subject of several recent studies in the statistical mechanics literature~\cite{Gershenson05,HelbingSiegmeierLammer07,LammerHelbing08,GershensonRosenblueth11}.
In general, each node in a network has a number of signal {\em phases}\footnote{Not to be confused with the statistical-mechanical meaning of {\em phase}.}, which determine which incoming links 
have right of way at a given instant. 
For example, for our implementation of SOTL on the square lattice, each node has four phases, corresponding to two through phases ($\mP_A$ and $\mP_C$) and two turning phases ($\mP_B$ and $\mP_D$).
See Fig.~\ref{fig:phases}.
The SOTL system is based on the simple principle that each node should choose its current phase to be the one with the highest demand.
The SOTL system that we study here uses density as the demand function. A precise algorithmic description of SOTL is given in \ref{app:solt}.
We note that the NetNaSch model takes into account the fact that even if facing a green light, right turning vehicles in a through phase must give way to oncoming traffic from the opposite direction.

We also study an idealized model, which is designed to act as an intermediary between the more realistic SOTL system, and the one-dimensional system studied in Section~\ref{sec:1D}.
We shall refer to this model as the {\em Overpass} system. In this model, each node simply alternates between the two through phases, $\mP_A$ and $\mP_C$. 
Each phase lasts only a single time step.
This is reminiscent of a {\em four-way stop} intersection.
In addition however, we also remove the constraint that right turning vehicles give way to oncoming traffic.
This can be viewed as connecting each right-turning lane to an {\em overpass}.

\subsection{Link-dependent domain wall model}
\label{ssec:2D_dwm}
We now slightly extend the domain wall model discussed in Section~\ref{ssec:1D_transient}, to apply it to the defect route in a two-dimensional network.
Typically, for two-dimensional networks, the densities of different links may differ, even at stationarity and with homogeneous boundary conditions.
Therefore, instead of having subsystems $\mU$, $\mC$, $\mF$ and $\mM$ with homogeneous densities, 
the two-dimensional domain wall model assumes homogeneous density regions $\mU_l$, $\mC_l$, $\mF_l$ and $\mM_l$ for each link $l$.
As a result, the hopping rates (\ref{eqn:hoppingrates}) and drift velocity (\ref{eqn:wallspeed}) become link-dependent.
With these modifications, we apply the domain wall model presented in Section~\ref{sec:1D} to the defect route of our two-dimensional network.

\subsection{Simulations: Overpass Network}
\label{ssec:fc}
We begin by discussing the results for the overpass network. In this case, the defect is located in the middle of the defect link.
\subsubsection{Stationary state -- Fundamental diagram}
\begin{figure}[!t]
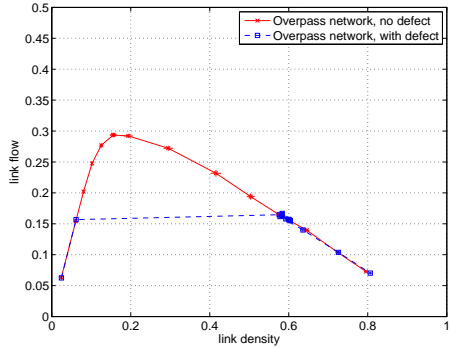

\centering
\CRfds
\caption{Fundamental diagram for the link immediately upstream of the defect in the overpass network from simulations.
  The solid curve corresponds to the unperturbed FD, whilst the dashed curve illustrates the unachievable region in the perturbed FD.
  Error bars corresponding to one standard deviation are shown.}
\label{fig:OP_fds}
\end{figure}

Fig.~\ref{fig:OP_fds} shows the aggregated FD of all the links upstream of the defect in the defect route, for both the perturbed and unperturbed cases.
Compared with the one-dimensional system shown in Fig.~\ref{fig:1d_fds}, the capacity of the overpass network is slightly reduced, and the critical density is slightly increased.
In addition, for both the perturbed and unperturbed cases, the high density branch of the overpass system lies below that of the one-dimensional FD. The shape is qualitatively similar however.
We note that the FD we used as input for the DW model was constructed by overlaying all the link FDs and constructing a linear interpolation of the data.

\subsubsection{Domain wall model}
As for the one-dimensional system, see Fig.~\ref{fig:disruption_recovery_demo}, we assume that the state of $\mU_l$ is given by the state of link $l$ in the unperturbed stationary state,
while the states of $\mC_l$ and $\mF_l$ are given by the perturbed stationary states.
Given the homogeneous spatial distribution of density in the unperturbed system, we assume the state of $\mU_l$ is independent of $l$ in the defect route.
Since the FD of the overpass system is again unimodal, the argument presented in Section~\ref{recovery process} implies that the recovery domain $\mM_l$ will be in the maximum flow regime for all $l$.
We shall return to this point in Section~\ref{ssec:turns}.
We do assume the states of regions $\mC_l$ and $\mF_l$ are link-dependent however.

\subsubsection{Moderately low density}
\label{overpass low density}
\begin{figure}[!t]
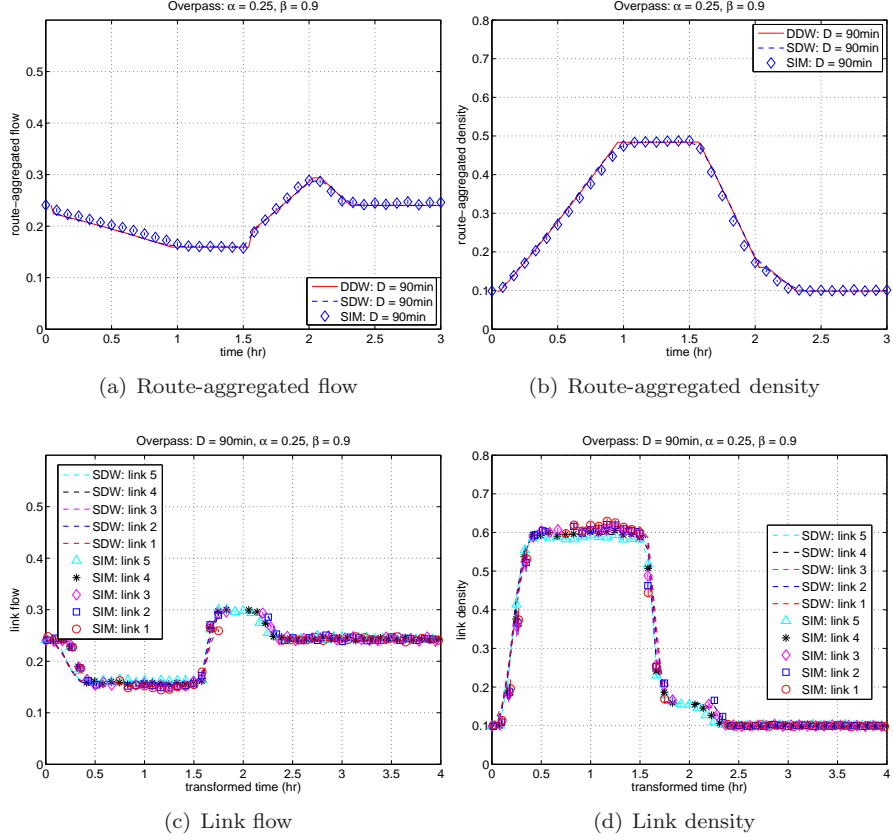

\centering
\subfigure[Route-aggregated flow]\dwmodelCRrouteflowLD \subfigure[Route-aggregated density]\dwmodelCRroutedensityLD\\
\subfigure[Link flow\label{overpass_link_flow_ld}]\dwmodelCRlinksflowLD \subfigure[Link density\label{overpass_link_density_ld}]\dwmodelCRlinksdensityLD\\
\caption{
Comparison of flow and density derived from the domain wall model and CA simulations for the two-dimensional overpass network, initially in the moderately low density regime
with $\alpha=0.25$, $\beta=0.9$ and $D=90$min.
Error bars corresponding to one standard deviation are shown.}
\label{fig:fc_dwm_dj_ld}
\end{figure}

Fig.~\ref{fig:fc_dwm_dj_ld} shows the link observables $J_l(t)$ and $\rho_l(t)$ for all links upstream of the defect, as well as the route aggregated quantities $J(t)$ and $\rho(t)$, 
when the unperturbed system is in the moderately low density regime.
For ease of comparison, for each link upstream of the defect we applied a link-dependent translation in the time variable $t\mapsto t'$ with
\be
t'=\left\{\begin{tabular}{ll}
$t-(k-l)|V_{\mU_l|\mC_l}|$,&\, $0 \leq t < D$,\\
$t-(k-l)|V_{\mC_l|\mM_l}|$,&\, $D \leq t < D+t_l^\ast$,\\
$t+(k-l)|V_{\mU_l|\mM_l}|$,&\, $t \geq D+t_l^\ast$,\\
\end{tabular}\right.
\label{eqn:time_transf_fc_ld}
\ee
where $k=5$ is the number of links upstream of the defect in the 8 by 8 network.
Here $t_l^\ast$ denotes the time when the wall $W_{\mC_l|\mM_l}$ has passed through link $l$, calculated via~(\ref{eqn:wallspeed}).
The excellent data collapse displayed in Figs.~\ref{overpass_link_flow_ld} and~\ref{overpass_link_density_ld} 
indicates that even the deterministic approximation to the domain wall model accurately predicts the transient behavior observed in the CA simulations.

The time translation (\ref{eqn:time_transf_fc_ld}) is also applied in the maximum flow and moderately high density cases considered in Sections~\ref{overpass capacity} and~\ref{overpass high density},
except that in the high density case the third expression becomes $t-(k-l)|V_{\mM_l|\mU_l}|$.

\subsubsection{Capacity region}
\label{overpass capacity}
\begin{figure}[!t]
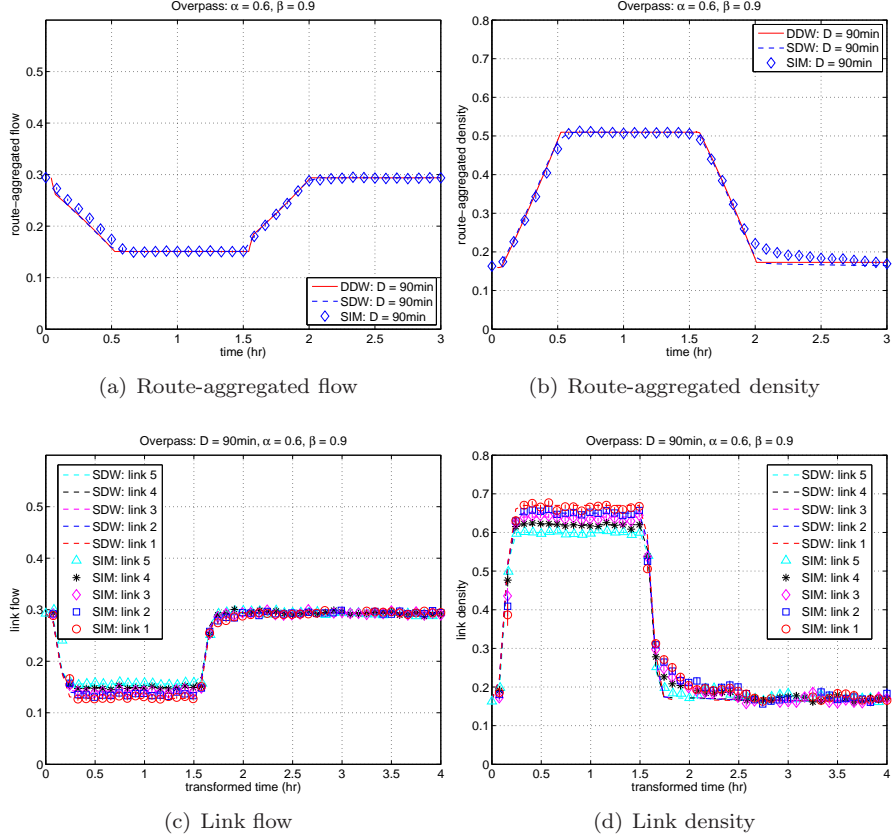

\centering
\subfigure[Route-aggregated flow]\dwmodelCRrouteflowMC \subfigure[Route-aggregated density]\dwmodelCRroutedensityMC\\
\subfigure[Link flow\label{overpass_link_flow_mc}]\dwmodelCRlinksflowMC \subfigure[Link density\label{overpass_link_density_mc}]\dwmodelCRlinksdensityMC\\
\caption{
Comparison of flow and density derived from the domain wall model and CA simulations for the two-dimensional overpass network, initially at capacity with $\alpha=0.6$, $\beta=0.9$ and $D=90$min. 
Error bars corresponding to one standard deviation are shown.}
\label{fig:fc_dwm_dj_mc}
\end{figure}
Fig.~\ref{fig:fc_dwm_dj_mc} shows the results for the case when the unperturbed system is at capacity.

A notable feature of these curves, compared with the low density curves discussed in Section~\ref{overpass low density},
is that the flows and densities in the perturbed stationary state (roughly $0.25\lesssim t' \lesssim 1.5$ hr in Figs.~\ref{fig:fc_dwm_dj_mc}(c) and  \ref{fig:fc_dwm_dj_mc}(d)) clearly have link-dependent values.
Specifically, in the perturbed stationary state, the link density increases with the distance of the link from the defect.
This can be understood by considering the flows between the defect route and the adjacent side streets.
Consider a link $l$ in the defect route.
The flow $J_l$ in $l$ is given by 
\begin{equation}
J_l = J_{l-1} - J_{l-1\mapsto s} + J_{s\mapsto l},
\label{overpass mc flows argument}
\end{equation}
where $J_{l-1\mapsto s}$ is the flow from link $l-1$ to the side streets, and $J_{s \mapsto l}$ is the flow from the side streets into link $l$.
Unless $l$ is extremely congested, the exogenous inflow $J_{s \mapsto l}$ from the side streets into $l$ is essentially governed by the flow on the side streets, $J_s$.
Given that the side streets will be less affected by the defect, to a first approximation we can assume that $J_s$ is approximately equal to the unperturbed stationary flow $\Jo$,
so that $J_{s\mapsto l} \approx \pt\, \Jo$.
On the other hand, the outflow $J_{l-1\mapsto s}$ to the side streets is essentially governed by the link flow $J_{l-1}$, so that $J_{l-1 \mapsto s} \approx \pt\, J_{l-1}$.
Since $J_{l-1} \leq \Jd < \Jo$, it follows that $J_{s \mapsto l} > J_{l-1 \mapsto s}$, and from (\ref{overpass mc flows argument}) it follows that $J_l>J_{l-1}$.
Since the region upstream of the defect is in high density, this then implies $\rho_l<\rho_{l-1}$.

Similar heterogeneities are also observed for the low and the high density cases, illustrated in Figs.~\ref{fig:fc_dwm_dj_ld} and \ref{fig:fc_dwm_dj_hd}, but are much less pronounced,
since $J_{s\mapsto l}- J_{l-1\mapsto s}$ is smaller in these cases.
The fact that these heterogeneities are more apparent for the low density case shown in Fig.~\ref{fig:fc_dwm_dj_ld} than the high density case shown in Fig.~\ref{fig:fc_dwm_dj_hd}, 
is simply because for the $\alpha,\beta$ values chosen, the low density case happens to be closer to capacity.

\subsubsection{Moderately high density}
\label{overpass high density}
\begin{figure}[!t]
\centering
\subfigure[Route-aggregated flow]\dwmodelCRrouteflowHD \subfigure[Route-aggregated density]\dwmodelCRroutedensityHD\\
\subfigure[Link flow\label{overpass_link_flow_hd}]\dwmodelCRlinksflowHD \subfigure[Link density\label{overpass_link_density_hd}]\dwmodelCRlinksdensityHD\\
\caption{
Comparison of flow and density derived from the domain wall model and CA simulations for the two-dimensional overpass network,
initially in the moderately high density regime with $\alpha=0.6$, $\beta=0.7$ and $D=90$min.
Error bars corresponding to one standard deviation are shown.}
\label{fig:fc_dwm_dj_hd}
\end{figure}

Fig.~\ref{fig:fc_dwm_dj_hd} compares the DW results with the CA simulations for the high density case.
We notice that the simulated recovery after the link flow begins decreasing (around $t=1.75$hr) is slightly slower than the domain wall prediction.
We shall argue in Section~\ref{ssec:turns} that this phenomenon is caused by congestion on the side streets.

\subsection{Simulations: SOTL Network}
\label{ssec:solt} 
We now consider the SOTL system.
\subsubsection{Mid-block defects}
\begin{figure}[!t]
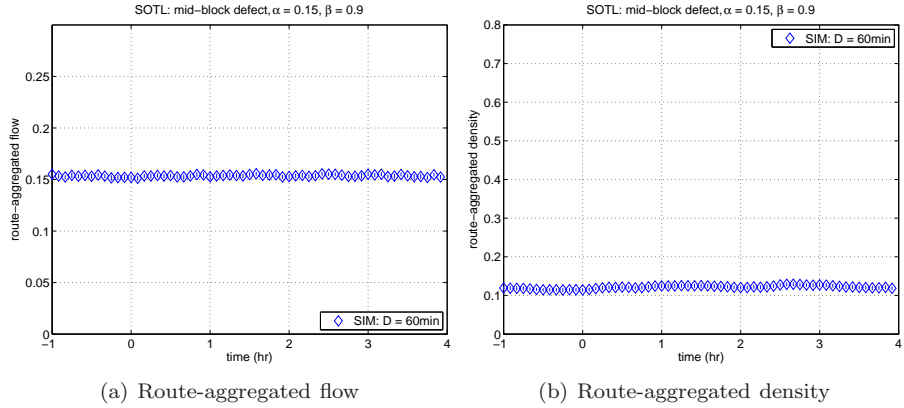

\centering
\subfigure[Route-aggregated flow]{\ANflowmidblock}
\subfigure[Route-aggregated density]{\ANdensitymidblock}
\caption{
$J(t)$ and $\rho(t)$ obtained from CA simulations of the two-dimensional SOTL network, initially in the moderately low density regime close to capacity
with $\alpha=0.15$, $\beta=0.9$, with a mid-block defect inserted at time $t=0$ for duration $D=90$min.
Error bars corresponding to one standard deviation are shown.}
\label{fig:an_sim_mid-block_dj}
\end{figure}

Fig.~\ref{fig:an_sim_mid-block_dj} displays the time evolution of $J(t)$ and $\rho(t)$ in the presence of a mid-block defect, when the unperturbed stationary state is moderately low density near capacity.
Contrary to the overpass model, the impact of the mid-block defect is clearly negligible.
Similar behavior holds for any initial stationary state, and the cause is easy to understand.
For road networks, intersections act as sitewise bottlenecks.
Given that the mid-block lane-reduction defect and the intersections are all well-separated, the overall capacity of the system is determined by the minimum capacity of these bottlenecks.
In Fig.~\ref{subfig:sotl_vs_link_fd}, we compare the FD for the unperturbed SOTL network (pink circles) with the FD for a single link with a lane-reduction defect.
We observe that the capacity for SOTL in the absence of the defect is close to, but slightly lower than, the capacity of a single link with lane-reduction defect. 
We would therefore not expect that introducing a lane-reduction defect that is well-separated from the intersections should affect the performance of the SOTL system.

The situation is qualitatively different when the lane-reduction defect is positioned close to an intersection however, since then the two defects cannot be considered independent.
Indeed, the lower curve (black triangles) in Fig.~\ref{subfig:sotl_vs_link_fd} corresponds to a SOTL network with a lane reduction defect positioned in the first cell of the sixth link in the defect route.
In this case, the defect and intersection combine to produce a noticeable impact on the network's behavior.
We will now turn our attention to understanding in more detail the impact of such arrival-side defects.

\subsubsection{Arrival-side defects: Fundamental Diagrams}
\begin{figure}[!t]
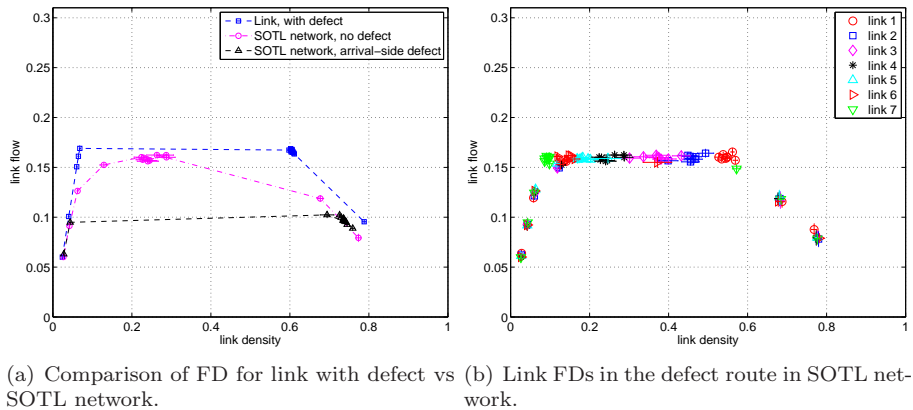

\centering
\subfigure[Comparison of FD for link with defect vs SOTL network.\label{subfig:sotl_vs_link_fd}]{\ANlinkfdscmp}
\subfigure[Link FDs in the defect route in SOTL network.\label{subfig:sotl_link_fds}]{\ANlinkfds}
\caption{
Fig.~\subref{subfig:sotl_vs_link_fd}: Fundamental diagram for the link immediately upstream of the defect in the unperturbed SOTL network (pink circles) from similations.
For comparison we also show the FD for a one-dimensional system with defect (blue squares),
and the FD for the link immediately upstream of the defect in the SOTL network with an arrival-side defect (black triangles).
Fig.~\subref{subfig:sotl_link_fds}: Link FDs for each link in the defect route of the SOTL system from similations.
Error bars corresponding to one standard deviation are shown.}
\label{fig:an_fds}
\end{figure}

Fig.~\ref{subfig:sotl_link_fds} plots the FDs of each link in the defect route for the unperturbed SOTL network.
In each case, the FD is trapezoidal. The capacity of each link is just less than half the capacity of an isolated link; see Fig.~\ref{fig:1d_fds}.
Rather than at a single density, capacity is achieved over the range $[\rho_{c-},\rho_{c+}] \approx [0.1,0.55]$.
In the presence of an arrival-side defect, the capacity drops further to around $0.1$.

We note that, as for the overpass network, the FD we used as input for the DW model was constructed by overlaying all the link FDs and constructing a linear interpolation of the data.

\subsubsection{Arrival-side defects: Transient processes}
\begin{figure}[p]
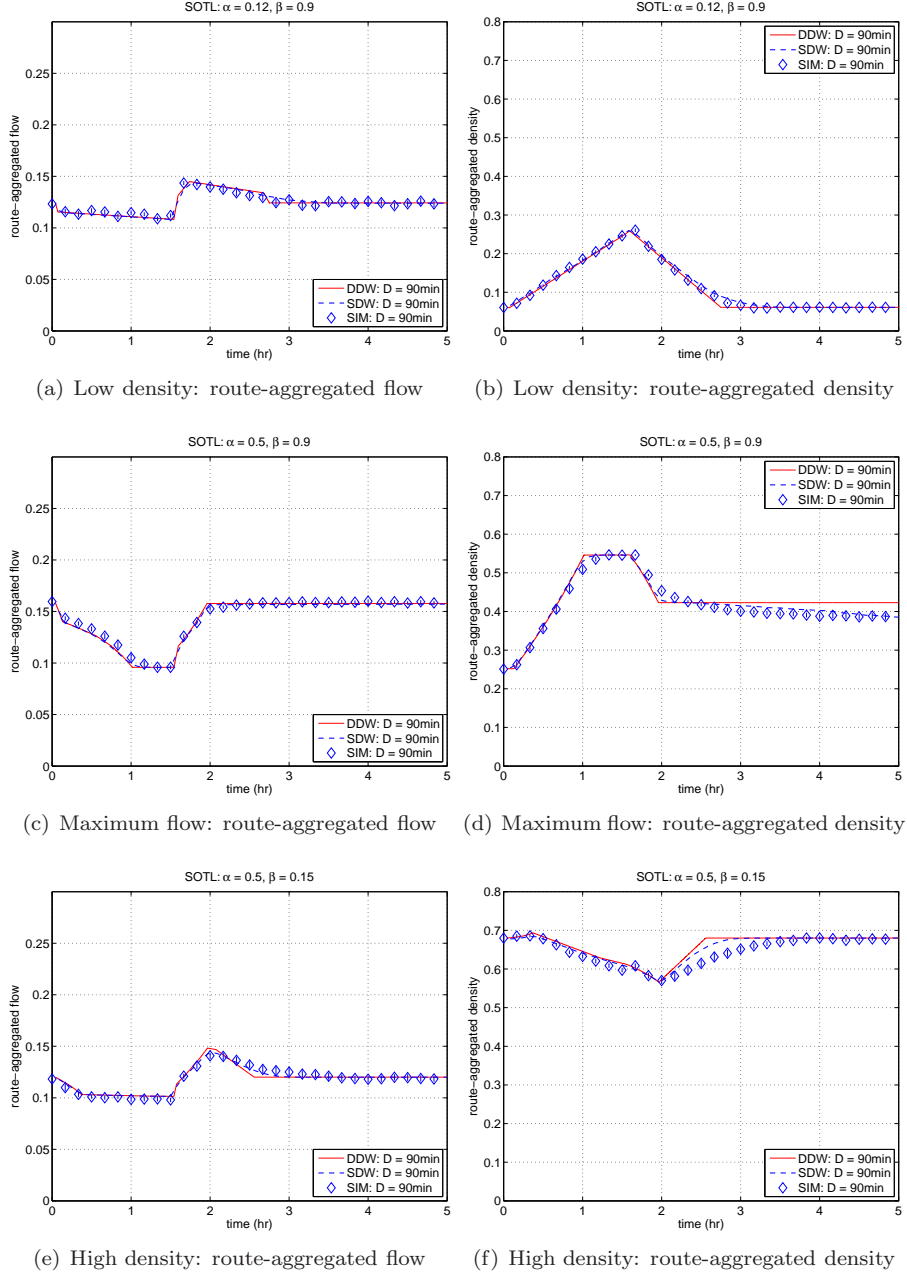

\centering
\subfigure[Low density: route-aggregated flow\label{subfig:sotl_ld_flow}]{\dwmodelANrouteflowLD}
\subfigure[Low density: route-aggregated density\label{subfig:sotl_ld_density}]{\dwmodelANroutedensityLD}\\
\subfigure[Maximum flow: route-aggregated flow\label{subfig:sotl_mc_flow}]{\dwmodelANrouteflowMC}
\subfigure[Maximum flow: route-aggregated density\label{subfig:sotl_mc_density}]{\dwmodelANroutedensityMC}\\
\subfigure[High density: route-aggregated flow\label{subfig:sotl_hd_flow}]{\dwmodelANrouteflowHD}
\subfigure[High density: route-aggregated density\label{subfig:sotl_hd_density}]{\dwmodelANroutedensityHD}
\caption{Time evolution of route-aggregated flow and density derived from the stochastic domain wall model and the CA simulation for the SOTL network with arrival-side defect of duration $D=90$min.
Top row: Moderately low density regime with $\alpha=0.12$, $\beta=0.9$.
Middle row: Maximum flow regime with $\alpha=0.5$, $\beta=0.9$.
Bottom row: Moderately high density regime with $\alpha = 0.5$, $\beta = 0.15$. 
Error bars corresponding to one standard deviation are shown.
}
\label{fig:an_dwm_dj}
\end{figure}

Fig.~\ref{fig:an_dwm_dj} shows a comparison of the route-aggregated flow and density predicted by the domain wall model and the CA simulations, when the initial unperturbed system is 
in three different regimes: moderately low density, maximum flow, and moderately high density.
In the moderately low and high density cases, the domain wall model applied to the SOTL network is essentially the same as that used for the overpass network, 
and we again assume each link in region $\mU$ is in the same unperturbed stationary state, and use link dependent states for $\mC_l$ and $\mF_l$ obtained from the perturbed stationary state.
For the case when the unperturbed state is at capacity, we also use link-dependent states for $\mU_l$.

The major difference between the overpass and SOTL networks however is that the overpass network has a unimodal FD whereas for the SOTL network capacity is not achieved at a unique density, 
but for a range of densities $[\rho_{c-},\rho_{c+}]$.
For this reason, it is necessary to assume link-dependent values of $(\rho_l,J_l)$ for the $\mM$ region during recovery.
Therefore, we assume that $\mM_l$ is in state $(\rho_{c+}, J_{c+})$ if link $l$ recovers from the high density regime, and otherwise assume $\mM_l$ is in state $(\rho_{c-}, J_{c-})$.
Consequently, for the simulation results shown in Fig. 16, we used $(\rho_{c+}, J_{c+})$ for the state of $\mM_l$ for all links upstream of the defect,
and $(\rho_{c-}, J_{c-})$ for the defect link and all links downstream of it.

When the unperturbed system is in either low or high density, the predictions of the domain wall model and its deterministic approximation are both in a good agreement with the simulated results.
When the unperturbed system is at capacity however, the situation is more subtle.
Since the FD for SOTL is trapezoidal, regions $\mU_l$ and $\mM_l$ can have the same flow but different densities.
However, (\ref{eqn:wallspeed}) implies that even though $\mU_l$ and $\mM_l$ are then in different states, the drift velocity of the domain wall separating them is precisely zero.
The domain walls $W_{\mU|\mM}$ and $W_{\mM|\mU}$ both perform symmetric random walks, and recovery will not occur until they merge.
Consequently, although the flows have already returned to their initial values, it can take a very long time for the system to recover to its unperturbed density distribution.
Since the drift velocity is zero, the deterministic approximation of the DW model is clearly incapable of modeling the recovery process. 
However Fig.~\ref{subfig:sotl_mc_density} shows that the stochastic DW model still provides a good approximation to the CA results in this case.

Finally, we note that in the high density case, the density for $t\gtrsim2$hr recovers somewhat more slowly than predicted by the domain wall model.
This was observed already for the overpass network in Section~\ref{overpass high density}, where we claimed the phenomenon is caused by congestion on the side streets.
Since the side streets along the defect route are more congested in the SOTL network that the overpass network, the effect is more pronounced. 
The effect also becomes more apparent by increasing the turning probabilities. We shall now discuss these issues in more detail.

\subsection{Turning probabilities}
\label{ssec:turns}
\begin{figure}[!t]
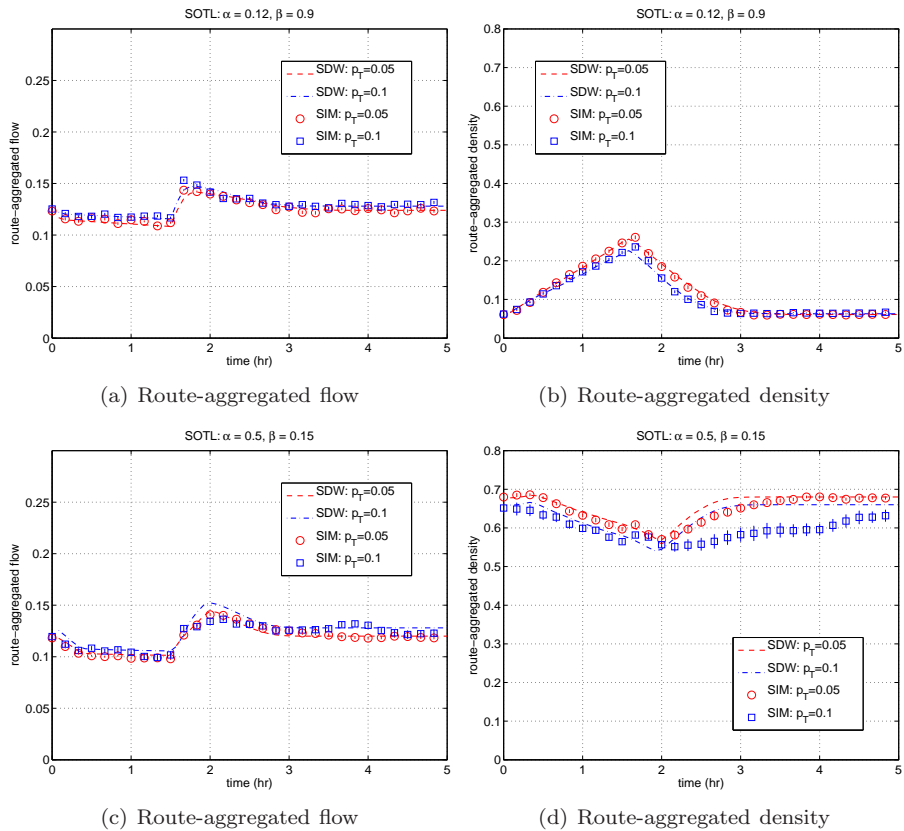

\centering
\subfigure[Route-aggregated flow]{\dwmodelANrouteflowLDT}
\subfigure[Route-aggregated density]{\dwmodelANroutedensityLDT}
\subfigure[Route-aggregated flow]{\dwmodelANrouteflowHDT}
\subfigure[Route-aggregated density]{\dwmodelANroutedensityHDT}
\caption{Time evolution of route-aggregated flow and density derived from the stochastic domain wall model and the CA simulation for the SOTL network with arrival-side defect of duration $D=90$min.
  $p_{\rm T}=0.05$, $0.1$ and $D=90$min. 
  Top row: Moderately low density regime with $\alpha = 0.12$, $\beta= 0.9$. 
  Bottom row: Moderately high density regime with $\alpha = 0.5$, $\beta= 0.15$.
  Error bars corresponding to one standard deviation are shown.}
\label{fig:var_turning}
\end{figure}

The simulations so far have used a turning probability of $p_{\rm T}=0.05$.
We now study the effect of increasing $p_{\rm T}$. 
Fig.~\ref{fig:var_turning} presents the route-aggregated observables for networks with turning probabilities $p_{\rm T} = 0.05$, $0.1$, for the SOTL network in the moderately-low and moderately-high density regimes.
We observe that for the low-density plots, the agreement between the CA simulations and the DW wall predictions is again excellent.

The high-density case deserves further comment however.
As noted already, in Figs.~\ref{fig:fc_dwm_dj_hd} and \ref{subfig:sotl_hd_density}, after the density starts to increase the DW model predicts the defect route recovers faster than actually observed in the CA simulations.
Fig.~\ref{fig:var_turning} shows that this effect becomes more pronounced for $p_{\rm T}=0.1$, suggesting it is caused by the interaction of the side streets with the defect route.
This portion of the recovery process corresponds to the drift of the walls $W_{\mU|\mM}$ and $W_{\mM|\mU}$ upstream.
In our discussion of the domain wall model thus far, we have made frequent use of the assumption that the regions $\mM_l$ in the recovery process are in maximum flow.
However, for two-dimensional networks initially in moderately high density, this can only be approximately correct.
Consider regions $\mM_l$ and $\mM_{l+1}$. The outflow from link $l$ is the sum of the outflow $J_{l\to l+1}$ to link $l+1$ and the outflow to the side streets $J_{l\to s}$.
Because the side streets are in high density, the outflow $J_{l\to s}$, which is proportional to $p_{\rm T}$, will not be high enough to maintain $\mM_l$ in a maximum current state.
As a consequence, the density of the $\mM_l$ regions becomes larger as $l$ becomes further upstream of the defect.
This explains the observed behavior in Figs.~\ref{fig:fc_dwm_dj_hd} and \ref{subfig:sotl_hd_density}, and also why the behavior is more pronounced in Fig~\ref{fig:var_turning}.

\section{Conclusions}
\label{sec:discussion}
We have studied the impact of a lane-reduction defect on the flows and densities of both one-dimensional and two-dimensional traffic networks, with a variety of boundary conditions.
We analyzed the transient behavior using an extended domain wall theory, in which multiple domain walls diffuse through the system, merging when they meet.
We then compared these predictions with simulations of a stochastic cellular automaton model.
The agreement between DW predictions and CA simulations is generally found to be excellent, in both one and two dimensions and for any initial state of the system.
Perhaps surprisingly, this is generically true even in realistic two-dimensional networks with adaptive traffic signal systems.

For one-dimensional systems, we  find that the defect reduces the capacity of the system by approximately half.
If the initial system is running below the capacity of the perturbed system, then the defect has almost no impact.
These observations have been made previously for ASEP with a sitewise defect~\cite{Kolomeisky98}.
We show that qualitatively the same behavior holds for a two-lane NaSch model with a lane-reduction defect.
The multiple domain wall model provides an accurate quantitative description of the disruption and recovery process in one dimension, even in its deterministic approximation.
We emphasize that this deterministic approximation can be expressed as a piecewise linear function of the drift velocities. 
It therefore provides an analytic expression for the transient profiles of flow and density, in terms of the stationary flows and densities in the unperturbed and perturbed systems.

For two-dimensional networks, the impact of the defect depends on both the traffic signal system as well as the location of the defect.
For the overpass network, whose capacity is close to that of the one-dimensional system, a mid-block defect reduces the network capacity significantly.
This is because the intersections themselves are much weaker defects than the lane reduction in this case.
By contrast, for the more realistic SOTL system which is governed by adaptive traffic signals, the unperturbed network capacity is approximately half of the one-dimensional system's capacity.
Introducing a mid-block lane-reduction defect in this case has negligible impact on the network, because the intersections act as the dominant defect.
The impact of the lane-reduction defect increases significantly when it is placed close to the intersection however. Likewise, we expect that spatially extended defects, not studied here, will have an increased impact.

The two-dimensional networks we studied were entirely {\em boundary loaded}; they included no internal sources and sinks.
Boundary loading generically leads to a link-dependent density distribution, even at stationarity.
By incorporating this link-dependence in the perturbed stationary densities the extended domain wall model is able to accurately reproduce the transient behavior of the defect route.
Perhaps surprisingly, unless the system is initially at capacity, 
we found that ignoring these heterogeneities in the unperturbed stationary states still led to very accurate predictions from the domain wall model.

Finally, we note that links in arterial networks generically have trapezoidal fundamental diagrams, as observed for SOTL.
As we have seen, the trapezoidal unperturbed fundamental diagram implies that if the system is initially at capacity, 
then its final recovery process corresponds to a domain wall performing a random walk with zero drift.
One would expect that this sort of behavior would therefore occur quite commonly in real arterial networks, and it would be of considerable interest for this question to be studied empirically.

\section*{Acknowledgment}
This work was supported under the Australian Research Council's Linkage Projects funding scheme (project number LP120100258),
and T.G. is the recipient of an Australian Research Council Future Fellowship (project number FT100100494).
This research was undertaken with the assistance of resources provided at the NCI National Computational Merit Allocation Scheme supported by the Australian Government. 
We also greatly acknowledge access to the computational facilities provided by the Monash Sun Grid.

\section*{References}
\bibliographystyle{elsarticle-num}
\bibliography{paper}

\appendix

\section{SOTL}
\label{app:solt}

SOTL is an acyclic signal system, in the sense that no fixed ordering of the phases is imposed.
Suppose we agree on a suitable demand function $d(\mP)$ which quantifies the demand of each phase $\mP$ of each given node.
Phases with large values of $d(\mP)$ should be candidates for being the next choice of the active phase.
However, one should also keep track of the time $\tau(\mP)$ that each phase has been idle, since we do not want a given phase to remain idle for too long, unless it has strictly zero demand.

The key idea behind SOTL is to compute a threshold function, $\kappa(\mP)$, for each phase $\mP$, which depends on both the phase's idle time and demand function,
and when $\kappa(\mP)$ reaches a predetermined threshold value,
$\kappa(\mP)>\theta,$
we consider making $\mP$ the active phase. For a detailed general discussion of the SOTL methodology, see~\cite{deGierGaroniRojas11}.

In the simulations performed in the current work, the demand $d(\mP)$ of phase $\mP$ was simply chosen to be the total number of vehicles over all its in-links, and the threshold function was
\be
\kappa(\mP)=\frac{d(\mP)\tau^c(\mP) e^{\frac{\tau(\mP)}{{\tau}_{\theta}}}}{\sum_{\mP'} d(\mP')},
\ee
This particular choice for the demand function implies that SOTL attempts, at each instant of time, to adaptively minimize the network's density heterogeneity. 
Instead of letting $\kappa(\mP)$ grow linearly with $\tau(\mP)$, we chose a threshold $\tau_\theta$ for the idle time. 
If $\tau(\mP)$ is larger than $\tau_\theta$, then the threshold function grows more quickly. 
This is intuitively reasonable, since after a driver has waited some {\em reasonable} amount of time for a green signal, the longer the driver has to wait further, the more impatient they become.
We used threshold values of $\theta=5$, $c=0.9$ and $\tau_\theta=50$.

\end{document}